\documentclass[traditabstract]{aa}  

\usepackage{amsmath}
\usepackage{fixltx2e}
\usepackage[english]{babel}
\usepackage{graphicx}
\usepackage{epstopdf}
\usepackage{epsf,color}
\usepackage[mathscr]{eucal}
\usepackage{amsmath}
\usepackage{amssymb,amsfonts}
\usepackage{natbib}
\usepackage{graphicx}
\usepackage{txfonts}
\usepackage{dsfont}
\definecolor{Mygreen}{rgb}{0.75, 0.0, 0.0}
\definecolor{Mypink}{rgb}{1.0, 0.0, 0.5}
\definecolor{Myred}{rgb}{0.7, 0.0, 0.0}
\usepackage[breaklinks, citecolor=blue, linkcolor=Myred, urlcolor=Myred, colorlinks=true]{hyperref}
\usepackage{float} 
\usepackage{color} 
\usepackage{ulem}
\usepackage{breqn}

\bibpunct{(}{)}{;}{a}{}{,}

\begin{document}

\title{{\tt MINOT}: Modeling the intracluster medium (non-)thermal content and observable prediction tools}
\author{R.~Adam \inst{\ref{LLR}}\thanks{Corresponding author: R\'emi Adam, \url{remi.adam@llr.in2p3.fr}}
\and H.~Goksu \inst{\ref{LLR}}
\and A.~Leing\"{a}rtner-Goth\inst{\ref{LLR}}
\and S.~Ettori\inst{\ref{INAF_bolo},\ref{INFN_bolo}}
\and R.~Gnatyk\inst{\ref{Kiev}}
\and B.~Hnatyk\inst{\ref{Kiev}}
\and M.~H\"utten\inst{\ref{MPI}}
\and J.~P\'erez-Romero\inst{\ref{IFT_UAM},\ref{DFT_UAM}}
\and M.~A.~S\'anchez-Conde\inst{\ref{IFT_UAM},\ref{DFT_UAM}}
\and O.~Sergijenko\inst{\ref{Kiev}}}

\institute{
Laboratoire Leprince-Ringuet, Ecole Polytechnique, CNRS/IN2P3, 91128 Palaiseau, France
\label{LLR}
 \and
INAF, Osservatorio di Astrofisica e Scienza dello Spazio, via Pietro Gobetti 93/3, 40129 Bologna, Italy
\label{INAF_bolo}
 \and
INFN, Sezione di Bologna, viale Berti Pichat 6/2, I-40127 Bologna, Italy
\label{INFN_bolo}
\and
Astronomical Observatory of Taras Shevchenko National University of Kyiv, 3 Observatorna Street, Kyiv, 04053, Ukraine
\label{Kiev}
\and
Max-Planck-Institut f\"ur Physik, F\"ohringer Ring 6, 80805 M\"unchen
\label{MPI}
\and
Instituto de F\'isica Te\'orica UAM-CSIC, Universidad Aut\'onoma de Madrid, C/ Nicol\'as Cabrera, 13-15, 28049 Madrid, Spain
\label{IFT_UAM}
\and
Departamento de F\'isica Te\'orica, M-15, Universidad Aut\'onoma de Madrid, E-28049 Madrid, Spain
\label{DFT_UAM}
}

\date{Last update: \today}
\abstract {
In the past decade, the observations of diffuse radio synchrotron emission toward galaxy clusters revealed cosmic-ray (CR) electrons and magnetic fields on megaparsec scales. However, their origin remains poorly understood to date, and several models have been discussed in the literature. CR protons are also expected to accumulate during the formation of clusters and probably contribute to the production of these high-energy electrons. In order to understand the physics of CRs in clusters, combining of observations at various wavelengths is particularly relevant.
The exploitation of such data requires using a self-consistent approach including both the thermal and the nonthermal components, so that it is capable of predicting observables associated with the multiwavelength probes at play, in particular in the radio, millimeter, X-ray, and $\gamma$-ray bands. We develop and describe such a self-consistent modeling framework, called {\tt MINOT} (modeling the intracluster medium (non-)thermal content and observable prediction tools) and make this tool available to the community.
{\tt MINOT} models the intracluster diffuse components of a cluster (thermal and nonthermal) as spherically symmetric. It therefore focuses on CRs associated with radio halos. The spectral properties of the cluster CRs are also modeled using various possible approaches. All the thermodynamic properties of a cluster can be computed self-consistently, and the particle physics interactions at play are processed using a framework based on the {\tt Naima} software. The multiwavelength observables (spectra, profiles, flux, and images) are computed based on the relevant physical process, according to the cluster location (sky and redshift), and based on the sampling defined by the user. With a standard personal computer, the computing time for most cases is far shorter than one second and it can reach about one second for the most complex models. This makes {\tt MINOT} suitable for instance for Monte Carlo analyses.
We describe the implementation of {\tt MINOT}  and how to use it. We also discuss the different assumptions and approximations that are involved and provide various examples regarding the production of output products at different wavelengths. As an illustration, we model the clusters Abell 1795, Abell 2142, and Abell 2255 and compare the {\tt MINOT} predictions to literature data.
While {\tt MINOT} was originally build to simulate and model data in the $\gamma$-ray band, it can be used to model the cluster thermal and nonthermal physical processes for a wide variety of datasets in the radio, millimeter, X-ray, and $\gamma$-ray bands, as well as the neutrino emission.}
\titlerunning{Modeling the ICM (non-)thermal content and observable prediction tools}
\authorrunning{R. Adam, H. Goksu and A. Leing\"{a}rtner-Goth}
\maketitle

\section{Introduction}
Galaxy clusters are the largest gravitationally bound structures that are decoupled from the expansion of the Universe. They form peaks in the matter density field. Their assembly has been driven by the gravitational collapse of dark matter \citep{Kravtsov2012}, which is thought to dominate the matter content of clusters (about 80\% in mass). Clusters also consist of baryonic matter, essentially in the form of hot ionized thermal plasma, called the intracluster medium (ICM; about 15\%), and of galaxies (about 5\%). While clusters are used to understand the formation of large-scale structures and to constrain cosmological models, they are also the place of very rich astrophysical processes and excellent targets for testing fundamental physics \cite[see, e.g.,][for a review]{Allen2011}.

Galaxy clusters form through the merging and accretion of other groups and surrounding material \citep{Sarazin2002}. This leads to the propagation of shocks and turbulences in the ICM \citep{Markevitch2007}, which can accelerate charged particles to very high energies. These cosmic rays (CRs) interact with the magnetized ICM, generating diffuse radio synchrotron emission \citep{Feretti2012,vanWeeren2019}, and they are expected to produce a $\gamma$-ray signal because of the inverse Compton interaction with background light or the decay of pions produced in proton-proton collisions \citep{Brunetti2014}. In addition, clusters also host active galactic nuclei (AGN), which are known to provide feedback onto the ICM \citep{Fabian2012}. This feedback is only poorly understood, but is expected to have a major effect on the formation and the evolution of galaxy clusters.

Diffuse radio synchrotron emission in galaxy clusters is generally classified as radio halos (including giant and mini-halos), radio relics, and revived AGN fossil plasma source \citep[see][for detailed discussions]{vanWeeren2019}. While relics are thought to be associated with the shock acceleration of electrons in the periphery of clusters, radio halos might originate from turbulent reacceleration of seed electrons and/or secondary electrons produced by hadronic interactions. $\gamma$-ray emission is also expected as a result of the inverse Compton emission that arises from the scattering of background photon fields onto relativistic electrons, or the hadronic interaction from CR protons (CRp) and the ICM \citep[see, e.g.,][for the signal expected based on numerical simulations]{Pinzke2010}. 

The annihilation or the decay of dark matter particles might also cause $\gamma$-ray emission from galaxy clusters \citep[see, e.g.,][]{Combet2018}, and many searches for this signal have been performed \citep[e.g.,][]{Ackermann2010,Aleksic2010,Arlen2012,Abramowski2012,Combet2012,Cadena2017,Acciari2018}. In the case of dark matter decay, galaxy clusters are particularly competitive targets because the signal scales linearly to the huge dark matter reservoirs in galaxy clusters. In the case of dark matter annihilation, clusters can be at the same flux level as dwarf galaxies when substructures are accounted for, and they are thus also highly relevant targets \citep{Sanchez2011,Moline2017}. However, the limits that can be set on the properties of dark matter depend on the uncertainties associated with the modeling of the background emission, so that accurate CR modeling is also essential for dark matter searches.

Many attempts to detect the cluster $\gamma$-ray emission have been made using ground-based \citep[e.g.,][at 50 GeV - 10 TeV energies]{Aharonian2009,Aleksic2012,Arlen2012,Ahnen2016} and space-based observations \citep[e.g.,][at about 30 MeV - 300 GeV]{Reimer2003,Huber2013,Prokhorov2014,Zandanel2014,Ackermann2014,Ackermann2015,Ackermann2016}. While unsuccessful so far, these searches were very useful to constrain the CR physics and particle acceleration at play in clusters, especially when combined with radio observations \citep[e.g.,][]{Vazza2015,Brunetti2017}. Recently, \cite{Xi2018} claimed the first significant detection of $\gamma$-ray signal toward the Coma cluster using data obtained with the \textit{Fermi}-Large Area Telescope (\textit{Fermi}-LAT). However, their results might be confounded by a possible point source because the signal-to-noise ratio and angular resolution of the observations are limited. While the \textit{Fermi}-LAT satellite is to continue to take data for several additional years (compared to about the 12 years of data collected so far), major discoveries concerning galaxy clusters are unlikely given the modest increase in statistics that is expected. From the ground, the Cherenkov Telescope Array \citep[CTA,][]{CTA2019} is expected to provide a major improvement in sensitivity in the 100 GeV - 100 TeV energy range.

 In order to address the CR physics in galaxy clusters, multiwavelength observations and analyses are becoming particularly relevant with the construction of such new facilities. While cluster CRs can essentially be accessed the radio and $\gamma$-ray bands, their physics is driven by the continuous interaction with the thermal plasma. When data are compared to modeling, or when mock observations are generated, the thermal and the nonthermal components should therefore be modeled together, in a self-consistent way, so that uncertainties and degeneracies between the two can be accounted for. The thermal emission can be probed in particular in the X-ray and at millimeter wavelengths through thermal Bremsstrahlung emission \citep{Sarazin1986,Bohringer2010} and the thermal Sunyaev-Zel'dovich (tSZ) effect \citep{Sunyaev1970,Sunyaev1972}. In addition to the primary components, modeling the particle interactions in the ICM relies on particle physics data from accelerators or a sophisticated Monte Carlo code \citep[see, e.g.,][for discussions]{Kafexhiu2014}, and they need to be accounted for carefully. Nevertheless, clusters are commonly modeled focusing on individual (or just a few) components, and no public self-consistent multiwavelength software exists in the literature. For instance, \cite{Li2019} and \cite{Bruggen2020} recently modeled the diffuse radio synchrotron emission of radio halos and radio relics, respectively, and employed the Press-Schechter formalism to estimate the statistical properties of the corresponding signal.

Here, we present a software dedicated to the self-consistent modeling of the thermal and nonthermal diffuse components of galaxy clusters, for which the main objective is computing accurate and well-characterized multiwavelength predictions for the radio, millimeter, X-ray, $\gamma$-ray, and neutrino emission. This software is called {\tt MINOT,} modeling the intracluster medium (non-)thermal content and observable prediction tools. It is based on the Python language and is available at the following url: \url{https://github.com/remi-adam/minot}. {\tt MINOT} includes various parameterizations for the radial profiles and spectral properties of the different cluster components. The code does not aim at computing the CR production rate from microphysics considerations (e.g., turbulence, shocks, or diffusion), but instead directly models the spatial and spectral distributions of the CRs and the thermal gas. The predictions for associated observables are available in the radio (synchrotron), millimeter (tSZ effect),  X-ray (thermal Bremsstrahlung), $\gamma$-ray (inverse Compton and hadronic processes), and also for neutrino emissions (hadronic processes). This includes surface brightness profiles or maps, spectra, and integrated flux computed with different options. For $\gamma$-rays, CR electrons (CRe), and neutrinos from hadronic origin, {\tt MINOT} includes the latest description of the hadronic interactions in the ICM, based on the {\tt Naima} software \citep{Zabalza2015}. The thermal modeling uses the {\tt XSPEC} software for X-ray predictions \citep{Arnaud1996}, and it includes an accurate description of the tSZ signal up to high plasma temperatures.

This article is organized as follows. In Section \ref{sec:overview} we provide a general overview of the code and discuss the different interfaces. Section \ref{sec:physical_modeling} discusses the physical modeling of the cluster components. The physical processes related to particle interactions are detailed in Section \ref{sec:particle_interactions}. In Section \ref{sec:observables} we discuss the prediction of observables in the relevant energy bands. The use of {\tt MINOT} is illustrated in Section~\ref{sec:comparison} for three nearby massive well-known clusters for which multiwavelength data are available in the literature. Finally, Section \ref{sec:conclusions} provides a summary and conclusion. Equations are given following the international system of units.

\section{General overview and structure of the code}\label{sec:overview}
\begin{figure*}
\centering
\includegraphics[width=0.85\textwidth]{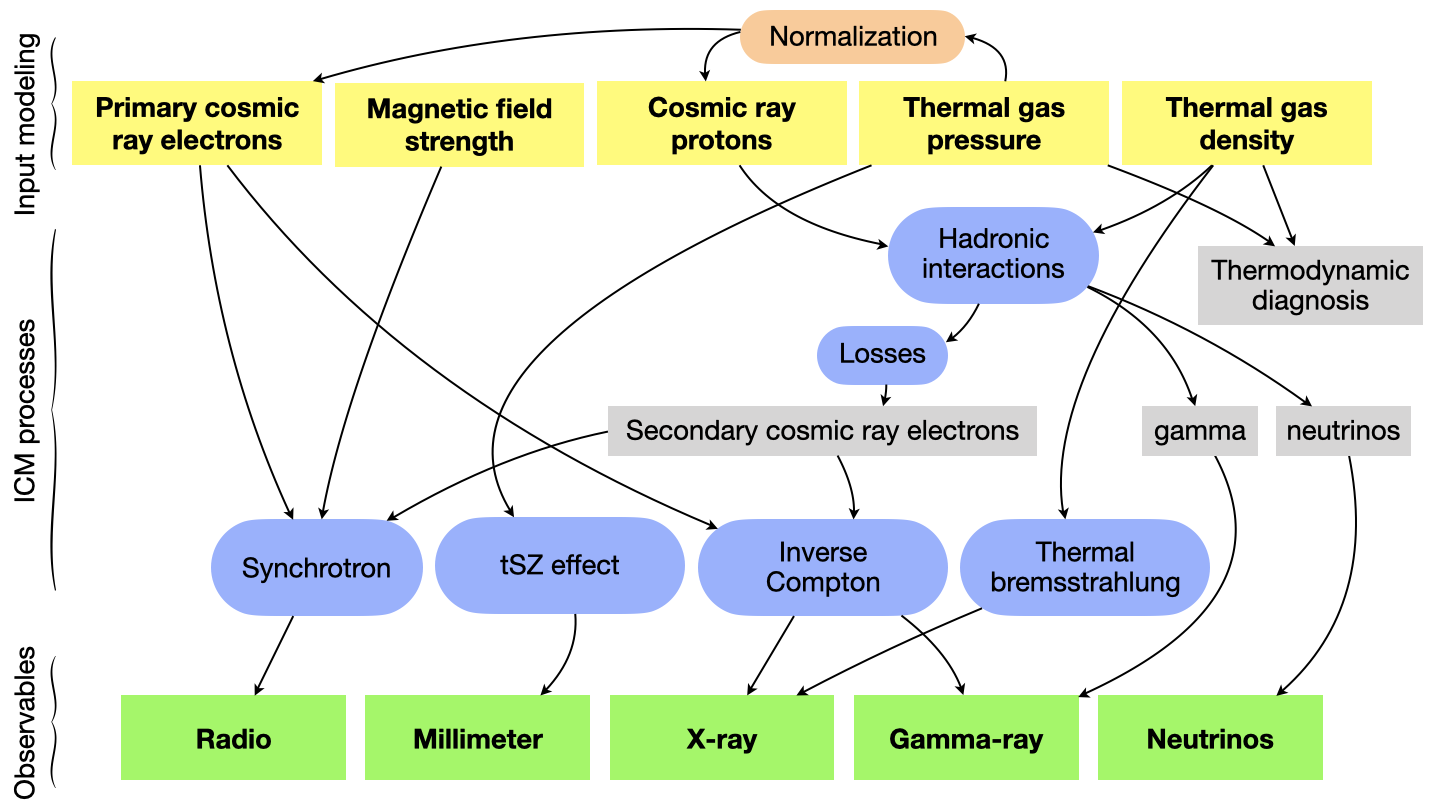}
\caption{Overview of the {\tt MINOT} input modeling, the considered physical processes at play in the ICM, and the observables that are computed. The interdependences are shown by the black arrows.}
\label{fig:code_overview1}
\end{figure*}

{\tt MINOT} is a Python-based code available at \url{https://github.com/remi-adam/minot}\footnote{Several Python notebook examples are also available. In particular, the notebook 'demo\_plot.ipynb' has been used to generate the figures of this paper.}. It essentially depends on standard Python libraries, but some functionalities require specific softwares and packages, as discussed below. In this section, we provide a general overview of the working principle of the code, of its structure, and the interactions between the different modules. The list of the code parameters is also discussed, as well as the available functional forms for the radial and spectral models. Figure~\ref{fig:code_overview1} highlights how the input modeling is used to generate observables via the different plasma processes considered in {\tt MINOT}, and the general overview of the code is illustrated in Fig.~\ref{fig:code_overview2}. 

\subsection{Overview of the physical modeling}
{\tt MINOT} was first developed to compute an accurate $\gamma$-ray prediction for galaxy clusters. As discussed below and shown in Fig.~\ref{fig:code_overview1}, this requires several key ingredients. Because the same ingredients also provide diagnoses for other observables at different wavelengths through various physical processes, {\tt MINOT} was further developed to account for them. This allowed us to provide external constraints to a given input modeling that is used to generate $\gamma$-ray observables, but also to provide further diagnosis of the physical state of the cluster. 

First, the spatial and spectral distributions of primary CRe (CRe$_1$) and protons are crucial. They generate $\gamma$-rays through inverse Compton scattering on the cosmic microwave background (CMB), or through hadronic interactions, respectively. Modeling the thermal gas is also essential because hadronic processes arise from the interaction between CRp and thermal plasma. As we show in Section~\ref{sec:physical_modeling}, the thermal component is based on the thermal electron pressure and density. Additionally, the normalization of the CR distributions is generally given relative to the thermal energy. The hadronic interactions also generate secondary CRe and positrons (CRe$_2$). Because they are affected by synchrotron losses, they require that the magnetic field is accounted for as another key ingredient of the input modeling (Section~\ref{sec:particle_interactions}). These electrons contribute to the inverse Compton emission (Section~\ref{sec:observables}). In summary, the necessary input modeling ingredients are CRe$_1$, magnetic field strength, CRp, and thermal electron pressure and density.

With these ingredients at hand, the radio emission that arises from CRe (primary and secondary) moving in the magnetic field can be modeled. Similarly, the thermal pressure and density allow us to compute the tSZ signal and the thermal Bremsstrahlung X-ray emission. They also provide a complete diagnosis of the thermodynamic properties of the cluster. Neutrinos are also produced during hadronic interactions, and their associated observable is thus available.

\subsection{Code structure}
\begin{figure*}
\centering
\includegraphics[width=0.7\textwidth]{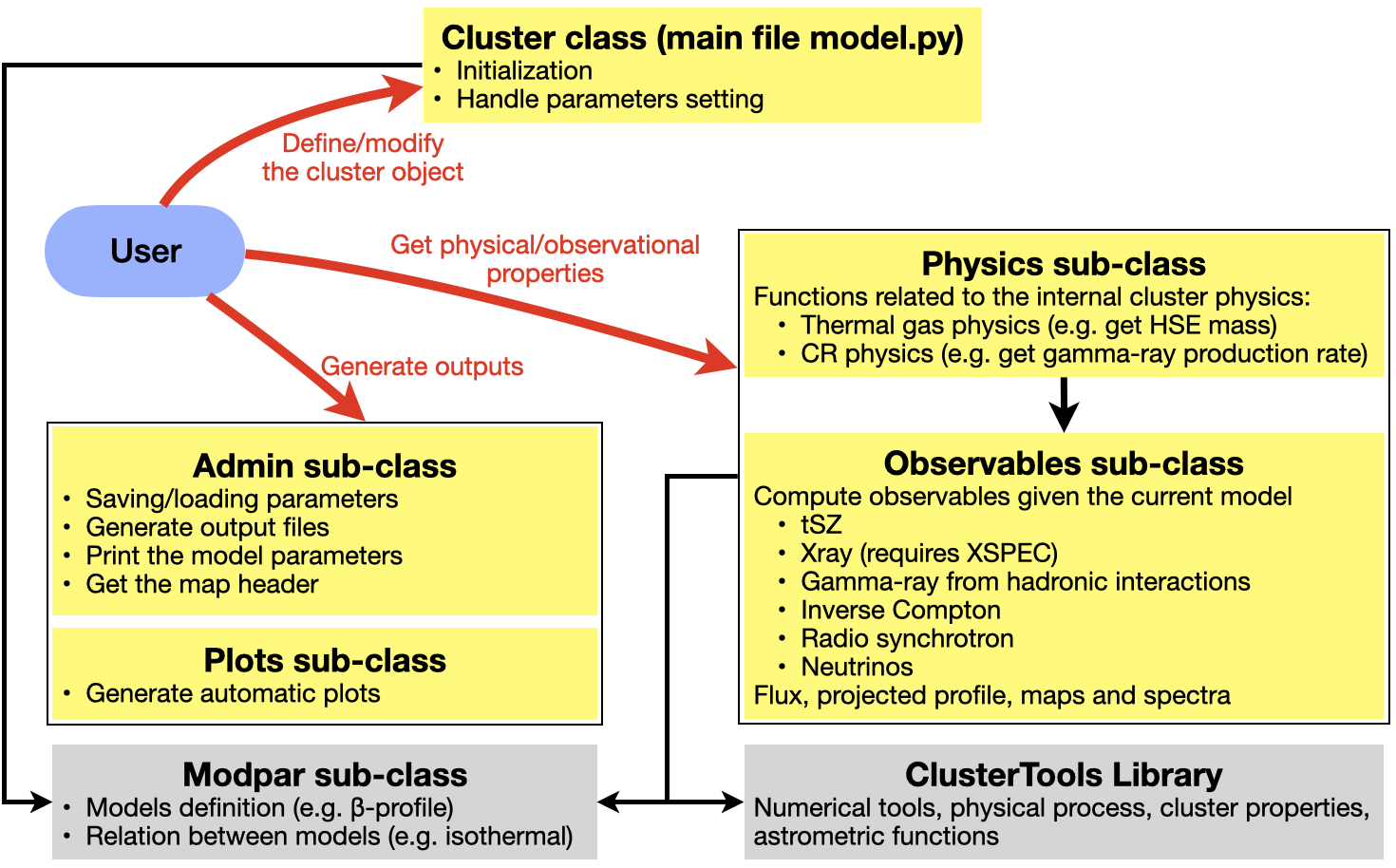}
\caption{Overview of the code structure and interfaces.}
\label{fig:code_overview2}
\end{figure*}
In order to model galaxy clusters and predict observables associated with the diffuse thermal and nonthermal components, {\tt MINOT} is organized in six main parts, each of which is related to specific functions and procedures.
We list these parts below.\begin{enumerate}
\item The main class, called {\tt Cluster}, is written in the file {\tt model.py} and provides an entry point for the user. It allows defining the model and solve entanglement between parameters.
\item A subclass called {\tt Admin} allows us to handle administrative tasks, in particular, input and output procedures.
\item The {\tt Modpar} subclass is dedicated to the model parameterization. It gathers a library of available radial and spectral models.
\item The physical modeling of the cluster is performed in the {\tt Physics} subclass. It includes many functions for retrieving the desired physical quantities.
\item The {\tt Observable} subclass allows us to extract the requested cluster observables based on the inner physics encoded in the model.
\item Finally, the subclass {\tt Plots} is designed for automated plots to provide a cluster diagnostic based on the current modeling.
\end{enumerate}
In addition to these six main parts, {\tt MINOT} also includes a library called {\tt ClusterTools} in which numerical tools and astrometric tools are defined. It also includes several classes that are used to compute various physical processes relevant for {\tt MINOT}. In the following subsections, we describe the working principle of the different functionalities. 

\subsection{Initialization and parameters}
As illustrated in Fig.~\ref{fig:code_overview2}, the user can directly define a cluster object by calling the {\tt Cluster} (main) class of {\tt MINOT}, 
\begin{verbatim}
cluster = minot.Cluster(optional parameters).
\end{verbatim}
Optional parameters, such as cluster name, coordinates, or redshift, can be passed directly to the initialization call. However, any parameter can be modified on the fly, for example,
\begin{verbatim}
cluster.redshift = 0.1
\end{verbatim}
The entanglement of parameters is solved in the code. For instance, changing the cluster redshift will automatically change the angular diameter distance of the cluster according to the current cosmological model. Information is provided to the user when the 'silent' parameter is set to 'False'. The list of parameters that describe the cluster object is available in Table~\ref{tab:table_parameters}. We note that whenever possible, the code uses {\tt Astropy} units for quantities\footnote{\url{https://docs.astropy.org/en/stable/units/}}.

We can distinguish four types of parameters, as listed in Table~\ref{tab:table_parameters}. The first type corresponds to administration-like parameters (e.g., output directory used to generate products), the second type concerns the global properties of the cluster object (e.g., the redshift), the third type is related to the radial and spectral modeling of the physical quantities of interest of the cluster (e.g., CR number density profile), and the last type allows the user to  sample the output observables. In particular, it is possible to set a map header (e.g., obtained for real data) on which the model prediction maps are projected, which facilitates a comparison of data and model.

\begin{table*}
        \caption{List of the parameters involved in the code.}
        \begin{center}
        \begin{tabular}{cccc}
        \hline
        \hline
        Parameter &Type & Default value & Description \\
        \hline
        \hline
        \multicolumn{4}{c}{Administrative parameters} \\
        \hline
        silent               & Boolean & False & Allows {\tt MINOT} to provide information when running \\
        output\_dir       & string     & './minot\_output' & Full path to the output directory for products saving \\
        \hline
        \multicolumn{4}{c}{Global physical properties} \\
        \hline
        cosmo            & cosmology$^{(a)}$ & Planck15 & Cosmological model \\
        name              & string & 'Cluster' & Name of the cluster \\
        coord              & SkyCoord$^{(b)}$ & $[0,0]$ deg & Coordinates of the cluster center \\
        redshift           & float & 0.01 & redshift of the cluster \\
        D\_ang             & quantity$^{(c)}$ & $(\star)$ & Angular diameter distance \\
        D\_lum             & quantity$^{(c)}$ & $(\star)$ & Luminosity distance \\
        M500               & quantity$^{(c)}$ & $10^{15}$ M$_{\sun}$ & Characteristic cluster mass \\
        R500                & quantity$^{(c)}$ & $(\star)$ & Characteristic cluster physical radius \\
        theta500           & quantity$^{(c)}$ & $(\star)$ & Characteristic cluster angular radius \\
        R\_truncation     & quantity$^{(c)}$ & $3 \times R_{500}$ & Physical extent (boundary) of the cluster \\
        theta\_truncation &quantity$^{(c)}$ & $3 \times \theta_{500}$ & Angular extent (boundary) of the cluster \\     
        helium\_mass\_fraction & float & 0.2735 & Helium mass fraction \\
        metallicity\_sol               & float & 0.0153 & Reference metallically of the Sun \\
        abundance                    & float & 0.3 & The metal abundance relative to the solar value \\
        EBL\_model                   & string & 'dominguez' & Name of the extragalactic background light model \\
        hse\_bias                       & float & 0.2 & Hydrostatic mass bias \\
        Epmin  &quantity$^{(c)}$ & $\sim 1.22$ GeV & Minimal energy of the CRp \\
        Epmax  &quantity$^{(c)}$ & 10 PeV & Maximal energy of the CRp \\ 
        Eemin  &quantity$^{(c)}$ & $m_e c^2$ & Minimal energy of the CRe$_1$ \\ 
        Eemax  & quantity$^{(c)}$ & 10 PeV & Maximal energy of the CRe$_1$ \\ 
        pp\_interaction\_model & string & 'Pythia8' &Name of the proton-proton interaction model \\
        cre1\_loss\_model & string & 'None' & Loss model to apply to the input primary CRe distribution \\
        \hline        
        \multicolumn{4}{c}{Radial and spectral modeling} \\
        \hline
        pressure\_gas\_model & dict & P13UPP$^{(d)}$ & Model to be used for the thermal gas pressure profile \\
        density\_gas\_model  & dict & $\frac{P_e(r)}{10 \ {\rm keV}}$ & Model to be used for the thermal gas number density profile \\
        magfield\_model      & dict & $\frac{P_e(r)}{P_e(10 \ {\rm kpc})} \times 10 \ \mu$G & Model to be used for the magnetic field profile \\
        X\_crp\_E                 & dict & 1\% within $R_{500}$ & CRp to thermal energy ratio and reference radius \\
        X\_cre1\_E               & dict & 1\% within $R_{500}$ & CRe$_1$ to thermal energy ratio and reference radius \\
        density\_crp\_model & dict & $\propto P_e(r)$ & Model to be used for the CRp number density profile \\
        density\_cre1\_model  & dict & $\propto P_e(r)$ & Model to be used for CRe$_1$ number density profile \\
        spectrum\_crp\_model & dict & Index 2.5 power law & Model to be used for the CRp spectrum \\
        spectrum\_cre1\_model & dict & Index 3.0 power law & Model to be used for CRe$_1$ spectrum \\
        \hline
        \multicolumn{4}{c}{Sampling parameters} \\
        \hline
        Rmin                              & quantity$^{(c)}$ & 1 kpc & Minimum radius used for log integration \\
        Npt\_per\_decade\_integ & int                                              & 30 & Number of points per decade \\
        map\_coord                    & SkyCoord$^{(b)}$ & $[0, 0]$ deg & Coordinates of the map center \\
        map\_reso                      & quantity$^{(c)}$ & 0.01 deg & Map resolution (pixel size) \\
        map\_fov                         & quantity$^{(c)}$ list & $[5, 5]$ deg & Map field-of-view size along R.A. and Dec. \\
        map\_header                  & string     & None & Header of the map \\
        \hline
        \end{tabular}
        \end{center}
        {\small {\bf Notes.} $^{(a)}$ From the {\tt astropy} package. $^{(b)}$ From the {\tt astropy.coordinates} package. $^{(c)}$ From the {\tt astropy.units} pacakge. $^{(d)}$ Universal pressure profile based on mass and redshift, from \cite{Planck2013V}. $(\star)$ Quantities that are computed from other parameters.}
        \label{tab:table_parameters}
\end{table*}

\subsection{Modeling the physical state of the cluster}
The parameters describing the cluster can be divided into two types: global properties that apply to the entire cluster (e.g., mass, redshift, and coordinates), and properties that vary as a function of radius or energy. This separation is highlighted in Table~\ref{tab:table_parameters}. Some parameters are assumed to be constant over the entire cluster volume, such as the hydrostatic mass bias or the metal abundances.

\begin{figure*}
\centering
\includegraphics[width=0.42\textwidth]{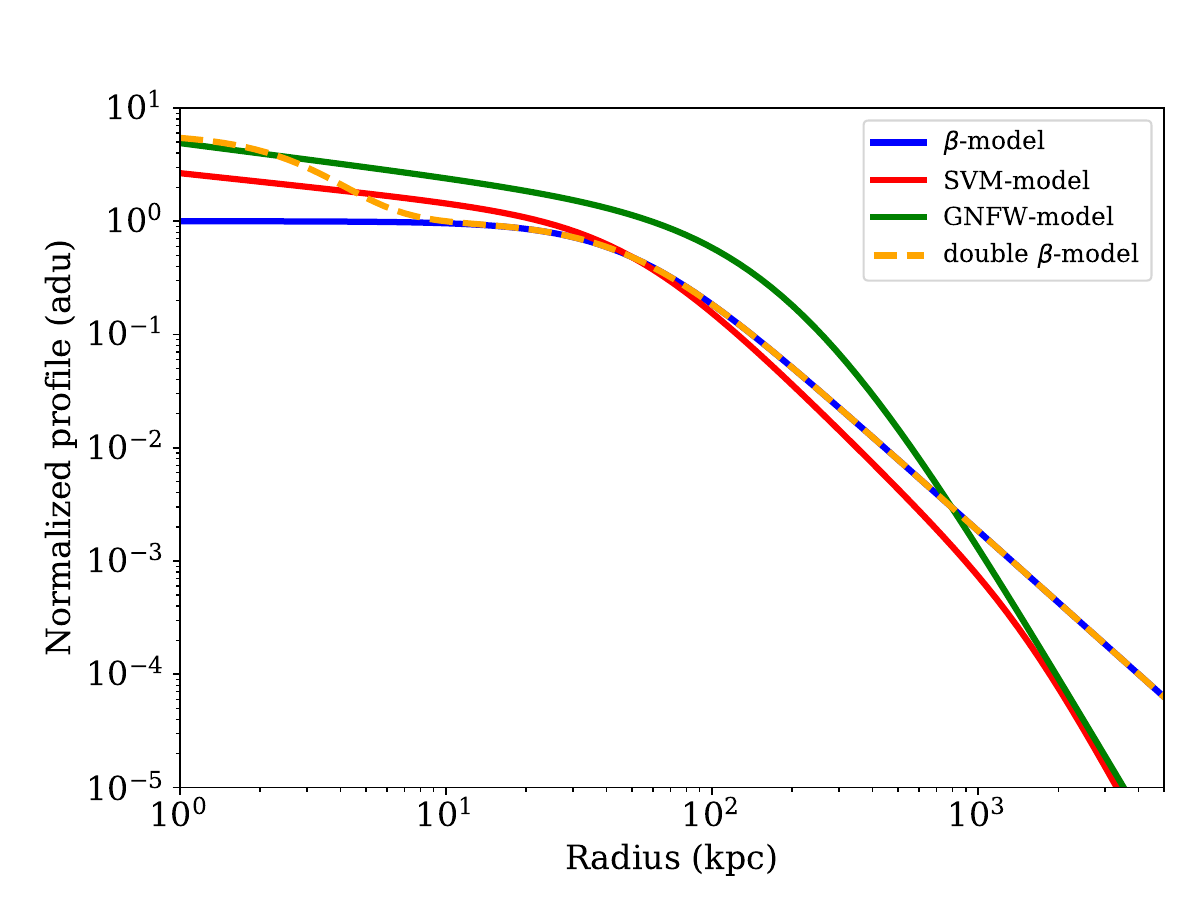}
\includegraphics[width=0.42\textwidth]{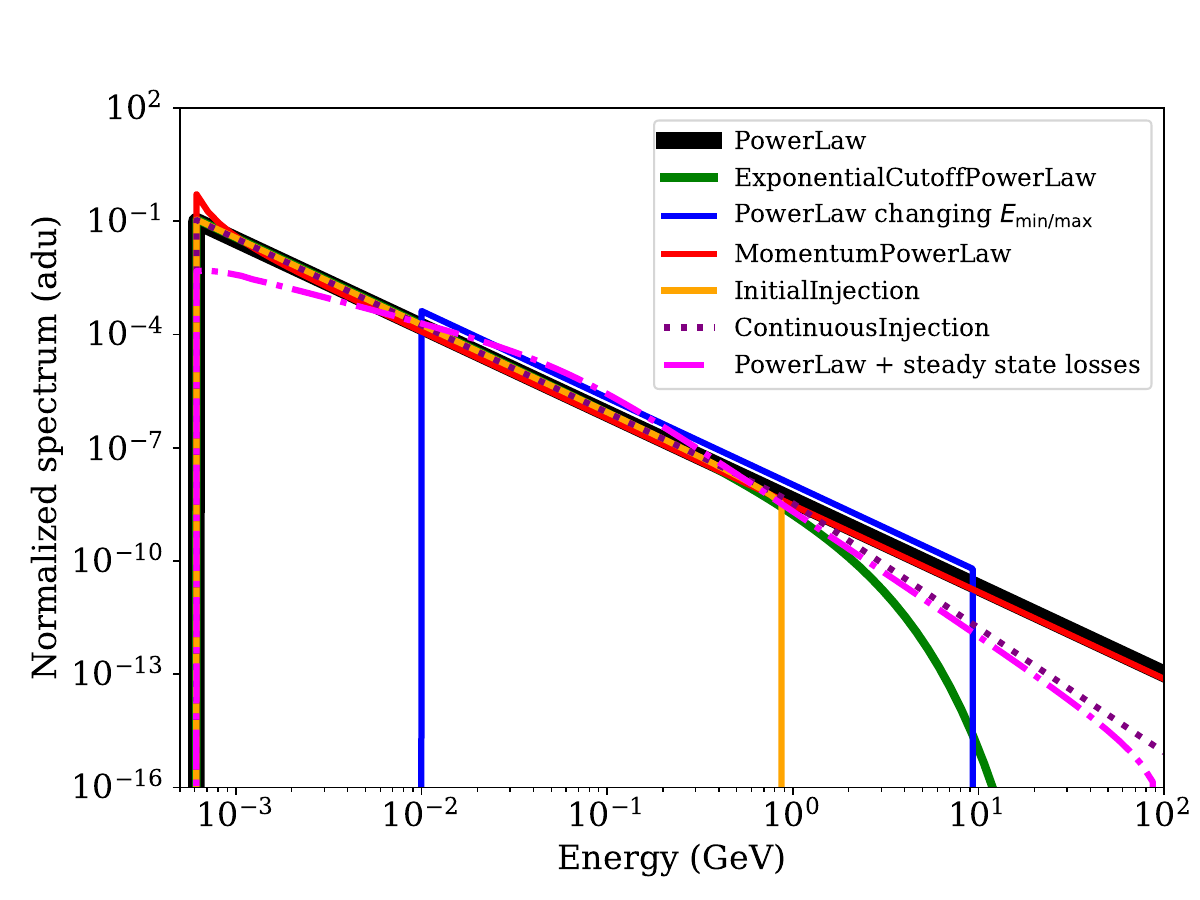}
\caption{{\bf Left}: Illustration of the different radial profiles available in the library. 
The $\beta$-model parameters are $(n_0, r_c, \beta) = (1, 50 \ {\rm kpc}, 0.7)$ ; the SVM model parameters are $(n_0, r_c, \beta, r_s, \gamma, \epsilon, \alpha) = (1, 50 \ {\rm kpc}, 0.7, 1500 \ {\rm kpc}, 3, 4, 0.5)$ ; the GNFW model parameters are $(P_0, r_p, a, b, c) = (1, 200 \ {\rm kpc}, 1.5, 4, 0.3)$ ; and the double $\beta$-model parameters are $(n_{01}, r_{c1}, \beta_1,(n_{02}, r_{c2}, \beta_2) = (1, 50 \ {\rm kpc}, 0.7, 5, 5 \ {\rm kpc}, 2)$. {\bf Right}: Illustration of the different spectral models available in the library. The index is set to $\alpha = 2.3$ for all models, and the cutoff or break energy to $E_{\rm break/cut} = 1$ GeV. In both cases, the 'User' model is not shown, but allows passing any arbitrary function that is interpolated in the code.}
\label{fig:profiles_spectra}
\end{figure*}

In addition to the global properties, the primary quantities that are used to define the physical state of the cluster are (see also Section~\ref{sec:physical_modeling} for further details) the gas pressure of thermal electrons, the gas number density of thermal electrons, the CRp number density profile and spectrum, the CRe$_1$ profile and spectrum, and the profile of the magnetic field strength. The CR distributions are normalized according to the ratio of CR and thermal energy enclosed within a given radius. The physical modeling of the radial and spectral properties of the cluster relies on a library of predefined models in the {\tt Modpar} subclass of {\tt MINOT}. The list of models that are currently available in the code is given in Tables~\ref{tab:table_spectral_models} and~\ref{tab:table_spatial_models} for the spectral and spatial component, respectively. Figure~\ref{fig:profiles_spectra}  illustrates the shape of the different models; they are further discussed in a more physical context in Section~\ref{sec:physical_modeling}. We note that the spatial and spectral parts of the modeling are currently decoupled (e.g., the spectrum of CRp does not change with radius), such that a physical quantity $f$ can be expressed as
\begin{equation}
        f(r, E) \propto f_1(r) f_2(E),
\end{equation}
where $E$ is the particle energy (only relevant for the CRs), and $r$ is the physical radius in three dimensions. However, functions that couple the radius and the energy dependence are ready to be implemented in the model library because any calculations relying on the modeling of $f(r, E)$ are made on 2D grids (energy versus radius) that are ignorant of the underlying parameterization of the distributions. In addition, it is possible to apply some losses, assuming a given scenario, to the input distribution. In this case, $f(r,E)$ is considered as an injection rate, and the output distribution is affected differently for different energy and radii. The implementation of the losses is discussed in detail in Section~\ref{sec:energy_loss}. 

A new model is set to a given physical property by passing a Python dictionary, such as
\begin{verbatim}
cluster.density_gas_model =
----- {'name':'beta', 
----- 'n_0':1e-3*u.cm**-3, 
----- 'beta':0.7, 'r_c':300*u.kpc}
\end{verbatim}
or
\begin{verbatim}
cluster.spectrum_crp_model =
----- {'name':'PowerLaw', 
----- 'Index': 2.5}
\end{verbatim}
It is also possible to automatically set a parameterization of several quantities to predefined physical states without directly setting the model parameters, for instance, forcing the CRp to follow the radial distribution of the gas density,
\begin{verbatim}
cluster.set_density_crp_isodens_scal_param()
\end{verbatim}
or to define the thermal electron number density based on the thermal electron pressure in the case of an isothermal cluster with a given temperature. These functions are written as part of the {\tt Modpar} subclass.

\begin{table*}
        \caption{List of spectral models.}
        \begin{center}
        \begin{tabular}{c|c|c}
        \hline
        \hline  
        Model name & Function & Dictionary keys \\
        \hline
        PowerLaw  & $f(E) = A \times \left(\frac{E}{E_0}\right)^{-\alpha}$ & 'name', 'Index' \\
        ExponentialCutoffPowerLaw & $f(E) = A \times \left(\frac{E}{E_0}\right)^\alpha \times \rm{exp} \left(-\frac{E}{E_{\rm cut}}\right)  $ & 'name', 'Index', 'CutoffEnergy\\
        MomentumPowerLaw  & $f(p) = A \times \left(\frac{p}{p_0}\right)^{-\alpha}$, with $E^2=p^2c^2+m^2c^4 = \left(E_{\rm kin}+mc^2\right)^2$ & 'name', 'Index', 'Mass' \\       
        InitialInjection & $f(E) = A \left(\frac{E}{E_0}\right)^{-\alpha} \begin{cases} (1 - E/E_{\rm break})^{\alpha - 2} &E < E_{\rm break}, \alpha \geqslant 2 \\
                                                                                                0 &E \geqslant E_{\rm break}
                                                                                                \end{cases}$ & 'name', 'Index', 'BreakEnergy' \\
        ContinuousInjection  & $f(E) = A \left(\frac{E}{E_0}\right)^{-(\alpha+1)} \begin{cases} 1-(1 - E/E_{\rm break})^{\alpha - 1} &E < E_{\rm break} \\
                                                                                                1 &E \geqslant E_{\rm break}
                                                                                                \end{cases}$ & 'name', 'Index', 'BreakEnergy' \\
        User  & $f(E) = {\rm anything}$ & 'name', 'User', 'energy', 'spectrum' \\
        \hline
        \end{tabular}
        \end{center}
        {\footnotesize Note: In addition to these models, the parameter cre1\_loss\_model allows applying an energy loss to the given parameterization, thus modifying its energy distribution (with a radial dependence). In this case, the parameterizations given here correspond to the injection rate $q(E,r)$ and not to the actual CR distribution $J_{\rm CR} \equiv \frac{dN_{\rm CR}}{dEdV}$ given in Eq.~\ref{eq:Jcr}. See Section~\ref{sec:secondary_electrons_in_steady_state} for further details, and in particular Eq.~\ref{eq:steady_state_computation_cre} for the steady-state scenario.}
        \label{tab:table_spectral_models}
\end{table*}

\begin{table*}
        \caption{List of spatial models.}
        \begin{center}
        \begin{tabular}{c|c|c}
        \hline
        \hline
        Model name & Function & Dictionary keys \\
        \hline
        \hline
        GNFW          & $f(r) = \frac{P_0}{\left(\frac{r}{r_p}\right)^c \left(1+\left(\frac{r}{r_p}\right)^a\right)^{\frac{b-c}{a}}}$, with $r_p = R_{500}/c_{500} $& 'name', 'P\_0', 'c500' or 'r\_p', 'a', 'b', 'c'\\
        SVM              & $f(r) = n_{0} \left[1+\left(\frac{r}{r_c}\right)^2 \right]^{-3 \beta /2}  \left(\frac{r}{r_c}\right)^{-\alpha/2} \left[ 1+\left(\frac{r}{r_s}\right)^{\gamma} \right]^{-\epsilon/2 \gamma}$  & 'name', 'n\_0', 'beta', 'r\_c', 'r\_s', 'alpha', 'gamma', 'epsilon' \\ 
        beta               & $f(r) = n_{0} \left[1+\left(\frac{r}{r_c}\right)^2 \right]^{-3 \beta /2} $  & 'name', 'n\_0', 'beta', 'r\_c' \\ 
        doublebeta    & $f(r) = n_{01} \left[1+\left(\frac{r}{r_{c1}}\right)^2 \right]^{-3 \beta_1 /2} + n_{02} \left[1+\left(\frac{r}{r_{c2}}\right)^2 \right]^{-3 \beta_2 /2}$ & 'name', 'n\_02', 'beta1', 'r\_c1', 'n\_02', 'beta2', 'r\_c2'\\ 
        User  & $f(r) = {\rm anything}$ & 'name', 'User', 'radius', 'profile' \\
        \hline
        \end{tabular}
        \end{center}
        \label{tab:table_spatial_models}
\end{table*}

\subsection{Derived physical properties and observables}
When the desired physical properties of the cluster are set, functions related to the physical description of the cluster, from the subclass {\tt Physics}, can be called to extract the thermodynamic and CR properties of the cluster, or the production rate of nonthermal particles (the physical modeling is further detailed in Section~\ref{sec:physical_modeling}), for example, extracting the hydrostatic mass profile, or the neutrino emission rate via
\begin{verbatim}
r, M = cluster.get_hse_mass_profile()
dN_dEdVdt = cluster.get_rate_neutrino().
\end{verbatim}

The user may also generate observables corresponding to the radio synchrotron emission, the tSZ signal in the millimeter, the thermal Bremsstrahlung in the X-ray, the inverse Compton emission, and the hadronic emission in the $\gamma$-rays, or the associated neutrino emission (see also Section~\ref{sec:observables} for more details). This is implemented in the subclass {\tt Observable} of {\tt MINOT}, for instance,\begin{verbatim}
E, dN_dEdSdt = cluster.get_gamma_spectrum().
\end{verbatim}

We note that many of the functions or subclasses related to the physical processes that are called when a function like this is run are gathered in the {\tt ClusterTools} subdirectory. This also includes many numerical tools.

\subsection{Administrative functions}
When it is defined, the cluster object also includes various administrative functions, gathered in the subclass {\tt Admin}. They can be used to display the current values of the parameters, to save the current status of the cluster object (or load a previously saved model), to generate output observable products automatically (maps, profiles, and spectra), or to get the header of the current map. The generation of automatic plots corresponding to the various observables included in {\tt MINOT} is also available using the subclass {\tt Plots}.

\subsection{Baseline model and test clusters}\label{sec:baseline_cluster_models}
\begin{table*}
        \caption{Main physical properties of the clusters used in this work for illustration. The 'Baseline' case does not correspond to a real cluster.}
        \begin{center}
        \begin{tabular}{c|c|c|c|c|c|c}
        \hline
        \hline
        Name & redshift & R.A., Dec. & $M_{500}$ & Dynamical state & Cool-core & Radio emission \\
        \hline
        Baseline & 0.02 & -- &  $7 \times 10^{14}$ M$_{\odot}$ & Disturbed & no & -- \\
        A1795 & 0.0622 & 207.21957, 26.589602 deg & $4.6 \times 10^{14}$ M$_{\odot}$ & Relaxed & yes & Mini-halo \\
        A2142 & 0.0900 & 239.58615, 27.229434 deg & $9.0 \times 10^{14}$ M$_{\odot}$ & Disturbed/elongated & yes & Giant halo \\
        A2255 & 0.0809 & 258.21604, 64.063058 deg & $5.3 \times 10^{14}$ M$_{\odot}$ & Disturbed & no & Giant halo + relic \\
        \end{tabular}
        \end{center}
        \label{tab:xcop_properties}
\end{table*}

In the following sections, we use different cluster models to illustrate the behavior of the {\tt MINOT} code. First, we define a baseline cluster model using parametric functions in order to show the effect of changes in the modeling on the observables. The baseline properties of the cluster were set using a generalized Navarro-Frenk-White profile \citep[GNFW,][]{Nagai2007} thermal electron pressure profile with $\left(P_0, c_{500}, a, b, c\right) = \left(2.2\times10^{-2} \ {\rm keV cm}^{-3}, 3.2, 1,5, 3.1, 0.0\right)$, and an SVM thermal electron number density profile with $\left(n_0, r_c, \beta, \alpha, r_s, \gamma, \epsilon\right) = \left(3 \times 10^{-3} \ {\rm cm}^{-3}, 290 \ {\rm kpc}, 0.6, 0.0, 1000 \ {\rm kpc}, 3, 1.7\right)$. This corresponds to a typical massive merging cluster. The redshift was set to $z=0.02$ and the mass to $M_{500} = 7 \times 10^{14}$ M$_{\odot}$, inspired by the Coma cluster. The CRp followed an exponential cutoff power-law spectrum, with spectral index 2.4 and a cutoff energy of 100 PeV. The normalization was set to have a CRp-to-thermal energy ratio within $R_{500}$ of $10^{-2}$, which corresponds to the typical expected values, according to \cite{Pinzke2010}. The CRe$_1$ followed a continuous injection spectrum, with an injection spectral index 2.3 and a break energy of 5 GeV. The normalization was set to have a CRe$_1$ -to-thermal energy ratio within $R_{500}$ of $10^{-5}$, that is, about the proton value scaled by the proton-to-electron mass ratio (i.e., a similar Lorentz factor distribution was assumed for the two). The spatial profiles of both CRp and CRe$_1$ were set to the same shape as the thermal gas density (see Section~\ref{sec:physics_thermal_component}). The magnetic field profile was set to follow the square root of the thermal gas density and was normalized to have an amplitude of 5 $\mu$G, assuming similar properties as those measured for the Coma cluster \citep{Bonafede2010}. This model, referred to as {\tt Baseline} in the following, was varied whenever the effect of relevant quantities to the physical state  of the cluster or observable are illustrated.

In addition to this baseline model, it is also useful to use real clusters, with the aim of comparing our model predictions to measurements available in the literature. To do so, we need clusters whose thermal properties have been measured and are available, over a wide range of spatial scales, in order to calibrate our model as well as possible. The sources targeted by the {\tt XMM Cluster Outskirt project}\footnote{XCOP, see \url{https://dominiqueeckert.wixsite.com/xcop/}} are perfectly suited for this purpose because the project allowed for the precise measurement of the thermal pressure and density profiles of nearby galaxy clusters from about 10 kpc to the cluster outskirts based on {\it XMM-Newton} and {\it Planck} data \citep{Tchernin2016,Eckert2017,Ghirardini2019}. Because we are interested in the nonthermal component of the ICM, we selected clusters from the 12 XCOP for which a diffuse radio halo has been observed, using the {\tt GalaxyCluster} database\footnote{\url{https://galaxyclusters.hs.uni-hamburg.de/}}, and for which \textit{Fermi}-LAT constraints have been obtained by \cite{Ackermann2014}. We found three objects: 1) Abell 1795, a relaxed cool-core system; 2) Abell 2142, an elongated, dynamically active cluster with a cool-core; and 3) Abell 2255, a merging cluster with a highly perturbed core \citep[see also the recent work by][]{Botteon2020}. We thus note that these objects also present the advantage of sampling different dynamical states that are generally observed in clusters. In order to model these three clusters, we extrapolated the precisely measured density profile of the thermal plasma with a high-order polynomial function. The pressure profiles were fit with a GNFW model, providing a good extrapolation to the data. The nonthermal properties were set following what we did for our baseline cluster model, except that the magnetic field was normalized to 5 $\mu$G at 100 kpc. The redshift, mass, and coordinates of the cluster were taken from the XCOP data. Table~\ref{tab:xcop_properties} summarize the properties of these clusters.

\section{Physical modeling of the primary components}\label{sec:physical_modeling}
In this section, we discuss the physical modeling of the cluster. First, the global cluster properties are briefly discussed, as well as several assumptions employed in the modeling. Then the properties of the thermal and nonthermal components are detailed. The cluster modeling relies on primary base physical quantities from which other cluster properties can be derived, in particular in the case of the thermal component. The choice of the base quantities is discussed. Then, the derivation of the secondary quantities that characterize the cluster are developed both for the thermal and nonthermal components.

\subsection{Global properties and assumptions}
Before we model the inner structure of the clusters, it is useful to characterize the global cluster properties, as listed in the second block of parameters in Table~\ref{tab:table_parameters}. The cluster location is defined in terms of redshift and sky coordinates. From the redshift, and given a cosmological model, the angular diameter, and luminosity distances are computed and used later in the code. The default cosmological model is based on \cite{Planck2016XIII}, but can be modified if necessary.

Even if it does not play a direct role in the modeling, the characteristic mass of the cluster $M_{500}$\footnote{$M_{500}$ is the mass enclosed within a radius $R_{500}$,  within which the mean cluster density reaches 500 times the critical density of the Universe at the cluster redshift.} is part of the global parameters. It can be used to set several internal properties of the cluster, to their universal expectation, according to the fact that clusters are at first order self-similar objects \citep[in particular for the thermal pressure, see][]{Arnaud2010}. The value of $M_{500}$ also allows us to set the characteristic radius, $R_{500}$ (see also Eq.~\ref{eq:mass_definition}). 

It is also worth emphasizing one of the global parameters, the truncation radius, which is used in {\tt MINOT} in order to set a physical boundary to the cluster, beyond which the density drops to zero. This is not only useful for numerical issues when the cluster properties are integrated, but might be associated with the accretion shock radius at which the kinetic energy from accreting structures is converted into thermal energy \citep[see, e.g.,][for the observation of such an accretion shock]{Hurier2019}.

In the modeling, the plasma is assumed to be fully ionized and to follow the ideal gas law. The ions and electrons are assumed to be in thermal equilibrium \citep[see][for discussions of the electron and ion temperatures]{Fox1997}. While it might in principle depend on radius, the hydrostatic mass bias, the helium mass fraction, and the metallicity of the cluster are assumed to be constant \citep[see, e.g.,][for measured cluster metallicity profiles and simulations of the nonthermal radial profile]{Leccardi2008,Nelson2014}. Some of these parameters, related to the global properties of the cluster, are further discussed in the following subsections.

\subsection{Thermal component}\label{sec:physics_thermal_component}
\begin{figure*}
\centering
\includegraphics[width=1\textwidth]{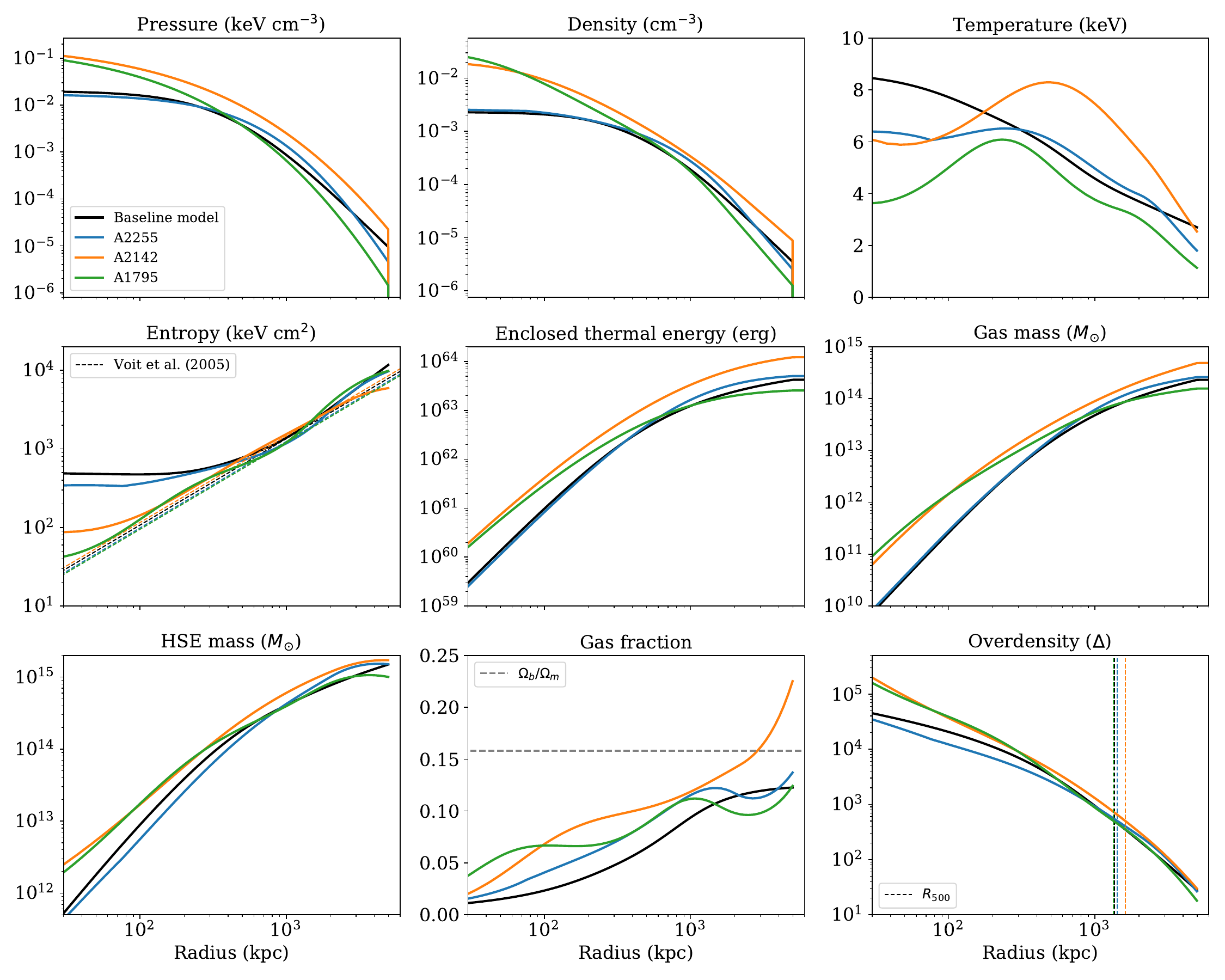}
\caption{Thermodynamic properties of our cluster sample defined in Table~\ref{tab:xcop_properties}, as computed based on the modeling of the thermal electron pressure and density (i.e., the base quantities used to derive the others). The truncation radius is shown at $r = 5$ Mpc, where all distributions drop, except for the integrated ones that remain constant afterward. For the entropy profile, we also show the expectation for purely gravitational collapse \citep{Voit2005}. For the gas fraction, the ratio between the mean baryon density and the mean matter density is shown as a dashed line. We also report the radius at which a density contrast of 500 (i.e., $R_{500}$) is reached in the overdensity profile.}
\label{fig:thermodynamics}
\end{figure*}

The base thermal properties are the electron number density and the electron pressure profiles. This choice is motivated by the fact the X-ray emission is directly sensitive to the electron number density, while the tSZ effect probes the electron pressure, but other choices could have been made (e.g., density and temperature). Generic literature parametric models are available, such as the $\beta$-model \citep{Cavaliere1978} or one of its extensions, the simplified Vikhlinin model \citep[SVM,][]{Vikhlinin2006}, which is generally used to describe thermal density cluster profiles. The GNFW profile is also available, and is generally used to describe the thermal pressure profile \citep{Arnaud2010,Planck2013V}. See Table~\ref{tab:table_spatial_models} for the parameterization of these models. In all cases, the different parameters can be used to control the amplitude, the characteristic or transition radius, and the slopes at different radii. In Fig.~\ref{fig:profiles_spectra} we show examples of these profiles for a given set of parameters.

With the electron pressure, $P_e(r)$, and electron number density, $n_e(r)$, profiles at hand, it is possible to compute the total gas pressure
\begin{equation}
        P_{\rm gas}(r) = \frac{\mu_e}{\mu_{\rm gas}} P_e(r),
\end{equation}
and the thermal proton number density profile as
\begin{equation}
        n_p(r) = \frac{\mu_e}{\mu_p} n_e(r).
\end{equation}
The mean molecular weights, $\mu_{\rm gas}$, $\mu_e$, $\mu_p$, and $\mu_{\rm He}$, are computed from the helium primordial abundance and the ICM metallicity, as
\begin{align} 
\begin{cases}
        \mu_{\rm gas} &= \frac{1}{2 \left(1-Y-Z\right) + \frac{3}{4} Y + \frac{1}{2}Z} \simeq 0.6 \\
        \mu_e              &= \frac{1}{(1-\frac{1}{2}Y-\frac{1}{2}Z) } \simeq 1.15 \\
        \mu_p              &= \frac{1}{1-Y-Z} \simeq 1.35 \\
        \mu_{\rm He}              &= \frac{4}{Y} \simeq 14.6,
\end{cases}
\end{align}
where $Y \simeq 0.27$ is the helium mass fraction and $Z \simeq 0.005$ is the heavy element mass fraction (defined through the solar reference metallicity multiplied by the metal abundance, see Table~\ref{tab:table_parameters}). Here, we used the approximation that $\frac{N_{\rm charge}+1}{N_{\rm nucleon}} \simeq 1/2$ for all metals, with $N_{\rm charge}$ the number of charge and $N_{\rm nucleon}$ the number of nucleons.

It is also straightforward to compute the temperature assuming the ideal gas law as
\begin{equation}
        k_{\rm B} \ T_e(r) = P_e(r) / n_e(r) \equiv k_{\rm B} \ T_{\rm gas}(r).
\end{equation}
Similarly, the electron entropy index, which records the thermal history of the cluster \citep{Voit2005}, can be defined as
\begin{equation}
        K_e(r) =  \frac{P_e(r)}{n_e(r)^{5/3}}.
\end{equation}
Temperature and entropy are useful diagnostics of the ICM. They can show the presence of a cool core \citep[e.g.,][]{Cavagnolo2009}, which is itself related to the central AGN activity and possibly to its CR feedback onto the surrounding gas \citep[e.g.,][]{Ruszkowski2017}.
They provide information on the dynamical state and accretion history of the cluster, which are connected to its CR content \citep[e.g., radio emission that begins during mergers,][]{Rossetti2011}.

The thermal energy density stored in the gas is given by
\begin{equation}
        u_{\rm th} = \frac{3}{2} n_{\rm gas} k_{\rm B} T = \frac{3}{2} P_{\rm gas}
\end{equation}
and can be integrated over the volume
\begin{equation}
        U_{\rm th} (R) = 4 \pi \int_0^R u_{\rm th} r^2 dr
\end{equation}
to obtain the total thermal energy up to radius $R$. This quantity is very useful when it is compared to the amount of energy stored in the CRs.

The cluster total mass within radius $r$, under the approximation of hydrostatic equilibrium, is given by
\begin{equation}
        M_{\rm HSE}(r) = -\frac{r^2}{\mu_{\rm gas} m_p n_e(r) G} \frac{dP_e(r)}{dr}
.\end{equation}
The hydrostatic mass is known to be biased with respect to the actual total mass \citep[see, e.g.,][for a review on the cluster mass scale]{Pratt2019}. The two can be related by
\begin{equation}
        M_{\rm tot}(r) = \frac{M_{\rm HSE}(r)}{\left(1 - b_{\rm HSE}\right)}
        \label{eq:hse_bias}
,\end{equation}
where $b_{\rm HSE}$ is the hydrostatic mass bias (see Table~\ref{tab:table_parameters}), which is assumed to be constant \citep[see][for detailed discussions of the bias value, which is expected to be $b_{\rm HSE} \sim 0.2$]{PlanckXX2014}.

Based on the electron number density, the gas mass within radius $R$ can be computed as
\begin{equation}
        M_{\rm gas}(R) = 4 \pi \int_0^R \mu_e m_p n_e(r) r^2 dr,
\end{equation}
and it provides a measurement of the available target mass for the interaction with CRp. The gas fraction can be derived using
\begin{equation}
        f_{\rm gas}(r) = \frac{M_{\rm gas}(r)}{M_{\rm tot}(r)}.
\end{equation}
The overdensity profile is computed using $\rho_c(z)$, the critical density of the Universe, by
\begin{equation}
        \Delta(R) = \frac{M_{\rm tot}(R)}{\frac{4}{3} \pi R^3 \rho_c(z)},
\end{equation}
which allows us to extract the value of the characteristic radius, $R_{\Delta}$, within which the density of the cluster is $\Delta$ times the critical density of the Universe at the cluster redshift. The value of $\Delta$ is generally taken to be $500$. The enclosed mass within $R_{\Delta}$ is then
\begin{equation}
        M_{\Delta} = \frac{4 \pi}{3} \Delta \rho_{\rm ref}(z) R_{\Delta}^3.
\label{eq:mass_definition}
\end{equation}

All the quantities defined here can be extracted as a function of radius from {\tt MINOT}, according to a given cluster model, using the dedicated functions that are located in the {\tt Physics} subclass. In Fig.~\ref{fig:thermodynamics} we  illustrate the main thermodynamic properties of our baseline cluster model and the three Abell cluster models discussed in Section~\ref{sec:baseline_cluster_models}. We show the thermal electron pressure, electron number density, gas temperature, entropy, enclosed thermal energy, enclosed gas mass, enclosed hydrostatic mass, gas fraction, and overdensity contrast. 

Depending on the dynamical state and the presence or absence of a cool core, the profiles are different. For instance, A1795 clearly presents a high-density cool core according to its temperature and entropy profiles. Its large-scale electron pressure and density fall quickly, consistent with a compact, relaxed morphology. In contrast, A2255 and our Baseline model show disturbed cores with a high entropy floor, and their pressure and density profiles are much flatter on large scales, consistent with a redistribution of the thermal energy in a merging event. A2142 is an intermediate case. It presents a peaked density profile (showing a compact core), but its pressure profile is relatively flat on large scales, typical of disturbed clusters. Because particle acceleration is expected to depend on the cluster dynamical state, these thermodynamic diagnosis are useful for characterizing individual clusters in the context of understanding cluster CR physics.

The enclosed thermal energy is directly related to the pressure profile. It is particularly relevant here because it provides a normalization for the number of CRs (see Section~\ref{sec:modeling_non_thermal}). Based on its high pressure profile and thus thermal energy, and based on its high density (which implies a high gas mass), A2142 would thus be the best target to search for $\gamma$-rays from proton-proton interaction in our sample, assuming the same CR distribution for all clusters. 

The hydrostatic mass provides a direct way to measure the cluster total mass profile (given a hydrostatic bias, Eq.~\ref{eq:hse_bias}). This can be particularly relevant for modeling the $\gamma$-ray signal associated with the decay or annihilation of dark matter particles (which is usually done assuming an NFW dark matter density profile with a given concentration and normalization). In addition, the gas fraction gives the ratio between the amount of dark matter and the amount of gas, which is expected to provide a proxy for the dark matter signal to CR background when indirect dark matter searches using clusters are performed, and thus they are an indication for the best targets. The overdensity contrast allows us to measure the radius $R_{500}$ (or $R_{200}$), and thus the corresponding mass.

For further discussions of the thermodynamic properties of galaxy clusters using a similar framework, we refer to the recent work by \citet{Tchernin2016}, \citet{Ruppin2018}, \citet{Ghirardini2019}, \citet{Ricci2020}.
 
\subsection{Nonthermal component}\label{sec:modeling_non_thermal}
\begin{figure*}
\centering
\includegraphics[width=0.42\textwidth]{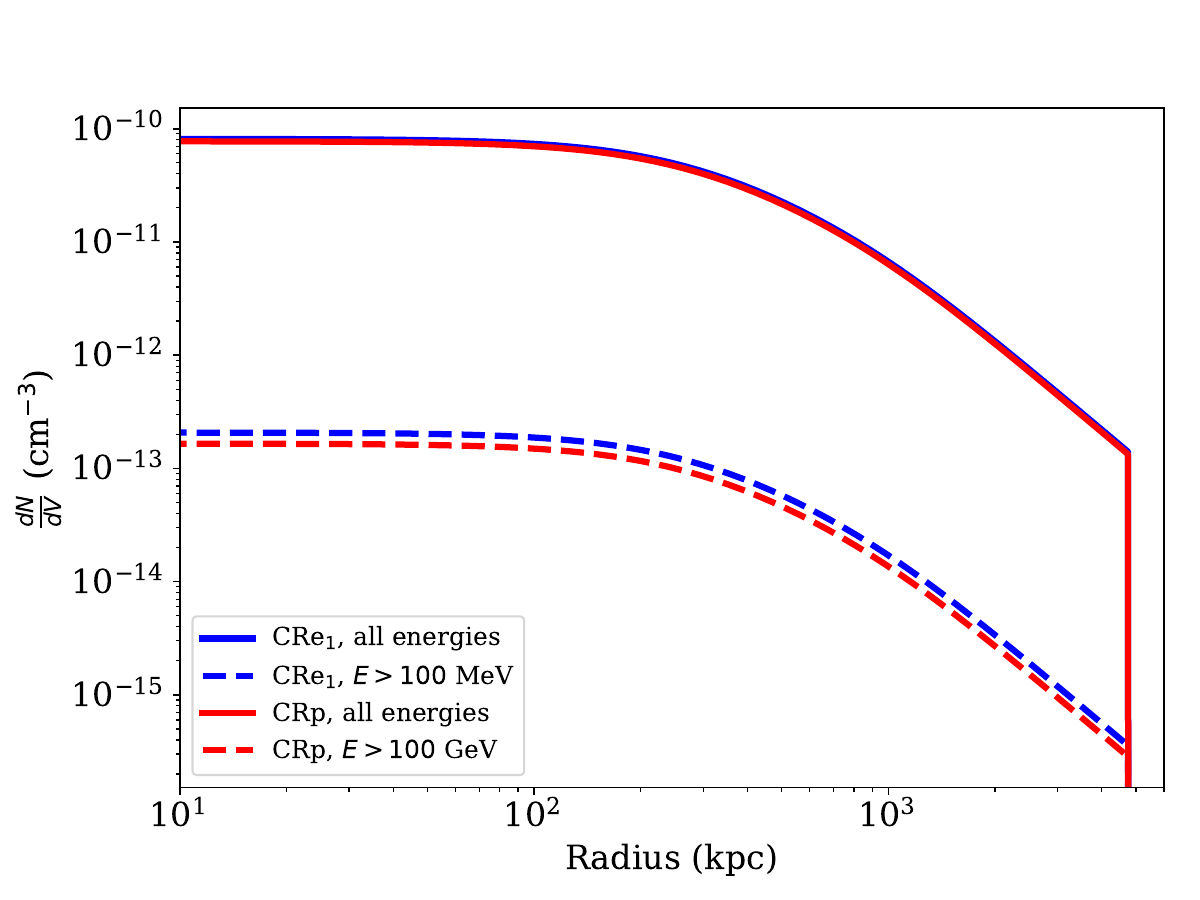}
\includegraphics[width=0.42\textwidth]{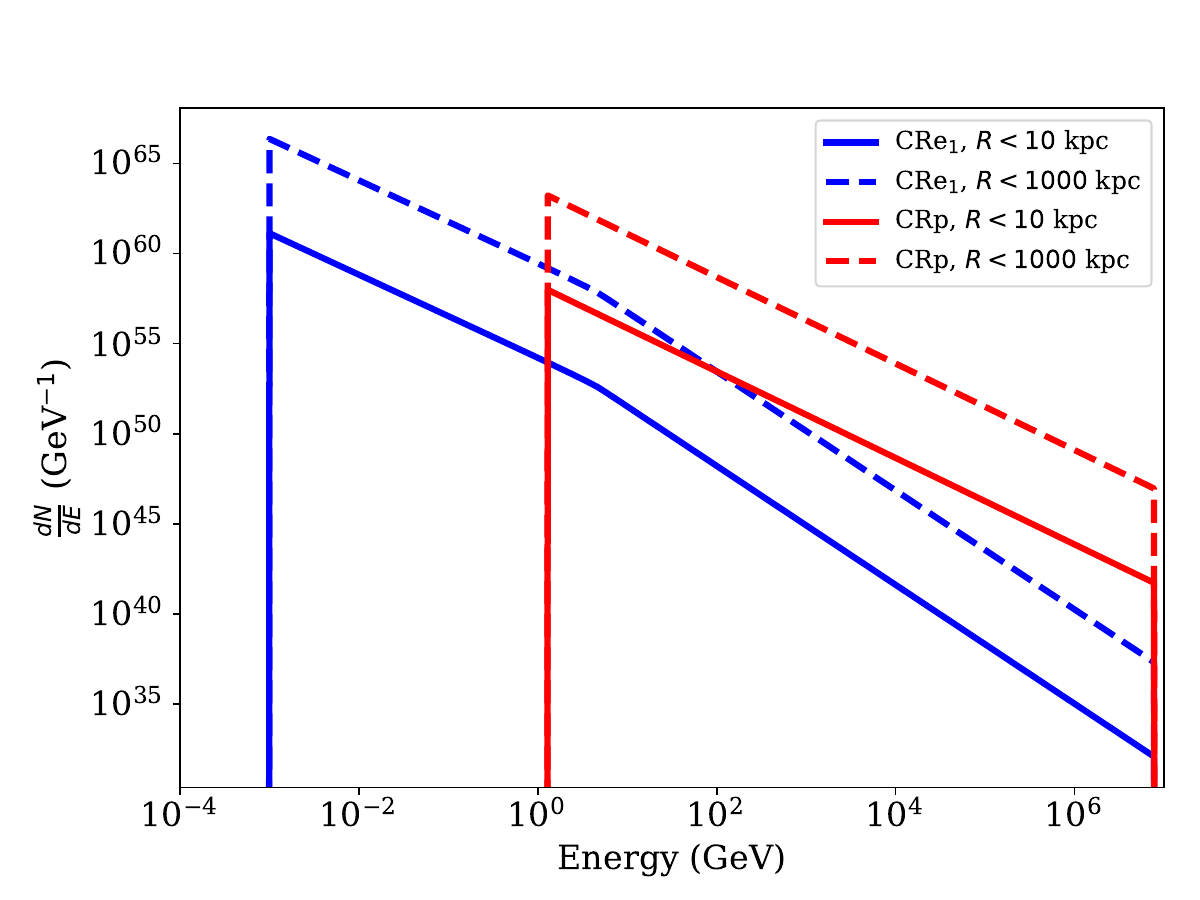}
\caption{CR properties of the baseline cluster model. {\bf Left}: CR number density profile (i.e., CR distribution integrated between $E_{\rm min}$ and $E_{\rm max}$, as indicated in the legend) for CRe$_1$ and CRp. {\bf Right}: CR spectrum integrated over the radius up to a maximum radius, as given in the legend, for CRe$_1$ and CRp.}
\label{fig:non_thermal}
\end{figure*}

\begin{figure*}
\centering
\includegraphics[width=0.42\textwidth]{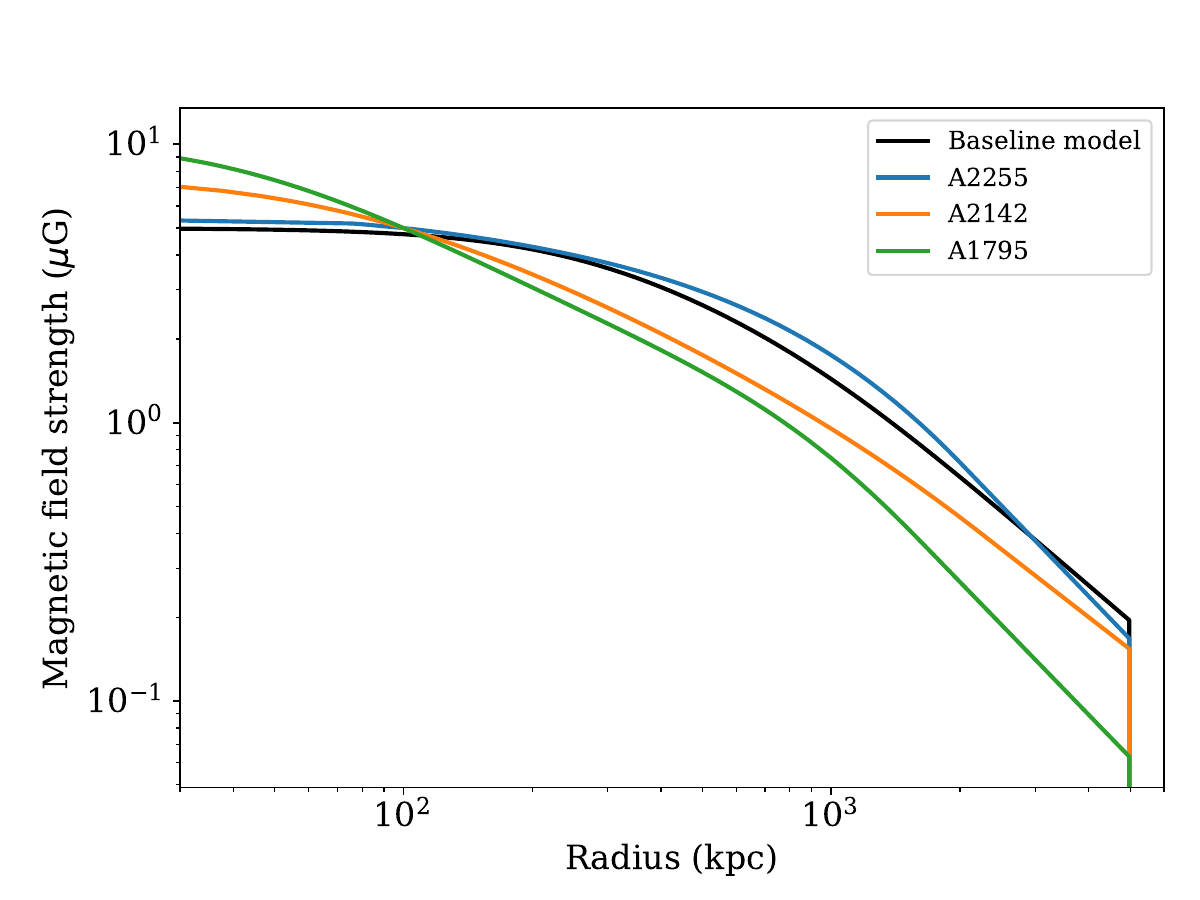}
\includegraphics[width=0.42\textwidth]{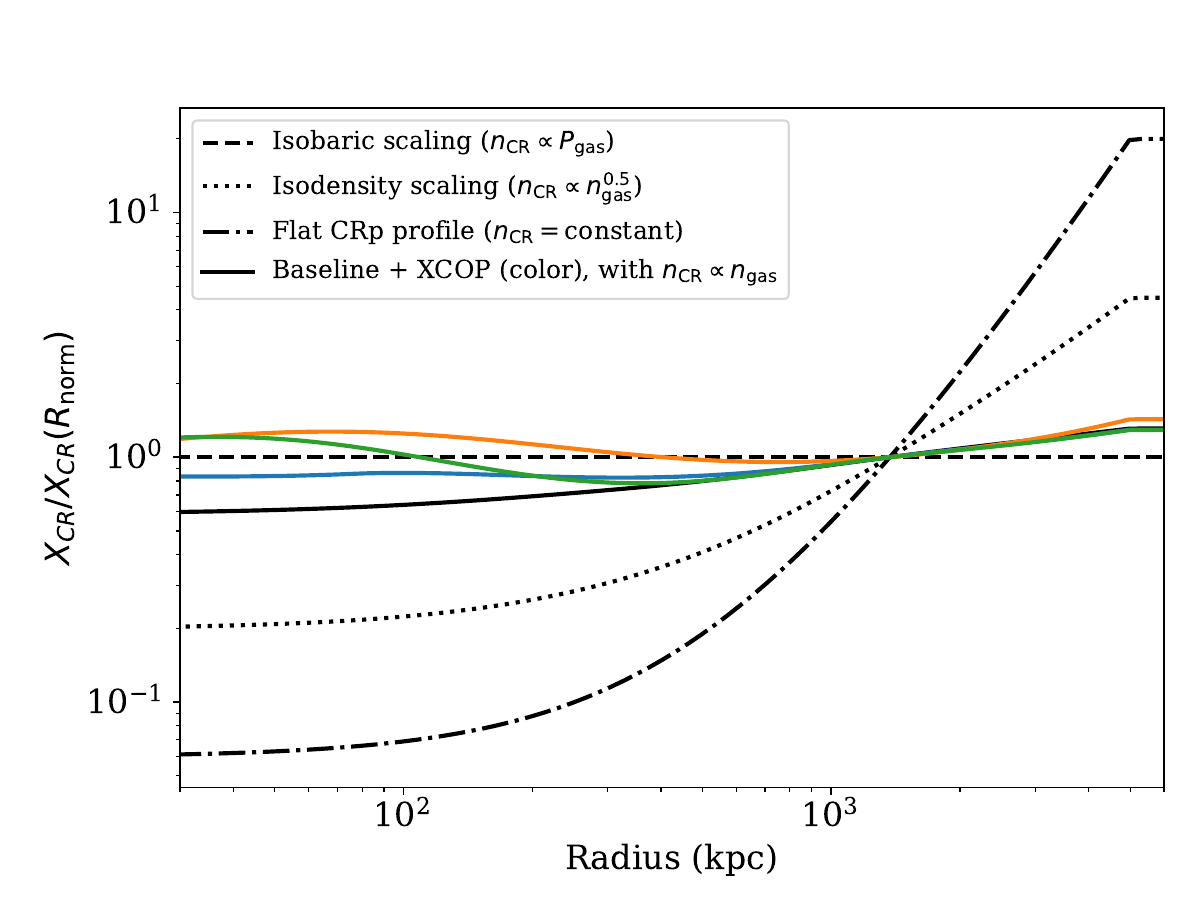}
\caption{Magnetic field and CRp-to-thermal energy properties of the baseline cluster model. {\bf Left}: Magnetic field profile obtained by assuming a fixed magnetic field strength of 5 $\mu$G at 100 kpc, and a scaling relative to the thermal density as $B \propto n_e^{0.5}$. {\bf Right}: CRp-to-thermal energy profile for different models of the CRp distribution. The color lines show each real cluster, using the same colors as in the left panel, and with $n_{\rm CRp} \propto n_{\rm gas}$. The black lines correspond to variation in the CRp scaling with respect to the thermal gas for the baseline cluster model, as indicated in the legend.}
\label{fig:non_thermal2}
\end{figure*}

The base nonthermal properties of the cluster are the magnetic field strength, the spectra and profile of the CRp, and the CRe$_1$. The CRe$_1$ differ from the CRe$_2$ because they correspond to a population that was accelerated from the plasma microphysics (e.g., shocks or turbulences, as for CRp), while the secondaries are the product of hadronic interactions (see Section \ref{sec:particle_interactions} for further details). The radial models available for the nonthermal component are the same as for the thermal component (see Table~\ref{tab:table_spatial_models}). The spectral distributions for currently available models are listed in Table~\ref{tab:table_spectral_models}. While all these models might be attributed to either the CRp or the CRe$_1$, the initial injection \citep{Jaffe1973} and continuous injection \citep{Pacholczyk1970} models are expected to account for electron losses and are thus poorly suited for CRp \citep[see][for discussions about the parameterization]{Turner2018}. We note that the minimum and maximum energy of the CRs is part of the parameters, and it is possible to use these parameters to truncate the spectra. By default, the minimum energy of CRp and CRe$_1$ corresponds to the energy threshold of the proton-proton interaction and the rest mass of the electrons, respectively. Finally, it is also possible to use this parametric function to inject CRe$_1$ and apply losses (see Section~\ref{sec:energy_loss} for details) to obtain the actual electron population. In Fig.~\ref{fig:profiles_spectra}, example spectra are shown for the CRe$_1$. The effect of these spectra on the observables is shown in Section~\ref{sec:observables}.

The radial and spectral distributions of the CRs are currently decoupled, and we can express the CR distribution (i.e., the CR number density per unit energy) as
\begin{equation}
        J_{\rm CR} (r, E) = A_{\rm CR} f_1(E) f_2(r)
        \label{eq:Jcr}
,\end{equation}
where $A_{\rm CR} $ is the normalization, and $f_1(E)$ and $f_2(r)$ are the spectral and radial distributions, respectively (see Table~\ref{tab:table_spectral_models} and~\ref{tab:table_spatial_models} for available models). In principle, functions $f_1$ and $f_2$ could be merged into $f_{1,2}(r,E)$ to include a radial dependence of the spectral component, but this function is not yet implemented. When losses are to be applied to the input distribution, a radial dependence affects the spectrum because the losses themselves depend on the radius (see Section~\ref{sec:energy_loss} for details).

In order to normalize the CR distribution, we compute the energy density that is stored between energy $E_1$ and $E_2$, which can be expressed by integrating over the energy as
\begin{equation}
        u_{\rm CR} (r) = 3 P_{\rm CR} (r)= \int_{E_1}^{E_2} E_{\rm CR} J_{\rm CR}(E_{\rm CR}) dE_{\rm CR},
\end{equation}
and it is related to the CR pressure, $P_{\rm CR}$. Here we assume that the CRs are ultrarelativistic particles, with adiabatic index $\Gamma = 4/3$. The result is only weakly sensitive to the upper bound, $E_2 = E_{\rm max, CRp/e}$, because the CR spectrum generally vanishes rapidly for a spectral index higher than 2. The default lower bound is set to the minimum proton energy necessary to trigger the pion production, $E_1 \equiv E_p^{\rm th}$, for protons and to the electron rest mass for the electrons. The total energy stored in CRs enclosed within the radius $R$ can then be computed as 
\begin{equation}
        U_{\rm CR} (R) = 4 \pi \int_0^R u_{\rm CR} r^2 dr.
\end{equation}
The CR-to-thermal energy density ratio is then given by 
\begin{equation}
        x_{\rm CR}(r) = \frac{u_{\rm CR} (r)}{u_{\rm th} (r)},
\end{equation}
or similarly,
\begin{equation}
        X_{\rm CR}(R) = \frac{U_{\rm CR} (R)}{U_{\rm th} (R)},
\end{equation}
when integrated over the volume up to the radius $R$.

In practice, the CR distribution, $A_{\rm CR} $, is normalized by setting the value of $X_{\rm CR}(R)$ at a given radius (e.g., $R_{500}$). We note that this fraction is defined relative to the enclosed energy here, while it is also common to find this definition in terms of pressure in the literature. The two differ by a factor of 2 because the thermal gas is nonrelativistic, while the CRs are in the relativistic regime.

The CR distributions can be integrated over energy as 
\begin{equation}
        n_{\rm CR}(r) \equiv \left.\frac{dN}{dV}\right|_{[E_1,E_2]} = \int_{E_1}^{E_2} J_{\rm CR} (r, E) dE
\end{equation}
or radius as
\begin{equation}
        \frac{dN}{dE}(<R) = \int_0^R 4 \pi r^2 J_{\rm CR} (r, E) dr
\end{equation}
to compute the number density profile within $E_1$ and $E_2$, or the spectrum enclosed within $R$, respectively.

In Fig.~\ref{fig:non_thermal} we illustrate the integrated CR number density profiles and spectra for the baseline cluster model. CRe$_1$ and CRp clearly follow the same profile because they are calibrated to follow the thermal electron number density. The number of electrons and protons is nearly the same given the chosen normalization. The number of CRs drastically decreases when a cut in energy is applied. The spectra show different shapes for the electrons and protons, reflecting our baseline choice (power law for the protons, and continuous injection with a break at 1 GeV for the electrons), and the minimum energy is also different for the two populations. The number of enclosed CRs naturally increases with increasing radius. We note that the figure would be very similar for the real cluster models because the underlying CR modeling is the same.

In Fig.~\ref{fig:non_thermal2} we show the magnetic field profiles of our cluster models in the left panel and the ratio between the CRp energy and the thermal energy in the right panel. The magnetic field was calibrated on the thermal density profile as $B \propto n_e^{0.5}$, and normalized to 5 $\mu$G (at the peak for the baseline model, and at 100 kpc for the real clusters). For the CR-to-thermal energy, we also varied the CR number density profiles using different scaling relations with respect to the thermal density and pressure to show the changes in the resulting profiles. However, we note that the exact shape also depends on the shape of the thermal pressure, which is kept fixed here. Because the pressure profile decreases with radius, setting the CR distribution to a flatter profile leads to a deficit in the center and an increase in the outskirt for the energy ratio $X_{\rm CR}$. When the CR number density profile follows the thermal density profile, the CR-to-thermal energy of cool-core clusters is boosted in the center because the thermal pressure is low relative to the thermal density in the core (see, e.g., the case of A1795). In all cases, the ratio was set to $X_{\rm CR}(R_{500}) = 10^{-2}$.

\section{Particle interactions in the ICM}\label{sec:particle_interactions}
The physical properties of the ICM for its thermal and nonthermal components have been defined in Section~\ref{sec:physical_modeling}. In this section, we model the hadronic interactions that take place in the plasma and generate secondary particles (see also Fig.~\ref{fig:code_overview1}). We also discuss the loss processes that affect them, in particular, the electrons.

\subsection{Production rate of secondary particles from hadronic interactions}\label{sec:prod_rate_of_secondaries}
\begin{figure*}
\centering
\includegraphics[width=0.33\textwidth]{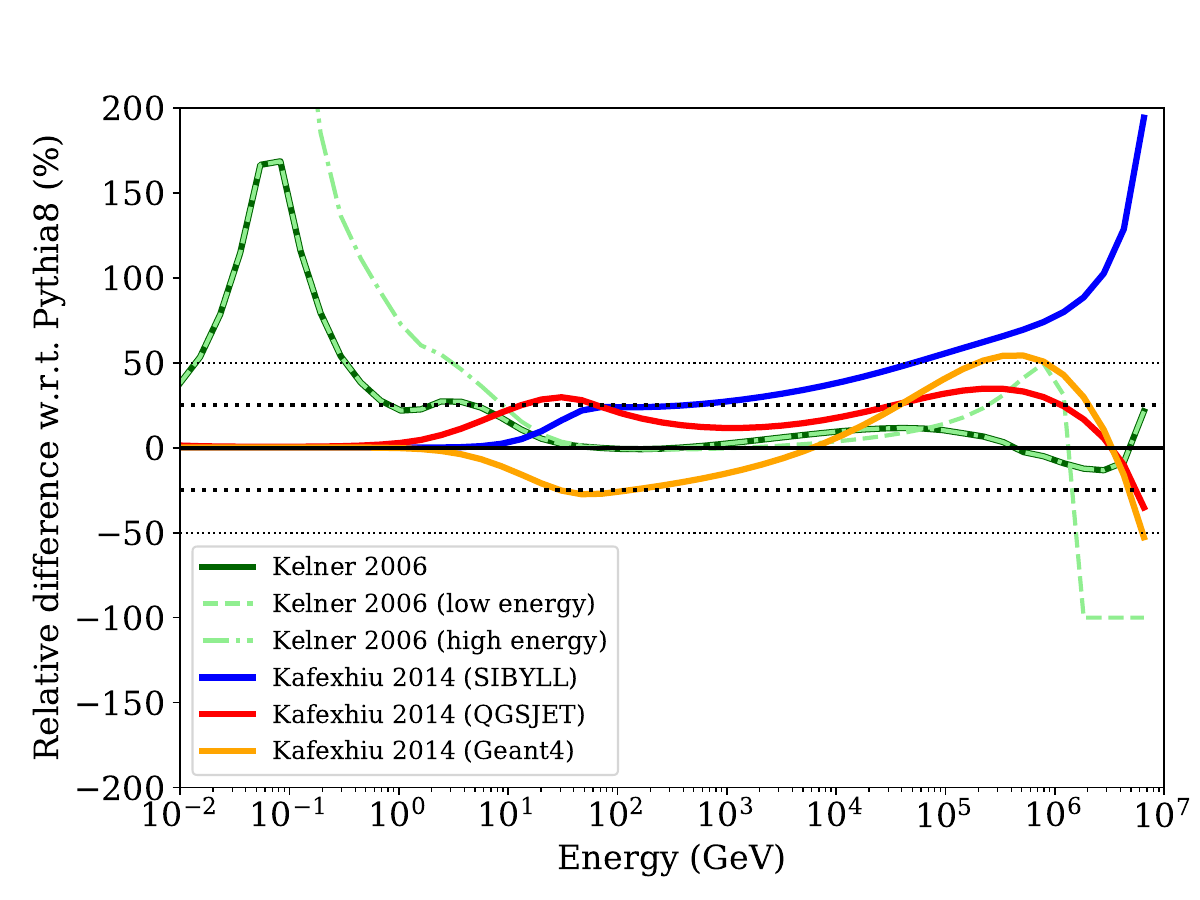}
\includegraphics[width=0.33\textwidth]{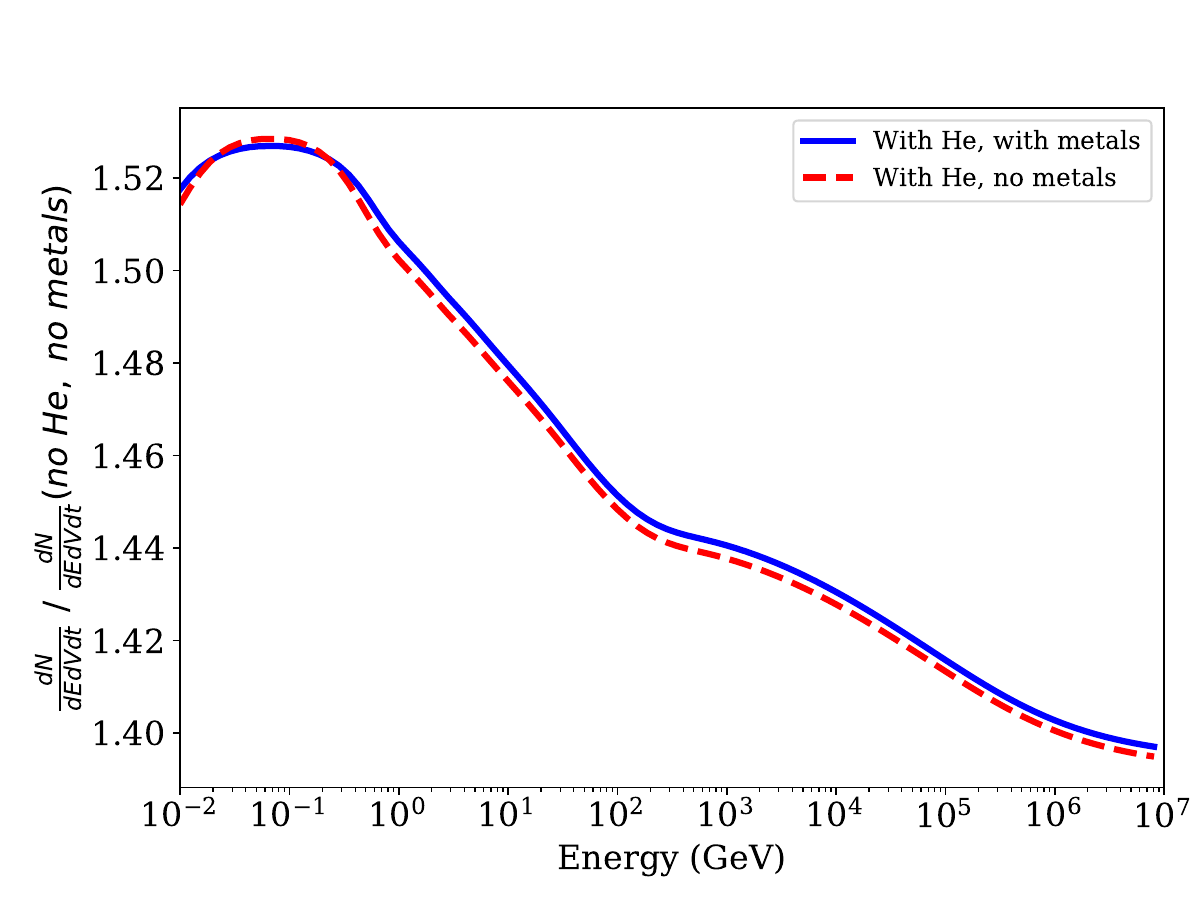}
\includegraphics[width=0.33\textwidth]{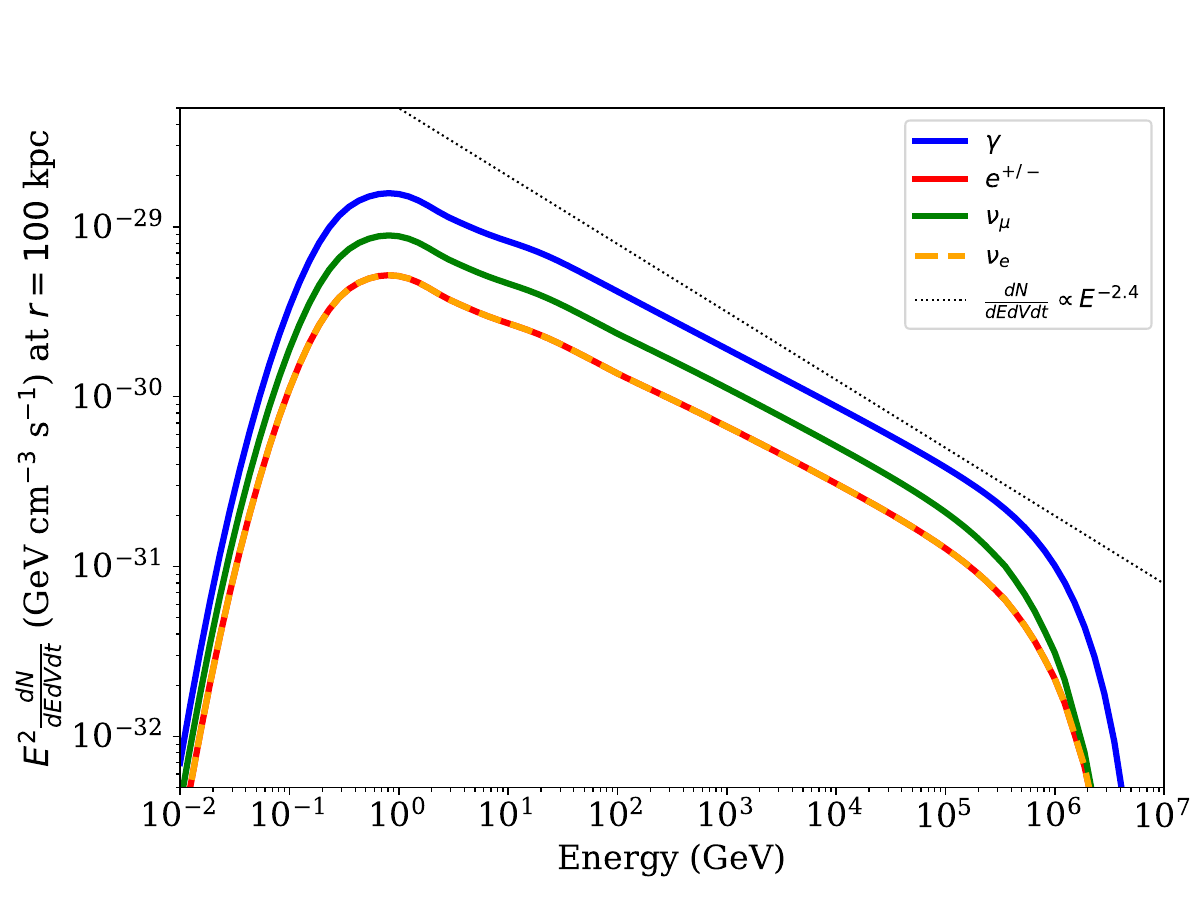}
\caption{Production rate of secondary particles from hadronic processes. 
{\bf Left}: Relative comparison of the production rate of $\gamma$-rays in the case of no helium and zero metallicity in the \cite{Kafexhiu2014} and the \cite{Kelner2006} parameterization, using the Pythia8 parameterization from \cite{Kafexhiu2014} as a reference. For \cite{Kelner2006}, we also show the low- and high-energy ($\delta$-approximation) limits as dashed lines, which were combined to compute the rate over the total energy range. 
{\bf Middle}: Effect of including helium and metals in the model. This was done with Pythia8 using the \cite{Kafexhiu2014} parameterization. 
{\bf Right}: Comparison of the production rate for $\gamma$-rays, electrons, positrons, and neutrinos, computed for a radius $r=100$ kpc. We also include the power-law function proportional to $E^{-2.4}$ for comparison because it corresponds to the injected CRp distribution.}
\label{fig:secondary_production_rate}
\end{figure*}

The collision between high-energy CRp and the thermal ambient gas produces $\gamma$-rays, electrons, positrons, and neutrinos, mainly through the production of pions,  following these interaction chains:
\begin{align}
\begin{cases}
        p + p \longrightarrow &\pi^0 + \pi^- + \pi^+ + {\rm others} \\
                                          &\pi^0 \longrightarrow 2 \gamma \\
                                        &\pi^{\pm} \longrightarrow \mu^{\pm} + \nu_{\mu}/\bar{\nu}_{\mu} \longrightarrow e^{\pm} + \nu_{e}/\bar{\nu}_{e} + \bar{\nu}_{\mu} /\nu_{\mu.}
\end{cases}
\end{align}

When the thermal plasma is considered to be at rest with respect to the CRs, the CRp-thermal proton collision rate per unit energy of CRp is given by
\begin{equation}
        \frac{dN_{\rm col}}{dE_{\rm CRp}dVdt} = \sigma_{\rm pp} \times v_{\rm CRp} \times n_{\rm p} \times J_{\rm CRp},
\end{equation}
where $\sigma_{\rm pp}$ is the proton-proton interaction cross section, $v_{\rm CRp} \simeq c$ is the speed of CRp, $n_{\rm p}$ the number density of thermal protons, and $J_{\rm CRp}$ the number density per unit energy of CRp. The production rate of secondary particles $X$ per unit volume, unit energy, and unit time can then be expressed as 
\begin{equation}
        \frac{dN_X}{dE_X dV dt} = \int_{E_X}^{+\infty} \frac{dN_{\rm col}}{dE_{\rm CRp}dVdt} F_X\left(E_X, E_{\rm CRp}\right) dE_{\rm CRp}.
\end{equation}
Two ingredients are thus necessary: 1) the total inelastic cross section of the proton-proton interaction and its evolution as a function of energy, $\sigma_{\rm pp}(E_{\rm CRp})$ ; and 2) the number of secondary particles produced in a collision per unit energy of the produced particle as a function of the initial energy of the CRp, namely $F_X(E_X, E_{\rm CRp})$. These ingredients are usually obtained by fitting parametric functions to accelerator data, together with Monte Carlo simulations performed with sophisticated codes \citep[e.g.,][]{Kelner2006,Kamae2006,Kafexhiu2014}. 

The \cite{Kafexhiu2014} parameterization was implemented in {\tt Naima} \citep{Zabalza2015}, a publicly available Python package dedicated to the computation of nonthermal radiation from relativistic particle populations. The work presented here is based on {\tt Naima}, to which the radial dimension was added, and thus, we also use the work by \cite{Kafexhiu2014} as our baseline. 

As we show below, heavy elements also contribute significantly to the particle production rate. This contribution is only available in the work by \cite{Kafexhiu2014}, which is also expected to have the most current data, in particular at the highest energies. However, \cite{Kafexhiu2014} only focused on the $\gamma$-ray production rate, while \cite{Kelner2006} also provided a parameterization for the leptons (electrons, positrons, and neutrinos), but did not include heavy elements. Therefore we employed a hybrid approach. We used the parameterization by \cite{Kafexhiu2014} for the $\gamma$-ray production and that from \cite{Kelner2006} for the leptons. To account for heavy elements in the case of leptons, we applied a rescaling of the production rate given by \cite{Kelner2006}. To do so, we assumed that the ratio between the production rate of leptons and that of $\gamma$-rays does not depend on whether heavy elements are included. This is motivated by the fact that \cite{Kafexhiu2014} accounted for heavy elements using a multiplicative correction to the cross section. The production rate of leptons is finally given by
\begin{equation}
        \frac{dN_{e^{\pm},\nu_{\mu,e}}}{dEdVdt} = \left.\frac{dN_{e^{\pm},\nu_{\mu,e}}}{dEdVdt}\right|_{\rm Kelner2006} \times 
        \frac{\left.\frac{dN_{\gamma}}{dEdVdt}\right|_{\rm Kafexhiu2014}}{\left.\frac{dN_{\gamma}}{dEdVdt}\right|_{\rm Kelner2006}}
        \label{eq:CRe_gamma_nu_injection}
.\end{equation}
While this approach allowed us to compute the electron, positron, and neutrino production rate in the presence of helium and nonzero metallicity of the ICM, it uses the so-called $\delta$-approximation for the ratio of leptons to $\gamma$-rays in the high-energy regime \citep[see][]{Kelner2006}, which is expected to be a relatively crude approximation \citep[see][]{Kafexhiu2014}. Nevertheless, the accuracy of the lepton-to- $\gamma$-ray ratio is expected to be a much better fit because biases in the spectra are expected to cancel each other out.

In Fig.~\ref{fig:secondary_production_rate} we illustrate the computation of the secondary particle production rate in the case of our baseline cluster model (see Section~\ref{sec:physical_modeling}), with a power-law model with index 2.4 for the CRp. The left panel shows the relative difference between the \cite{Kafexhiu2014} and \cite{Kelner2006} parameterizations for the $\gamma$-ray production rate when the helium and metal abundances were set to zero. We used the Pythia8 parametrization from \cite{Kafexhiu2014} as a reference. In practice, the \cite{Kelner2006} parameterization is the combination of a calculation at low energy and the use of the $\delta$-approximation at high energy, which are both shown as dashed lines. The agreement between Pythia8 and \cite{Kelner2006} is relatively good over most of the energy range (lower than 25\% for most of it, but a peak reaches more than 100\% around 100 MeV). The different high-energy parameterizations available in the work by \cite{Kafexhiu2014}, namely using the Monte Carlo codes Pythia8, SIBYLL, QGSJET, or Geant4, are also shown. As expected, the difference with respect to Pythia8 is only large at high energy. It remains below 25\% for energies below 1 TeV, and increases to more than 50\% above 100 TeV \citep[see][for further discussions]{Kafexhiu2014}. Based on the comparison of the top panels of Fig.~\ref{fig:secondary_production_rate}, systematic uncertainties in the modeling are expected to be about 30\% over most of the energy range probed here.

The middle panel quantifies the effect of accounting for helium and metals in the ICM. The helium mass fraction was chosen to be 0.27, and the metallicity was set to the the solar value. The helium can clearly boost the signal by more than 50\%, especially at low energies, but has a strong effect over the full energy range where its contribution remains higher than 40\%. The metals, on the other hand, only account for percent-level changes in the spectrum. We note that the ratio $\mu_e/\mu_p$ depends on the ICM composition, which affects the value of $n_p$ for fixed $n_e$, and explains why the $\gamma$-ray production rate can become lower when metals are included compared to the helium-only case; this is visible around 100 MeV. Based on these results, we expect that ignoring the helium contribution will lead to a strong systematic effect in the model amplitude that underestimates the signal by about 40-50\%.

Finally, the right panel of Fig.~\ref{fig:secondary_production_rate} provides the particle injection rate for $\gamma$-rays, electrons, and positrons, and both muonic and electronic neutrinos. The high-energy cutoff is due to the maximum energy of the CRp, which was set to 10 PeV, while the decrease below 1 GeV is due to the kinematic production threshold of the proton-proton interaction of about 1.2 GeV. Between these energies, the slope of the secondary particle production rate nearly follows that of the injected CRp, as shown by the dotted black line.

\subsection{Energy losses}\label{sec:energy_loss}
\begin{figure}
\centering
\includegraphics[width=0.42\textwidth]{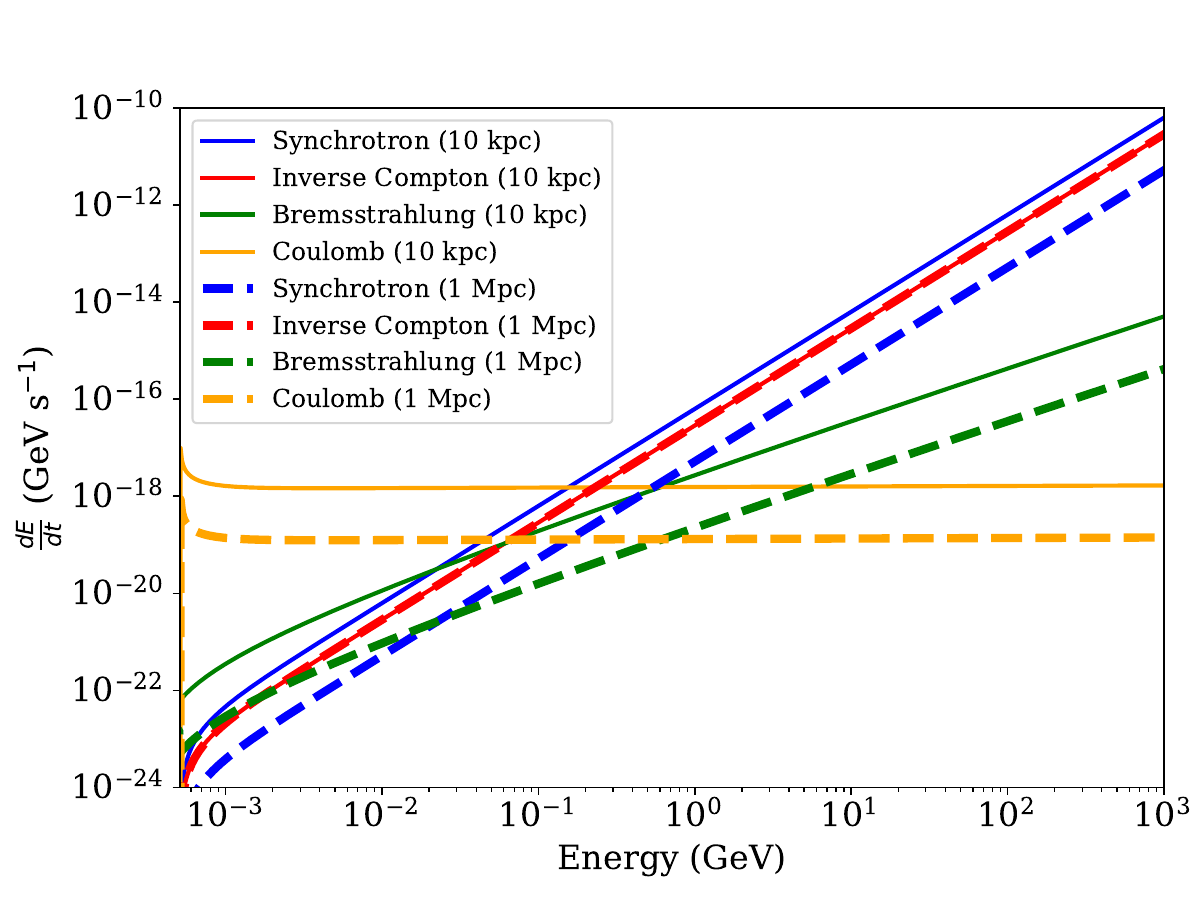}
\caption{Contributions to the energy loss rate as a function of energy for a radius $r=10$ kpc and $r=1000$ kpc.}
\label{fig:electron_losses}
\end{figure}

While the neutrinos and the $\gamma$-rays can escape the cluster and be detected by ground- and space-based instruments, the electrons evolve in the ICM and are affected by several sources of energy loss. We considered the main sources of energy loss: synchrotron radiation, inverse Compton interaction, Coulomb losses, and Bremsstrahlung radiation.

We first define the Lorentz factor of the electrons $\gamma = \frac{E}{m_e c^2}$, and the reduced speed of the electrons, $\beta = \sqrt{1-1/\gamma^2}$. The synchrotron radiation loss is given (in S.I. units) by \cite{Longair2011},
\begin{equation}
        \left.\frac{dE}{dt}\right|_{\rm sync} = -\frac{4}{3} \sigma_{\rm T} c \beta^2 \gamma^2 \frac{B^2}{2 \mu_0},
\end{equation}
with $\sigma_{\rm T}$ the Thomson cross section and $\mu_0$ the vacuum permeability. It is proportional to the amplitude of the magnetic field squared, $B^2$. We therefore expect it to be most efficient in the central regions of the cluster. Because of the $\gamma^2$ dependency, the synchrotron loss will be higher at high energy. 

Inverse Compton losses can be expressed in a very similar way \citep{Longair2011} as
\begin{equation}
        \left.\frac{dE}{dt}\right|_{\rm IC} = -\frac{4}{3} \sigma_{\rm T} c \beta^2 \gamma^2 u_{\rm CMB},
\end{equation}
where the dependence on magnetic field is replaced by the ambient photon field, assumed to be dominated by the CMB, whose energy density is given by
\begin{equation}
u_{\rm CMB} = \frac{8 \pi^5 \left(k_{\rm B} T_{\rm CMB} \left(1+z\right)\right)^4}{15 (h c)^3}.
\end{equation}
The inverse Compton energy losses do not depend on the cluster location, but increase with redshift because of the CMB dependence. 

The Coulomb losses are computed as \citep{Gould1972}
\begin{align}
        \left.\frac{dE}{dt}\right|_{\rm Coulomb} &= -\frac{3}{2} \sigma_{\rm T} n_e \frac{m_e c^3}{\beta} \left({\rm ln} \left(\frac{m_e c^2 \beta \sqrt{\gamma-1}}{\hbar \omega_{\rm p}}\right)\right. \notag\\ &\left.\quad- {\rm ln}\left(2\right) \left(\frac{\beta^2}{2} + \frac{1}{\gamma}\right) + \left(\frac{\gamma-1}{4 \gamma}\right)^2 + \frac{1}{2}\right),
\end{align}
where $\omega_{\rm p} = \sqrt{\frac{e^2 n_e}{m_e \epsilon_0}}$ is the plasma frequency. The Coulomb losses are proportional to the thermal  electron number density and are therefore more effective in the cluster core. They are almost independent of energy.

Finally, the Bremsstrahlung losses are computed as \citep{Blumenthal1970}
\begin{equation}
        \left.\frac{dE}{dt}\right|_{\rm Brem.} =  -8 \alpha c r_0^2 E(n_p+3 n_{\rm He}) \left({\rm ln}\left(2 \gamma\right)+\frac{1}{3}\right) 
,\end{equation}
where $\alpha = \frac{e^2}{4 \pi \epsilon_0 \hbar c}$ is the fine-structure constant and $r_0 = \frac{e^2}{4 \pi \epsilon_0 m_e c^2}$ is the classical electron radius. We neglected elements heavier than helium and used the completely unscreened limit that is appropriate for low-density plasma. As for the Coulomb losses, the Bremsstrahlung losses depend on the number density of thermal nuclei, but they increase with energy.

In Fig.~\ref{fig:electron_losses} we provide the loss function for the synchrotron, inverse Compton, Bremsstrahlung, and Coulomb contributions for two different radii from the center, 10 and 1000 kpc, for our baseline model. At low energy, the Coulomb losses are expected to dominate, while at high energy, the synchrotron and inverse Compton losses dominate, depending on the relative value of the magnetic field and the CMB photon field. The contribution by Bremsstrahlung is always subdominant.

\subsection{Secondary electrons in the steady-state approximation}\label{sec:secondary_electrons_in_steady_state}
\begin{figure*}
\centering
\includegraphics[width=0.42\textwidth]{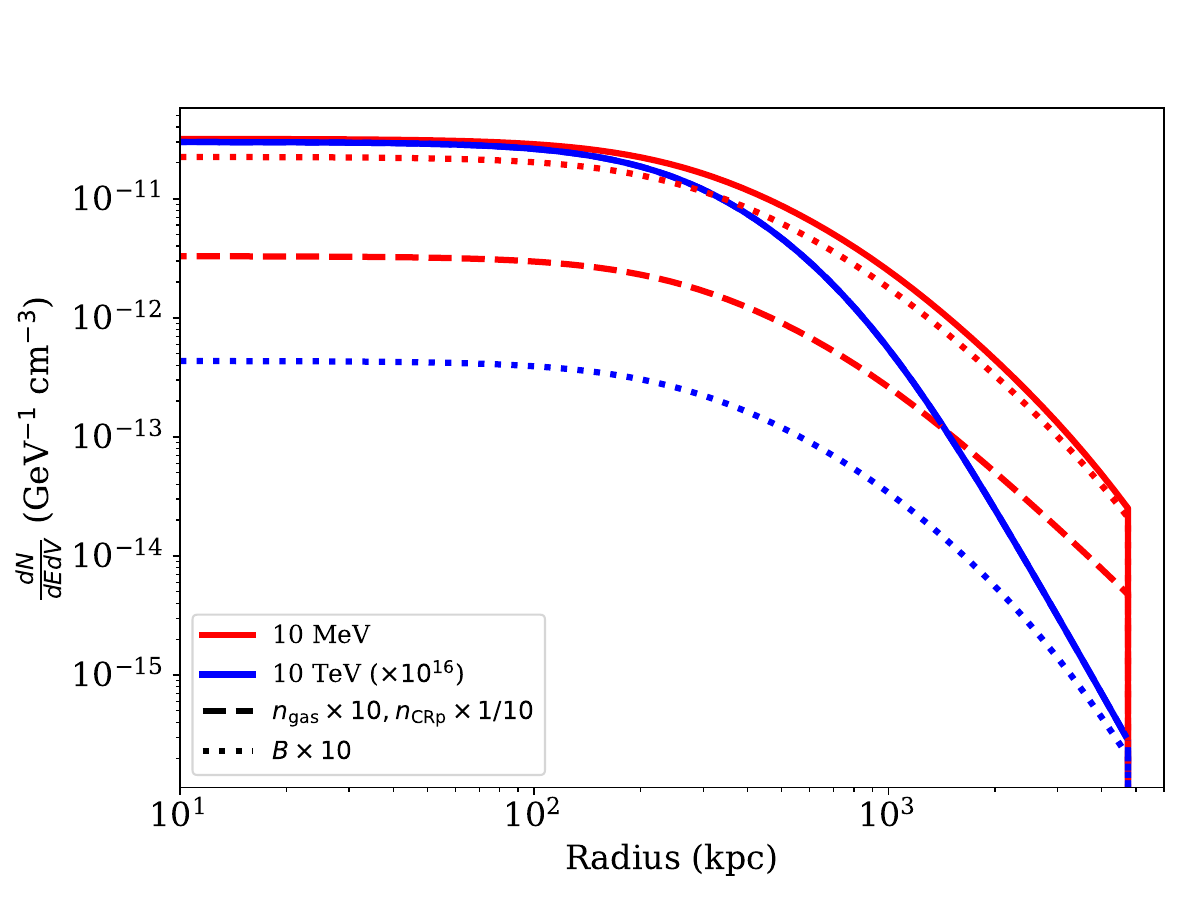}
\includegraphics[width=0.42\textwidth]{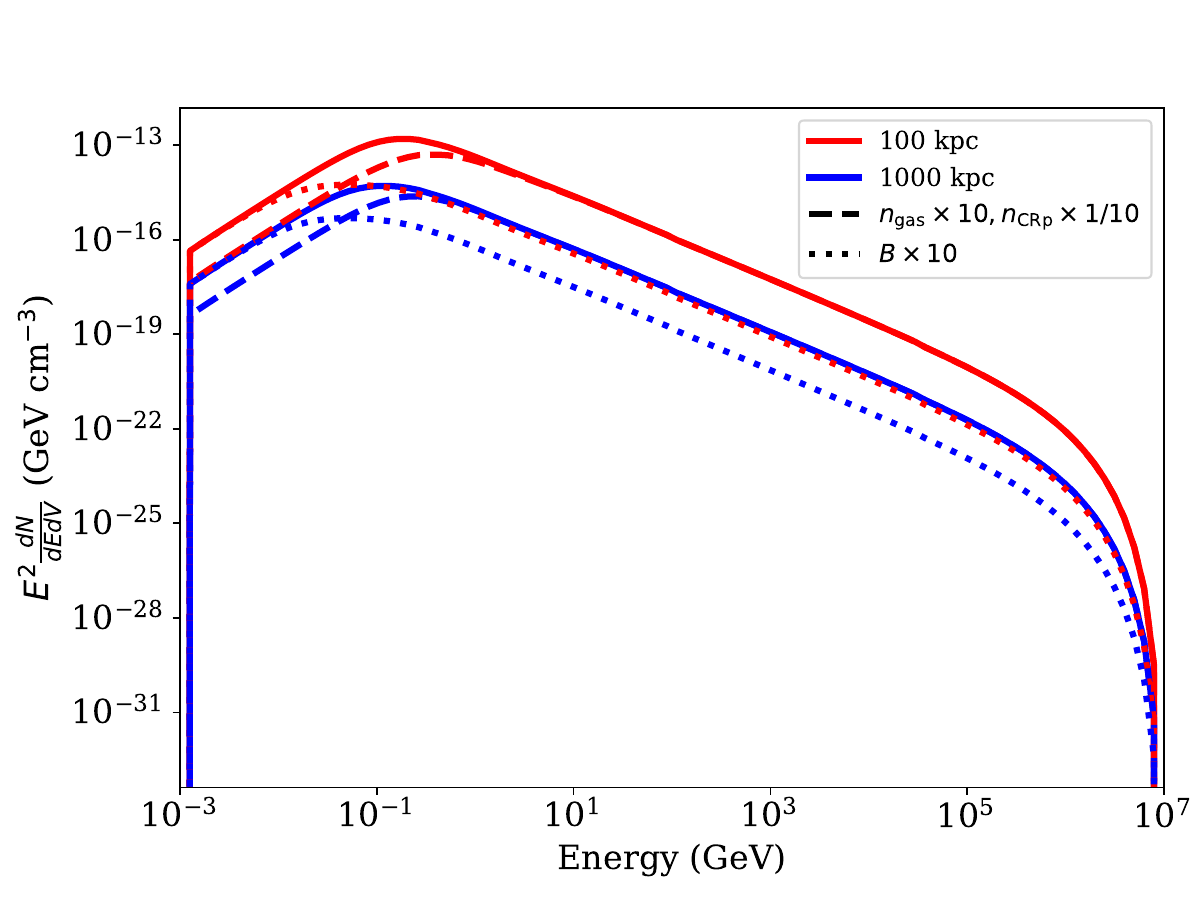}
\caption{{\bf Left}: Profile of the CRe$_2$, taken at 10 MeV and 10 TeV (after applying a multiplicative factor of $10^{15}$). {\bf Right}: Spectrum of the secondary electron at 100 and 1000 kpc.}
\label{fig:CRe2_steady_state}
\end{figure*}

When CRs are injected into the ICM, their evolution is expected to follow the diffusion-loss equation~\citep{Berezinskii1990},
\begin{equation}
        \frac{\partial n(E,r,t)}{\partial t} = -\vec{\nabla} \left(n(E,r,t) \ \vec{v}\right) + \nabla \left[D(E) \nabla n(E,r,t)\right] + \frac{\partial}{\partial E} \left[\ell(E, r)n(E,r)\right] + q(E,r,t),
\end{equation}
where $n(E,r,t) \equiv \frac{dN_{\rm CRe}}{dEdV}$ is the number of CRe per unit volume and energy, $q(E,r,t)$ is the injection rate, $\vec{v}$ is the ICM velocity, and $D(E)$ is the diffusion coefficient, and where the loss function, $\ell(E,r),$ is given by
\begin{equation}
\ell(E,r) = -\left(\left.\frac{dE}{dt}\right|_{\rm sync} + \left.\frac{dE}{dt}\right|_{\rm IC} + \left.\frac{dE}{dt}\right|_{\rm Coulomb} + \left.\frac{dE}{dt}\right|_{\rm Brem.}\right).
\end{equation}
Assuming that the CRe do not significantly diffuse, and assuming steady-state condition, we write the number density of CRe at equilibrium as
\begin{equation}
        \frac{dN_{\rm CRe}}{dE dV}(E, r) = \frac{1}{\ell(E,r)}\int_E^{\infty} q(\epsilon, r) d\epsilon,
        \label{eq:steady_state_computation_cre}
\end{equation}
where $q(\epsilon, r)$ is computed as the output of Eq.~\ref{eq:CRe_gamma_nu_injection}.

We note that we did not account explicitly for the possible reacceleration of seed electrons by ICM turbulences \citep[e.g.,][and references therein]{Brunetti2017}. However, because we did not model the details of the microphysics here, such a population might be included in the CRe$_1$, which are modeled independently, as discussed in Section~\ref{sec:modeling_non_thermal}. By doing so, we would implicitly assume that the reacceleration process, which would effectively contribute to a loss for our secondary electron population (i.e., a population transfer) is subdominant with respect to the losses. In this case the physical consistency between the different CR populations will not necessarily be verified. While reacceleration models are beyond the scope of the current work, we leave room for implementing reacceleration options in the {\tt MINOT} code in the future.

In Fig.~\ref{fig:CRe2_steady_state} we present the radial profile and the spectra of the CRe$_2$ in the steady-state approximation, with no diffusion. The profile becomes steeper at higher energy because inverse Compton and synchrotron losses become more important in this regime relative to the Coulomb loss, which is more efficient in the core. Moreover, electrons accumulate around 100 MeV because lower energy electrons quickly disappear because of Coulomb losses and higher energy electrons are more affected by inverse Compton and synchrotron losses. We also provide the same profiles and spectra when the magnetic field and thermal density are higher by a factor of 10. In the latter case, we also decreased the number of CRp by a factor of 10 so that the rate of proton-proton collisions was conserved. An increase in thermal plasma density leads to a much flatter profile because Coulomb losses are far higher in the core where the density is high. An increased magnetic field also flattens the profile, but less drastically because the magnetic field profile itself is flatter. In the spectrum, the magnetic field has a stronger effect at high energy, while the increase in thermal plasma density leads to more losses at low energy. The peak of the secondary electron spectrum therefore depends on the competition between the energy losses in the magnetic field and in thermal plasma.

\section{Multiwavelength observables}\label{sec:observables}
In this section, the physical properties of the cluster (Section~\ref{sec:physical_modeling}) and the production of secondary particles in the ICM (Section~\ref{sec:particle_interactions}) are used to compute the observables of galaxy clusters related to the diffuse gas component. This includes the tSZ effect, the thermal X-ray emission, the radio synchrotron emission, the inverse Compton emission, and the $\gamma$ and neutrino emission from hadronic processes. We focus here on showing the effect of model changes on the observables using our baseline cluster model.

\subsection{General considerations}\label{sec:general_considerations}
In general, the cluster observables are associated with physical processes at play in the ICM, which can be described in terms of production rate (this does not strictly apply to the tSZ signal because it is a spectral distortion, as we discuss in Section~\ref{sec:obs_tSZ}). We define $Q(r, E) \equiv \frac{dN}{dEdVdt}$, the emission rate associated with the physical process considered. For instance, in the case of X-ray emission, $Q$ would be the number of X-ray photons emitted per unit volume, per unit of time, and per unit energy in the ICM. 

The surface brightness (or flux per solid angle) at a projected distance $R$ from the center is therefore given by integrating $Q(r,E)$ over the line of sight as 
\begin{align}
        \frac{dN}{dEdSdtd\Omega}(R, E) &= \frac{D_{A}^2}{4 \pi D_{L}^2} \int_{-\infty}^{+\infty} Q(r) d\ell \notag\\
&=  \frac{D_{A}^2}{4 \pi D_{L}^2} \int_R^{R_{\rm max}} \frac{2 r Q(r)}{\sqrt{r^2-R^2}} dr,
\label{eq:los_integ}
\end{align}
where the factor $D_{A}^2$ accounts for the conversion from physical area into solid angle, and the normalization by $4 \pi D_{L}^2$ assumes that the emission is isotropic. We note that Eq.~\ref{eq:los_integ} is valid in the small-angle approximation (i.e., assumes the cluster size to be small against the distance to the observer), which is expected to be accurate for our purpose because the extent of clusters never exceeds a few degrees. It also neglects the redshift extent of the cluster.

Based on Eq.~\ref{eq:los_integ}, we now wish to compute several quantities accessible from observations: 1) the surface brightness profile; 2) the spectrum, by integrating the signal over the cluster volume or the solid angle; 3) the total flux, by integrating over both the volume and the energy (or at fixed energy); and 4) the map of the signal, which in our case is equivalent to the surface brightness profile because of azimuthal symmetry, but allows us to generate spatial templates for dedicated analysis.

The surface brightness profiles (and maps), are computed by log-log integration of the quantity $\frac{dN}{dEdSdtd\Omega}(R, E)$ over the requested energy range, or simply by fixing the energy to the one required by the observation (as is done, e.g., in the radio and millimeter domain). In the case of the map, the signal is projected on a grid corresponding to the header (or sampling properties) set by the user. An option allows the user to normalize the map to the total flux so that the map only accounts for the spatial dependence of the signal in units proportional to the inverse solid angle. This proves useful for a $\gamma$-ray analysis, for example, in which image templates are needed.

The spectra are computed in a similar way, by log-log integration over the volume. In this case, two possibilities are available. 1) Integration over the solid angle within a circle of radius $R_{\rm max}$, so that the total integration volume is a cylinder. 2) The emission rate $Q(r,E)$ can be integrated spherically up to $R_{\rm max}$ before normalization by $4 \pi D_{L}^2$. The two quantities only differ by the definition of the integration volume, and should converge when all the cluster emission is accounted for with increasing $R_{\rm max}$. The cylindrical integration resembles more what would be accessible directly from observations, while the spherical integration is a more natural from a physical point of view because it returns a quantity that is computed in a single physical (3D) radius.

In order to compute the flux, we integrated over the cluster volume and the energy. Alternatively, the energy might be fixed, as discussed above. By definition, the luminosity of a given source within the energy band $\Delta E \equiv [E_1, E_2]$ is given by
\begin{equation}
        L_{\Delta E} =  4 \pi D_L^2 F_{\Delta E}.
\end{equation}

We also applied the redshift stretching to the energy of the photons. This function can be switched off by the user.

\subsection{Thermal X-ray emission}
\begin{figure*}
\centering
\includegraphics[width=0.42\textwidth]{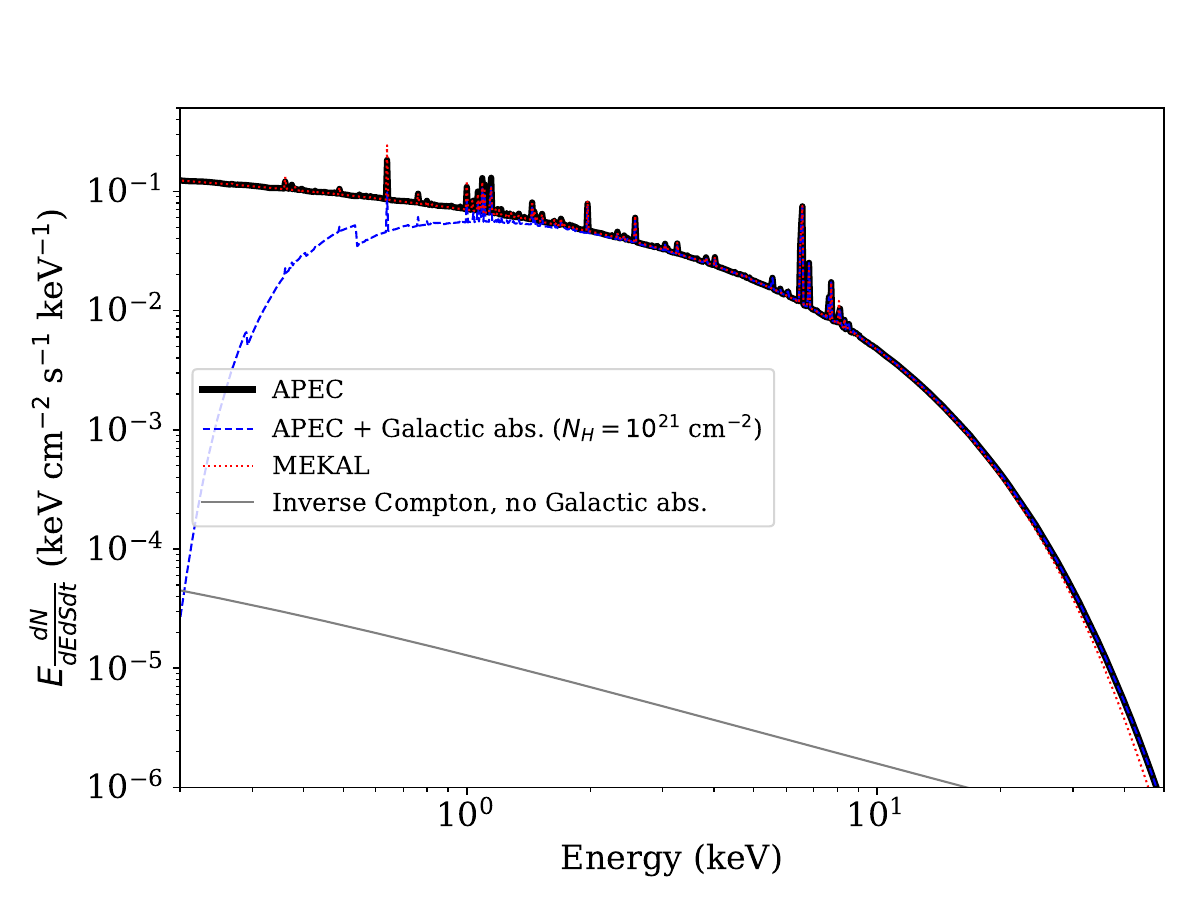}
\includegraphics[width=0.42\textwidth]{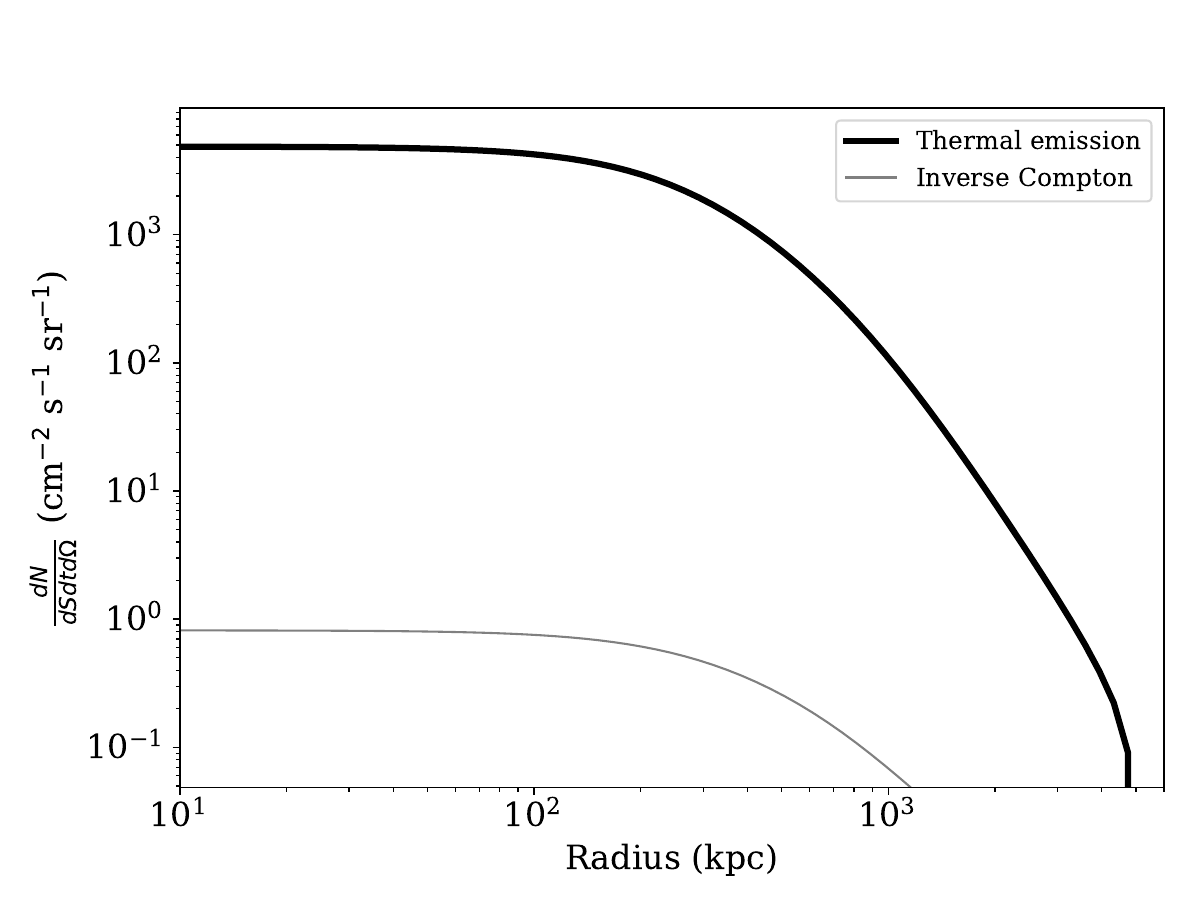}
\caption{Observables associated with the X-ray emission. {\bf Left}: X-ray spectrum within $R_{500}$ shown for both MEKAL and APEC models, as well as with and without galactic photoelectric absorption. The MEKAL model is difficult to distinguish because it coincides very well with the APEC model. {\bf Right}: Surface brightness profile in the band 0.1-2.4 keV. The dynamical range of the profile amplitude has been set to the same value for all observables.}
\label{fig:obs_xray}
\end{figure*}

Because of the gas temperature (a few keV) and the density (typically $10^{-3}-10^{-5}$ cm$^{-3}$), the leading emission process at X-ray energies is thermal Bremsstrahlung \citep[see][for reviews]{Sarazin1986, Bohringer2010}. The X-ray emission is thus a direct probe for the thermal gas density. It presents a characteristic exponential cutoff at high energies, determined by the gas temperature. Heavy elements also induce a large number of spectral lines. The X-ray surface brightness is generally expressed as
\begin{equation}
        S_{\rm X} = \frac{1}{4 \pi (1+z)^4} \int n_e^2 \Lambda(T_e, Z) dl,
\end{equation}
where $\Lambda(T_e, Z)$ is the cooling function, which varies with temperature.

In practice, {\tt MINOT} uses the {\tt XSPEC} software \citep{Arnaud1996} to directly compute the counts using either the MEKAL or APEC X-ray plasma spectral models. These models require the ICM abundance, the redshift, the temperature, the energy range, and a normalization defined as
\begin{equation}
        {\rm norm} = \frac{10^{{-14}}}{4 \pi \left(\left(\frac{D_A}{1 \ {\rm cm}}\right) (1+z)\right)^2} \int \left(\frac{n_e}{1 \ {\rm cm}^3}\right) \left(\frac{n_H}{1 \ {\rm cm}^3}\right)  dV.
\end{equation}
{\tt MINOT} also accounts for the foreground photoelectric absorption using the value of the hydrogen column density at the cluster location. The {\tt XSPEC} outputs are then normalized to compute the emission rate (counts or energy) per unit volume and time. It is also possible to account for the response function of X-ray satellites, so that the outputs are normalized by the effective area of the observation. The spectrum, surface brightness profile and maps, and flux are then extracted as discussed in Section~\ref{sec:general_considerations}.

In Fig.~\ref{fig:obs_xray} we illustrate these X-ray observables in the case of our baseline cluster. The two models MEKAL and APEC agree well. The dashed blue spectrum shows the effect of the photoelectric absorption from the foreground, leading to low energy cuts. At high energy, the exponential cutoff is clearly visible, and the spectral lines are also visible below 10 keV. The raw signal associated with the inverse Compton emission is also shown; it is discussed in Section~\ref{sec:obs_inverse_compton}. We note that it is well below the thermal X-ray emission, but might become significant with increasing energy. The surface brightness profile, computed between 0.1 keV and 2.4 keV, drops very rapidly in the outskirt because the signal is proportional to the density squared. The integrated flux reaches about 1 ph cm$^{-2}$ s$^{-1}$ at large radii. 

\subsection{Thermal Sunyaev-Zel’dovich signal}\label{sec:obs_tSZ}
\begin{figure*}
\centering
\includegraphics[width=0.42\textwidth]{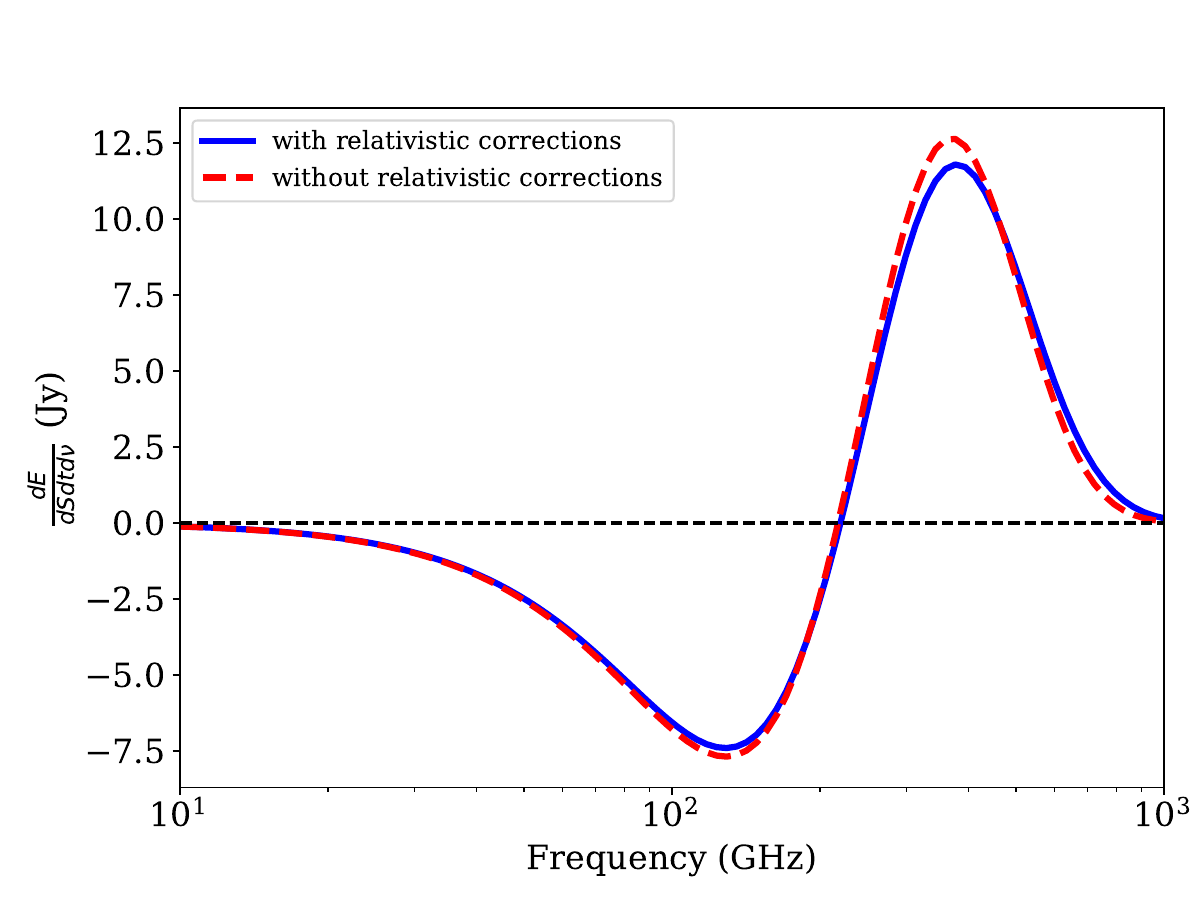}
\includegraphics[width=0.42\textwidth]{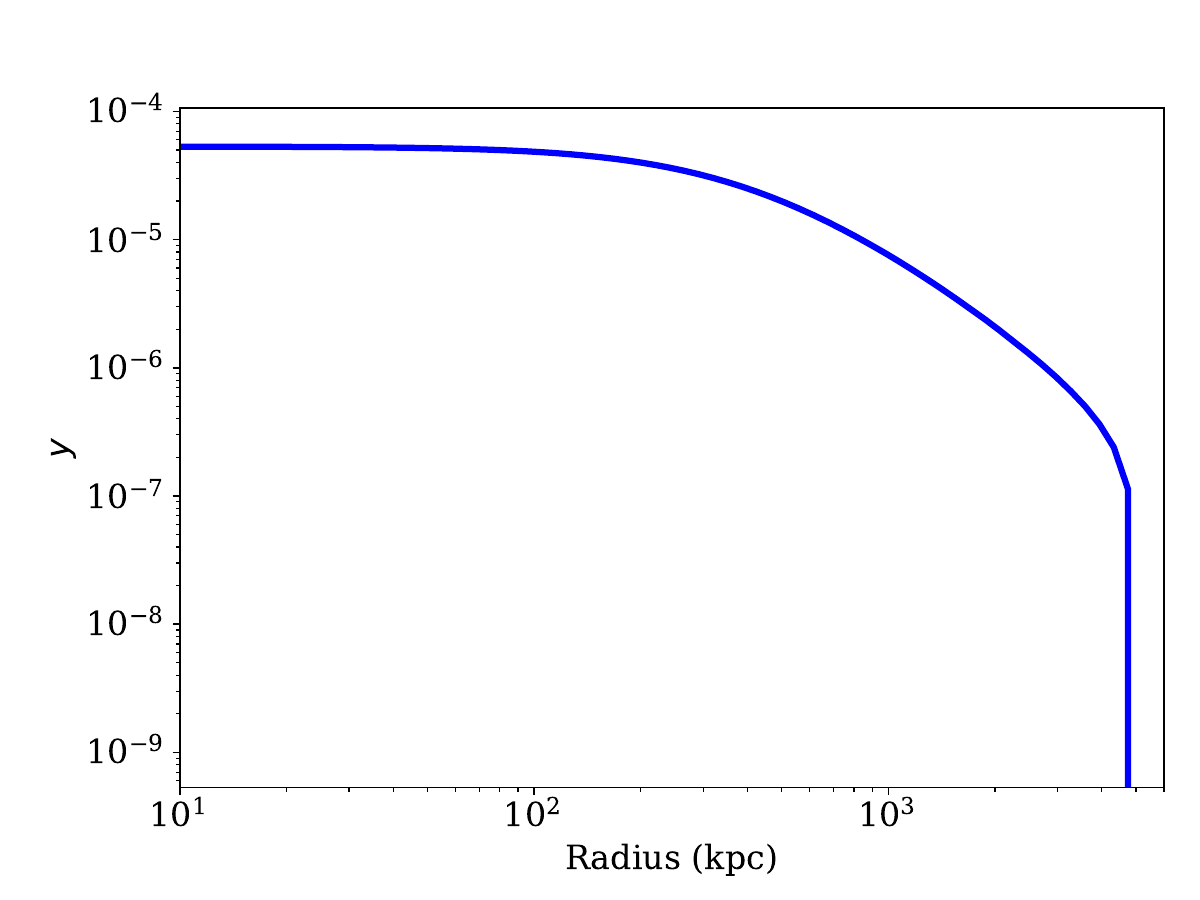}
\caption{Observables associated with the tSZ effect. {\bf Left}: tSZ spectrum within $R_{500}$. We also provide the spectrum in the case when relativistic corrections are neglected for illustration. {\bf Right}: Compton parameter profile. The dynamical range of the profile amplitude has been set to the same value for all observables.}
\label{fig:obs_tsz}
\end{figure*}

The tSZ effect distorts the CMB blackbody spectrum as a result of inverse Compton scattering onto energetic thermal electrons \citep[see][for reviews]{Birkinshaw1999,Mroczkowski2019}. 
Because it is a spectral distortion, it does not suffer from redshift dimming, and the general considerations of Section~\ref{sec:general_considerations} do not strictly apply here. The change in surface brightness is expressed as
\begin{equation}
        \frac{\Delta I_{\rm tSZ}}{I_0} = y \ f(x, T_e)\end{equation}
with respect to the CMB, $I_0 = \frac{2 (k_{\rm B} T_{\rm CMB})^3}{(hc)^2} \simeq 270.1$ MJy sr$^{-1}$. The parameter $y$ is the so-called Compton parameter, which gives the normalization of the tSZ effect. It provides a measurement of the thermal electron pressure integrated along the line of sight as
\begin{equation}
        y = \frac{\sigma_{\mathrm{T}}}{m_e c^2} \int P_e d\ell.
        \label{eq:y_compton}
\end{equation}
The frequency dependence of the tSZ effect is given by
\begin{equation}
        f(x, T_e) = \frac{x^4 e^x}{\left(e^x-1\right)^2} \left(x  \ \mathrm{coth}\left(\frac{x}{2}\right) - 4\right) \left( 1 + \delta_{\rm tSZ}(x, T_e) \right), 
\end{equation}
where $x = \frac{h \nu}{k_{\rm B} T_{\rm CMB}}$. The term $\delta_{\rm tSZ}(x, T_e)$ is a relativistic correction that introduces a small temperature dependence to the tSZ effect, and becomes important when the temperature becomes higher than about 10 keV, depending on the frequency. The relativistic correction was implemented following  \cite{Itoh2003}, who are expected to be accurate at the percent level up to 50 keV. When relativistic corrections are neglected, the tSZ spectrum is null at 217 GHz, negative below (with a minimum around 150 GHz), and positive above (peaking at about 350 GHz).

The tSZ integrated flux, often used to track the cluster mass, can be expressed as
\begin{equation}
        Y_{\rm cyl}(R) = \int_0^R 2 \pi r y dr
\end{equation}
in the case of cylindrical integration, or as
\begin{equation}
        Y_{\rm sph}(R) = \frac{\sigma_{\mathrm{T}}}{m_e c^2} \int_0^R 4\pi r^2 P_e dr
\end{equation}
for the spherically integrated flux. The integrated flux, $Y_{\rm cyl,sph}$, can be expressed in units of surface or normalized by $D_{A}^2$ to be homogeneous to solid angle, as is commonly done in the literature.

In Fig.~\ref{fig:obs_tsz} we illustrate the spectrum and profile (shown in terms of the Compton parameter) of our reference cluster. The tSZ flux can be significant down to a few GHz, and could thus affect radio synchrotron observations. On the other hand, the synchrotron emission is not shown here, but is far lower than the tSZ signal in the considered frequency range. Given the linear sensitivity of the tSZ signal to the pressure, the profile is relatively flat (e.g., compared to the X-ray surface brightness).

\subsection{$\gamma$-ray hadronic emission}\label{sec:obs_hadronic}
\begin{figure}
\centering
\includegraphics[width=0.42\textwidth]{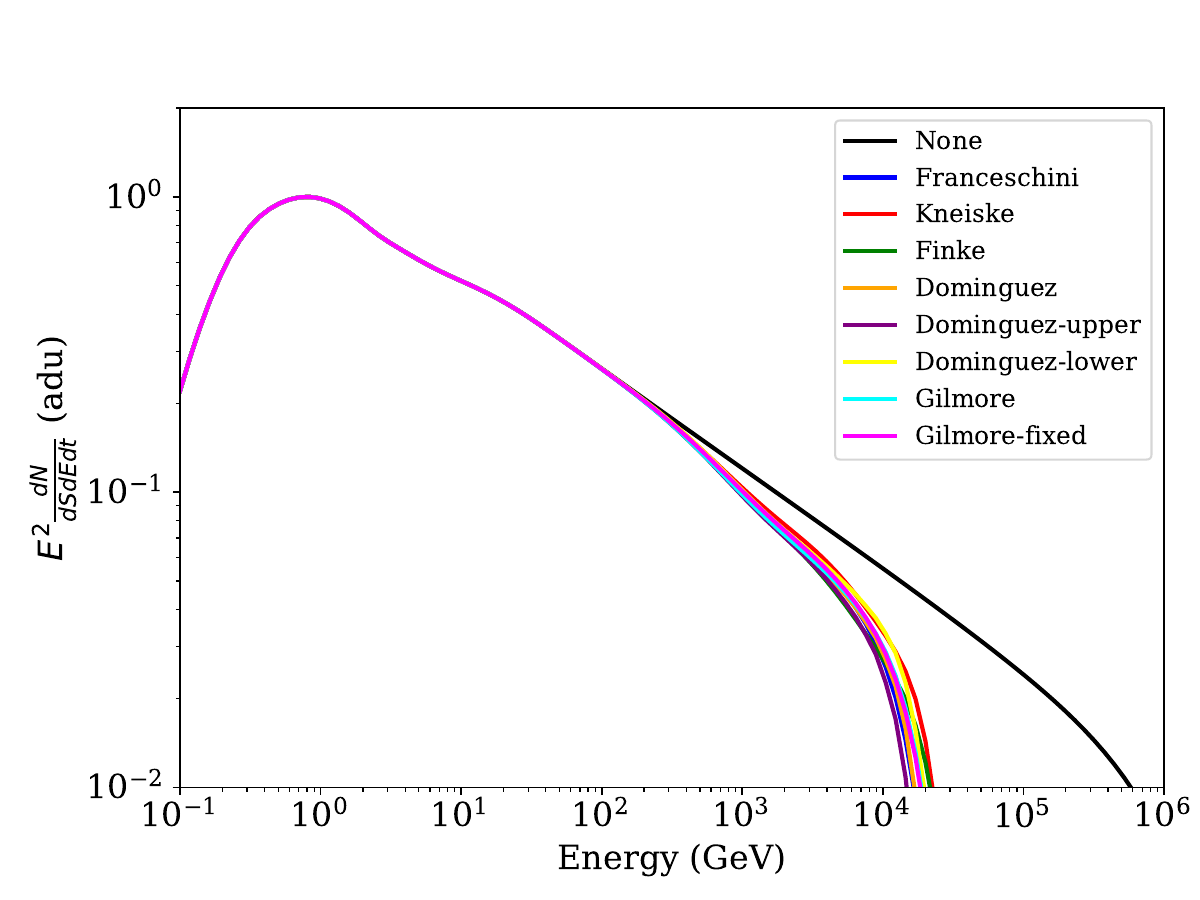}
\caption{Effect of the EBL on the normalized spectra of a baseline model cluster at $z=0.02$. All the available models are shown as indicated in the legend. Because of the redshift stretching, the shape of the $\gamma$-ray spectrum slightly changes even in the case without EBL absorption ('none').}
\label{fig:ebl_absorb}
\end{figure}

\begin{figure*}
\centering
\includegraphics[width=0.42\textwidth]{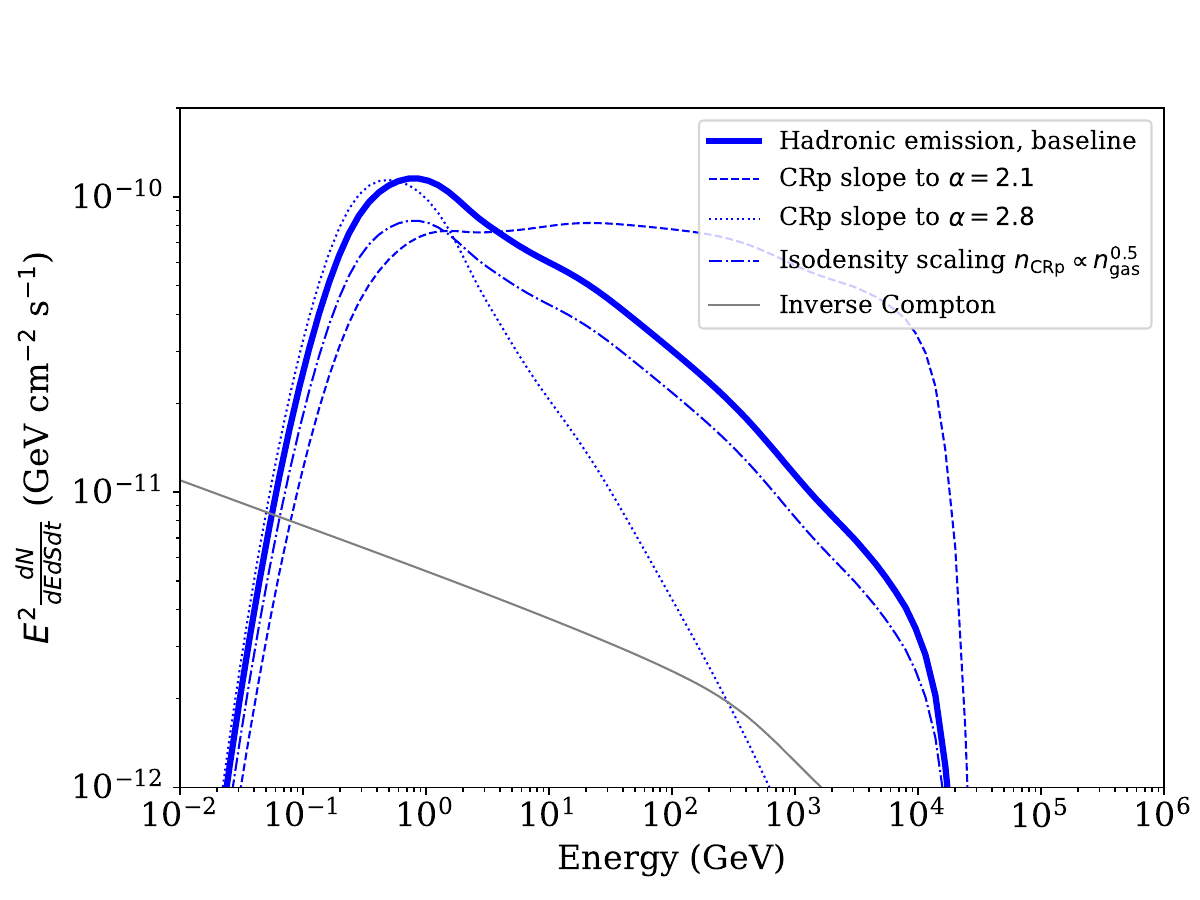}
\includegraphics[width=0.42\textwidth]{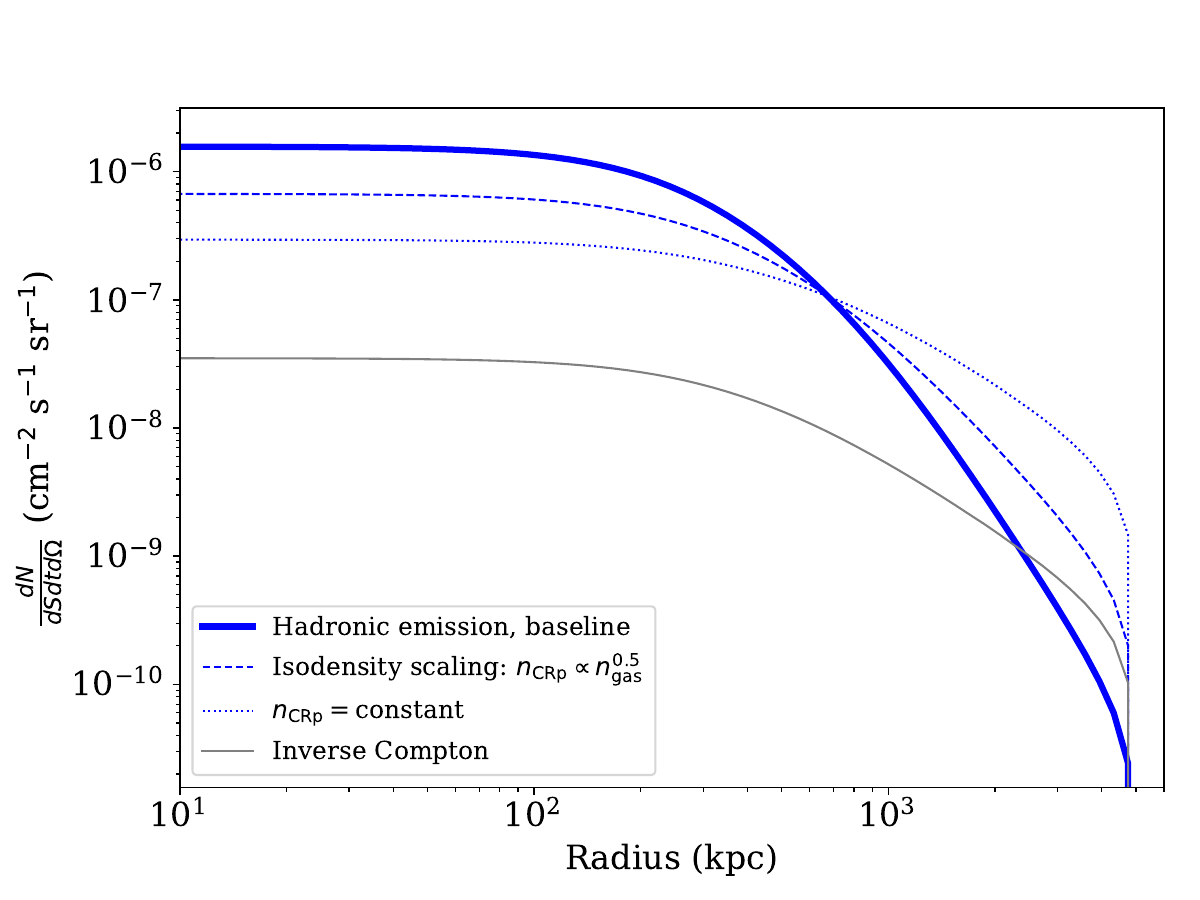}
\caption{Observables associated with the $\gamma$-ray hadronic emission. The signal coming from inverse Compton interactions is also shown for comparison. {\bf Left}: $\gamma$-ray spectrum within $R_{500}$. {\bf Right}: $\gamma$-ray profile integrated between 1 GeV and 1 TeV. The dynamical range of the profile amplitude has been set to the same value for all observables.}
\label{fig:obs_gamma}
\end{figure*}

In the case of high-energy photons, the absorption by the extragalactic background light \cite[EBL,][]{Dwek2013} needs to be accounted for, which is thought to have been produced by the sum of all light contributions (e.g., starlight or dust reemission) at all epochs in the Universe. While traveling from the cluster to the Earth, $\gamma$-rays may interact with the EBL through electron-positron pair production, and thus be effectively absorbed along the way as 
\begin{equation}
        \frac{dN}{dEdSdt} \longrightarrow \frac{dN}{dEdSdt} \times {\rm exp} \left( -\tau(E) \right),
\end{equation}
with $\tau(E)$ the optical depth. EBL absorption depends on redshift and on the energy of the $\gamma$-rays. To account for EBL absorption, we used the {\tt ebltable} Python package \footnote{\url{https://github.com/me-manu/ebltable/}}, which reads in and interpolates tables for the photon density of the EBL and the resulting opacity for high-energy $\gamma$-rays. This package provides different models for the EBL based on \citet{Franceschini2008}, \citet{Kneiske2010}, \citet{Finke2010}, \citet{Dominguez2011}, and \citet{Gilmore2012}. We illustrate the effect of the EBL for various models available in Fig.~\ref{fig:ebl_absorb} for a cluster redshift $z=0.02$ . While it is crucial to account for the EBL, especially at very high energies, the uncertainties associated with the EBL models are expected to be small.

The intergalactic magnetic fields probably also affect the predictions for the $\gamma$-ray observables \citep[e.g.,][]{Neronov2009}. However, because of the current uncertainties on the properties of the magnetic field, its effect remains uncertain. While this effect is not yet implemented, it is being considered for the future improvement of the {\tt MINOT} code.

The $\gamma$-ray production rate resulting from hadronic interaction is computed following \cite{Kafexhiu2014}, as described in Section~\ref{sec:prod_rate_of_secondaries}. We compute the spectrum (within $R_{500}$, using spherical integration) and profile as detailed in Section~\ref{sec:general_considerations}. These quantities are displayed in Fig.~\ref{fig:obs_gamma} for our baseline cluster model. The spectrum peaks at GeV energies, quickly vanishes at lower energies, and is affected by a cutoff at high energy due to the EBL. We also show the effect of the change in CRp slope on the $\gamma$-ray spectrum for a fixed-normalization $X_{CRp}$. In the case of a flatter CRp profile, the amplitude of the spectrum is reduced because the number of proton-proton collisions is reduced by the lower spatial coincidence of thermal and CRp. The inverse Compton signal computed for our baseline model is also shown for comparison. It is below the hadronic emission except at low energy. The profile presents a compact signal because it arises from the product of the thermal electron number density and the CR density. When the CR number density profile is flattened, the $\gamma$-ray signal itself becomes flatter. Nevertheless, it still decreases with radius even for a completely flat CR number density profile because the thermal density profile remains peaked. The inverse Compton signal is expected to be flatter than the signal of the hadronic emission, but it is also lower in amplitude in our baseline model. The integrated spectrum between 1 GeV and 1 TeV almost reaches $10^{-9}$ photon cm$^{-2}$ s$^{-1}$ in the case of this baseline model. 

\subsection{Neutrino (hadronic) emission}
\begin{figure*}
\centering
\includegraphics[width=0.42\textwidth]{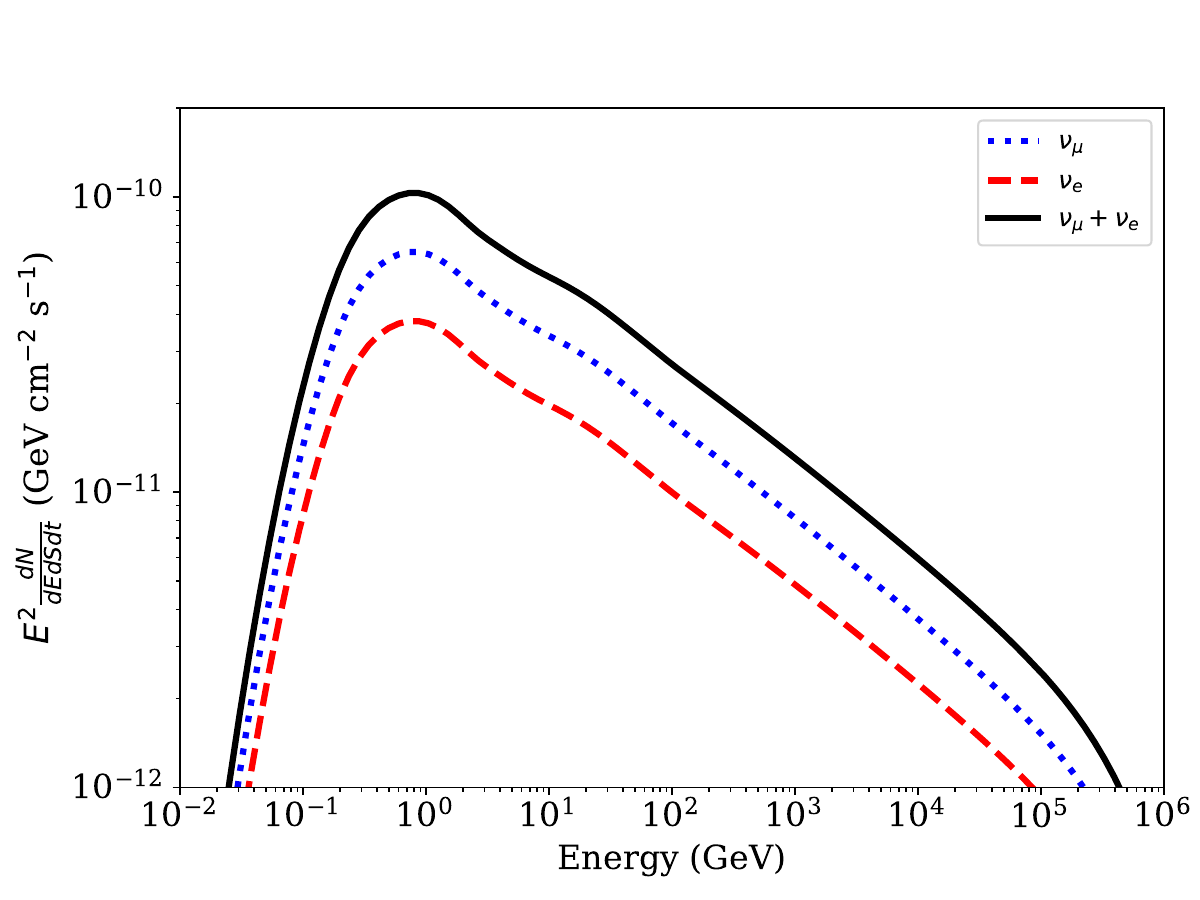}
\includegraphics[width=0.42\textwidth]{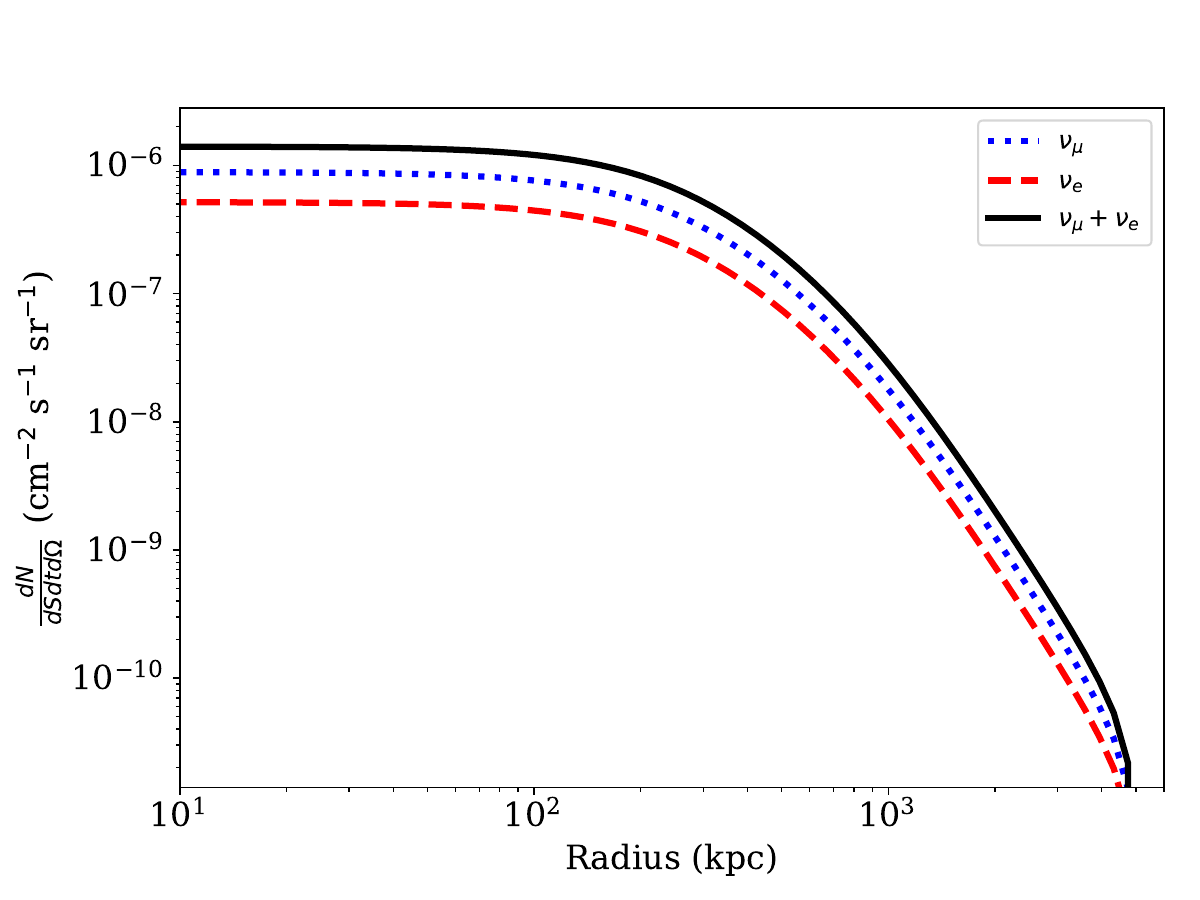}
\caption{Observables associated with neutrino hadronic emission. {\bf Left}: Neutrino spectrum within $R_{500}$. {\bf Right}: Neutrino profile integrated between 1 GeV and 1 TeV. The dynamical range of the profile amplitude has been set to the same value for all observables.}
\label{fig:obs_neutrino}
\end{figure*}

In contrast to the $\gamma$-rays, the neutrinos are not affected by the EBL absorption. Except for the EBL, their observables are computed in the same way as the $\gamma$-rays associated with hadronic interactions (Section~\ref{sec:obs_hadronic}); the production rate is computed following a combination of \cite{Kelner2006} and \cite{Kafexhiu2014}, as described in Section~\ref{sec:prod_rate_of_secondaries}.

Figure~\ref{fig:obs_neutrino}  illustrates the neutrino observable both for the muonic and electronic neutrinos in the case of the spectrum and the profile, and for the sum of the two for the flux. Because of the neutrino oscillation, we would in practice expect a mix between the ratio of neutrinos of different flavors. Because the processes associated with the neutrino is the same as that of the $\gamma$-rays, their observables are very similar to one another, except for a small difference in the normalization.

\subsection{Inverse Compton emission}\label{sec:obs_inverse_compton}
\begin{figure*}
\centering
\includegraphics[width=0.42\textwidth]{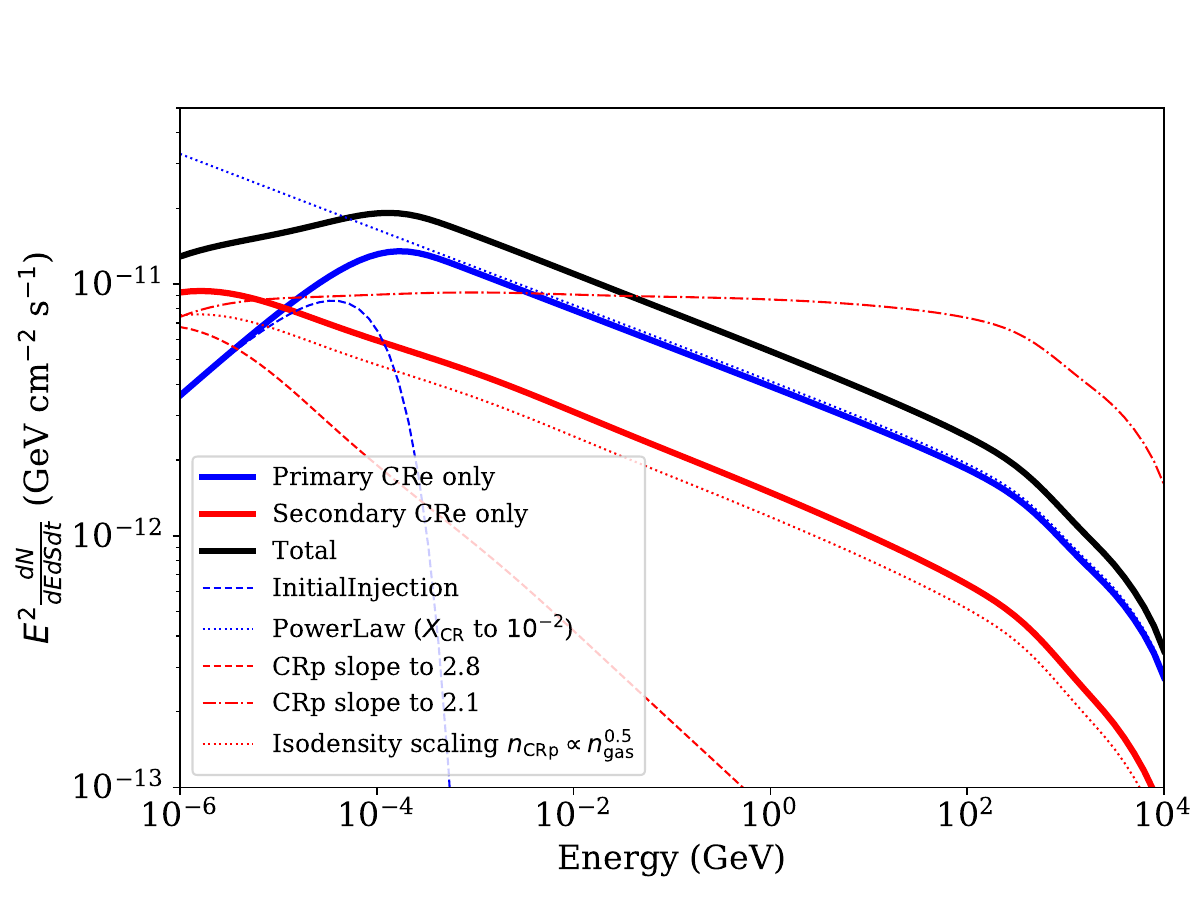}
\includegraphics[width=0.42\textwidth]{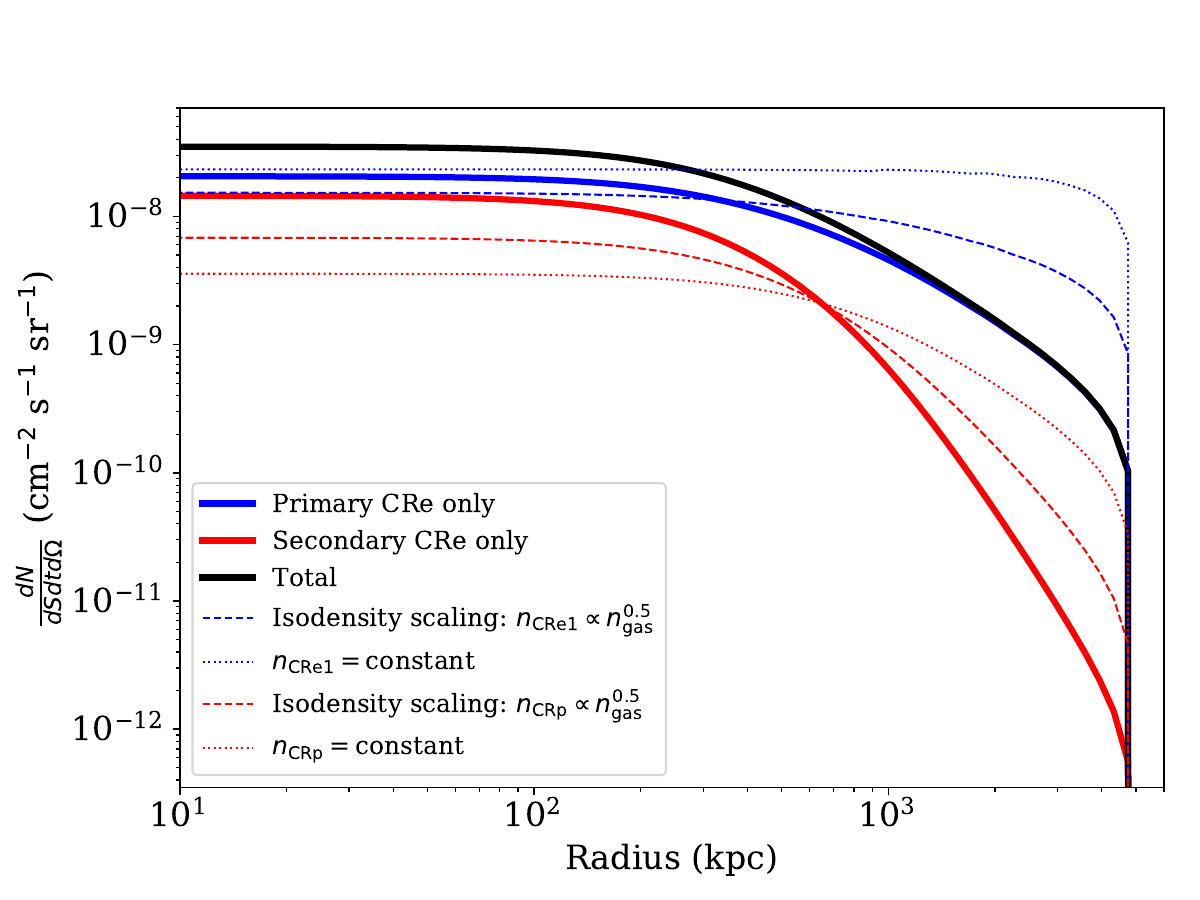}
\caption{Observables associated with inverse Compton emission. {\bf Left}: Inverse Compton spectrum within $R_{500}$, with the contribution from both primary and secondary electrons. {\bf Right}: Inverse Compton profile integrated between 1 GeV and 1 TeV. The dynamical range of the profile amplitude has been set to the same value for all observables.}
\label{fig:obs_ic}
\end{figure*}

The inverse Compton emission is also affected by the EBL absorption in the high-energy limit. We refer to Section~\ref{sec:obs_hadronic} and Fig.~\ref{fig:obs_gamma} for this effect. 

We use the analytical approximation for the treatment of inverse Compton scattering of relativistic electrons in the CMB blackbody radiation field given by \cite{Khangulyan2014}, which is expected to be accurate within 1\% uncertainty throughout the application range. In particular, their Eq.~14 gives us the number of inverse Compton photons produced per unit energy and time per CRe as a function of the CRe energy, $\frac{dN_{\rm IC}}{dEdt}$, which we express as
\begin{equation}
        \frac{dN_{\rm IC}}{dE_{\rm IC}dt} = \frac{2 r_0^2 m_e^3 c^4}{\pi \hbar^3} \left(\frac{k_{\rm B} T_{\rm CMB} (1+z)}{E_{\rm CRe}}\right)^2 G_{\rm IC}\left(E_{\rm IC}, E_{\rm CRe}\right),        
\end{equation}
with $G_{\rm IC}\left(E_{\rm IC}, E_{\rm CRe}\right)$ an analytical function computed following \cite{Khangulyan2014}, using the approximation given by their Eq.~24. We thus integrate this quantity over the electron energy, accounting for the amount of CRe in the ICM. The emissivity is expressed as
\begin{equation}
        \frac{dN_{\rm IC}}{dE_{\rm IC}dVdt} = \int J_e(E_{\rm CRe}) \frac{dN_{\rm IC}}{dE_{\rm IC}dt} dE_{\rm CRe},
\end{equation}
where $J_e(E_{\rm CRe}) \equiv \frac{dN_{\rm CRe}}{dE_{\rm CRe}dV}$ is the CRe number density, summing the contributions from primary and secondary electrons.

In Fig.~\ref{fig:obs_ic} we illustrate the observables associated with inverse Compton emission for our baseline cluster model. The spectrum shape reflects the complex processing of secondary electrons by their production rate in hadronic interaction, their losses in the ICM, and the production of inverse Compton after having also summed the CRe$_1$. 
In particular, changing the distribution of CRe$_1$ to an initial injection model removes relativistic electrons with energy higher than $E_{\rm break}$, and thus does not remove the inverse Compton emission at energies above $\sim E_{\rm break}/100$. In contrast, when a power-law model is used, more CRe are present at low energy, which increases the inverse Compton emission in this regime. The secondary electrons change the slope of the inverse Compton emission according to the CRp slope. In addition, as in the case of hadronic emission, flattening the CRp profile decreases the production of secondary particles, and thus that of inverse Compton associated with secondary electrons. The inverse Compton profile is relatively flat for primary electrons because the signal depends linearly on the CRe spatial distribution. For the secondary electrons, the profile is more compact because the secondary electrons are produced proportionally to the product of the thermal gas and the CRp number density. For both populations, we also show the effect of flattening the profiles of either the CRe$_1$ or the CRp. The integrated flux reaches more than $10^{-11}$ ph cm$^{-2}$ s$^{-1}$ between 1 GeV and 100 TeV in our baseline model.

\subsection{Radio synchrotron emission}
\begin{figure*}
\centering
\includegraphics[width=0.42\textwidth]{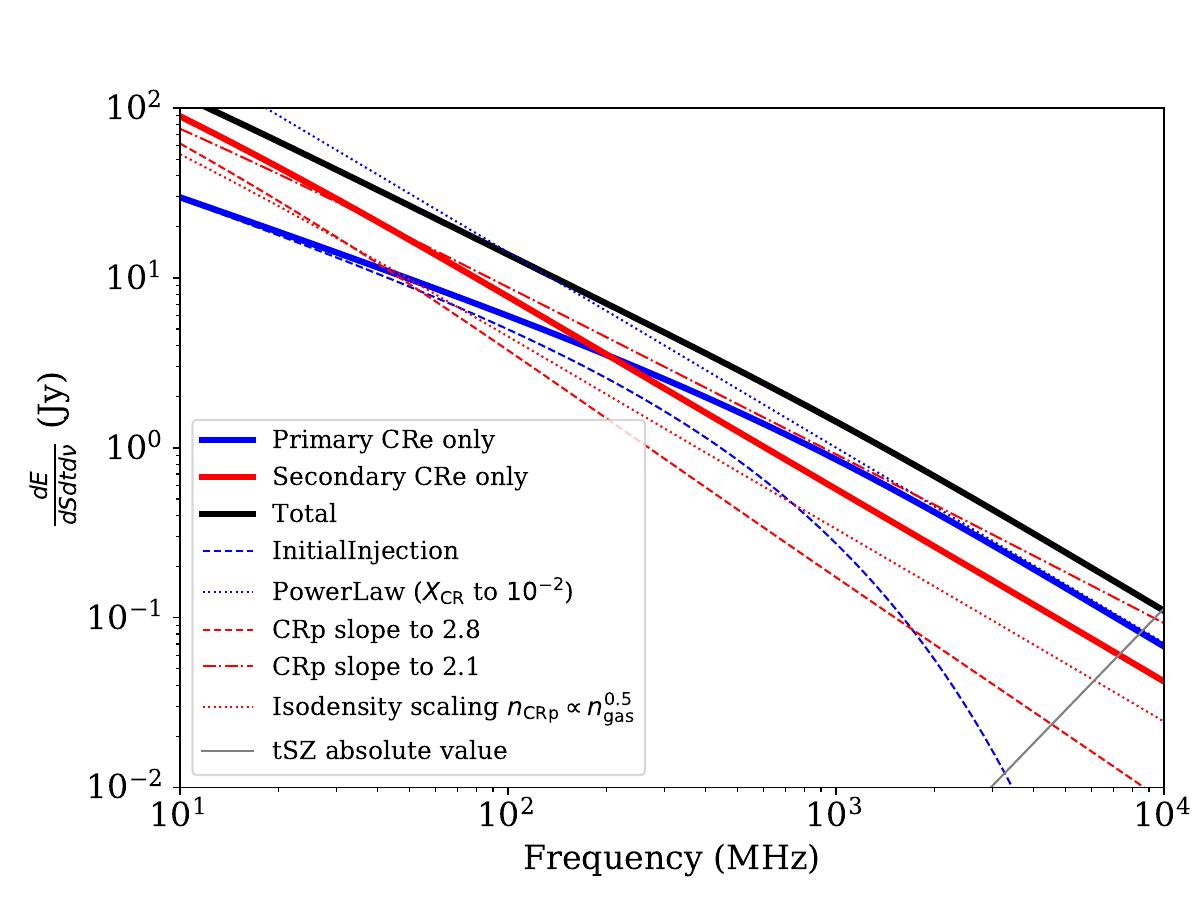}
\includegraphics[width=0.42\textwidth]{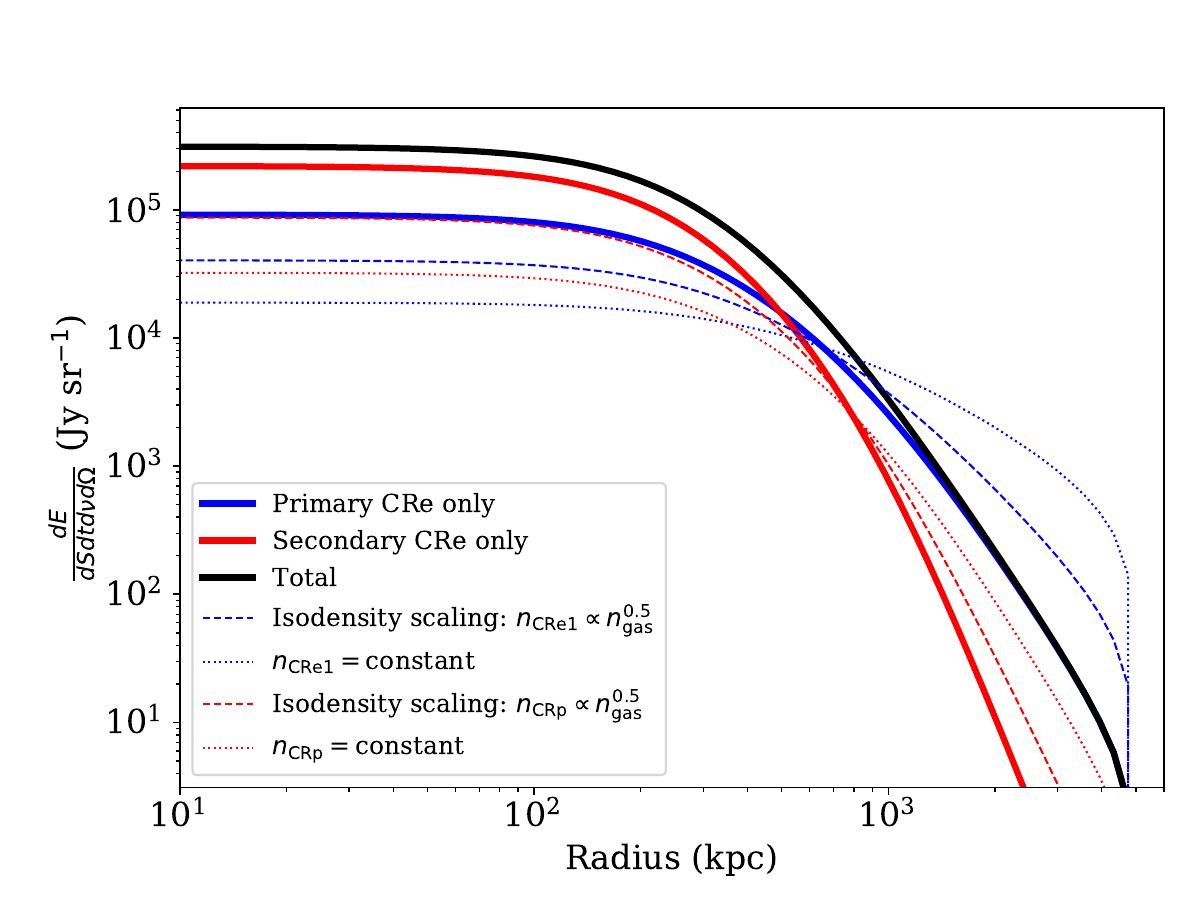}
\caption{Observables associated with the synchrotron emission. {\bf Left}: Synchrotron spectrum within $R_{500}$, with the contribution from both primary and secondary electrons. {\bf Right}: Synchrotron emission profile at 100 MHz. The dynamical range of the profile amplitude has been set to the same value for all observables.}
\label{fig:obs_sync}
\end{figure*}

The {\tt MINOT} code focuses on the diffuse galaxy cluster emission associated with the bulk of the X-ray emitting ICM. In the case of the diffuse synchrotron emission, we therefore focus on the emission associated with radio halos and leave radio shocks \citep[or relics, see][]{vanWeeren2019} aside. Because the orientation of the magnetic field is expected to be chaotic in the bulk ICM regions of a galaxy cluster, we need to average the standard energy distribution of the synchrotron emission over the directions of the field orientation. To do so, we follow the results of \cite{Aharonian2010}, Appendix D, in which the orientation of the magnetic field is assumed to be randomized. This provides a convenient approximation with an accuracy better than 0.2\% over the entire energy range \cite[see][for more details on the approximation and its accuracy]{Aharonian2010}.

As in the case of inverse Compton emission, we express
\begin{equation}
        \frac{dN_{\rm sync}}{dE_{\rm sync}dVdt} = \int J_e(E_{\rm CRe}) \frac{dN_{\rm sync}}{dE_{\rm sync}dt} dE_{\rm CRe},
\end{equation}
where
\begin{equation}
\frac{dN_{\rm sync}}{dE_{\rm sync}dt} = \frac{3 e^3 B}{8 \pi^2 \epsilon_0 m_e c \ \hbar E_{\rm sync}} \tilde{G}(E_{\rm sync} / E_c).
\end{equation}
The quantity $E_c = \frac{3 e B \ \hbar \gamma^2}{2 m_e}$ is the synchrotron characteristic energy and $\tilde{G}(x)$ the emissivity function of synchrotron radiation, which quickly increases from $x=0$ to $x \simeq 0.23$ and smoothly vanishes for increasing $x$ \citep[see][]{Aharonian2010}.

In Fig.~\ref{fig:obs_sync} we illustrate the observables associated with synchrotron emission, including the contribution from primary and secondary electrons. As for the inverse Compton case, the emission reflects the complex processing of secondary electrons, while it is more direct for the primary electrons. In particular, the curvature in the synchrotron emission is due to the losses of CRe at high energies. When the CRe$_1$ population model is changed to an initial injection scenario, the high-frequency curvature is significantly enhanced by the lack of very high energy electrons. In the case of a power-law CRe population, which extends to a lower energy, the spectrum is enhanced at low frequency. The slope of the CRp for secondary electrons is directly reflected in the synchrotron spectrum slope. Similarly as in the case of hadronic $\gamma$-ray emission, flattening the CRp profile decreases the amount of synchrotron emission because fewer secondary electrons are produced. As highlighted in the figure, the tSZ contribution might be high at high frequencies, and it might mimic a curved synchrotron spectrum if not accounted for. The synchrotron profile is more compact than that of the inverse Compton emission because it also depends on the magnetic field, which decreases with radius. For the inverse Compton emission, the profile is more compact for secondary electrons because it depends on the product of the thermal density and the CRp number density. We also show that the flattening of the CR population is reflected in the synchrotron profile. The flux at 100 MHz reaches typical values of 10 Jy in our baseline model.

\section{Comparison to the literature}\label{sec:comparison}
In order to validate the modeling and to further illustrate the use of the {\tt MINOT} code, we compared the model predictions to results obtained in the literature and existing data. To do so, we used the models of the XCOP clusters defined in Section~\ref{sec:baseline_cluster_models} and Table~\ref{tab:xcop_properties}. We focused on millimeter and X-ray data for the thermal part and on $\gamma$-rays and radio data for the nonthermal component. Because the sensitivity of current neutrino telescopes is limited and because of the typical predicted fluxes, we leave the neutrino emission model predictions aside.

\subsection{Thermal gas}
\begin{figure*}
\centering
\includegraphics[height=7.5cm]{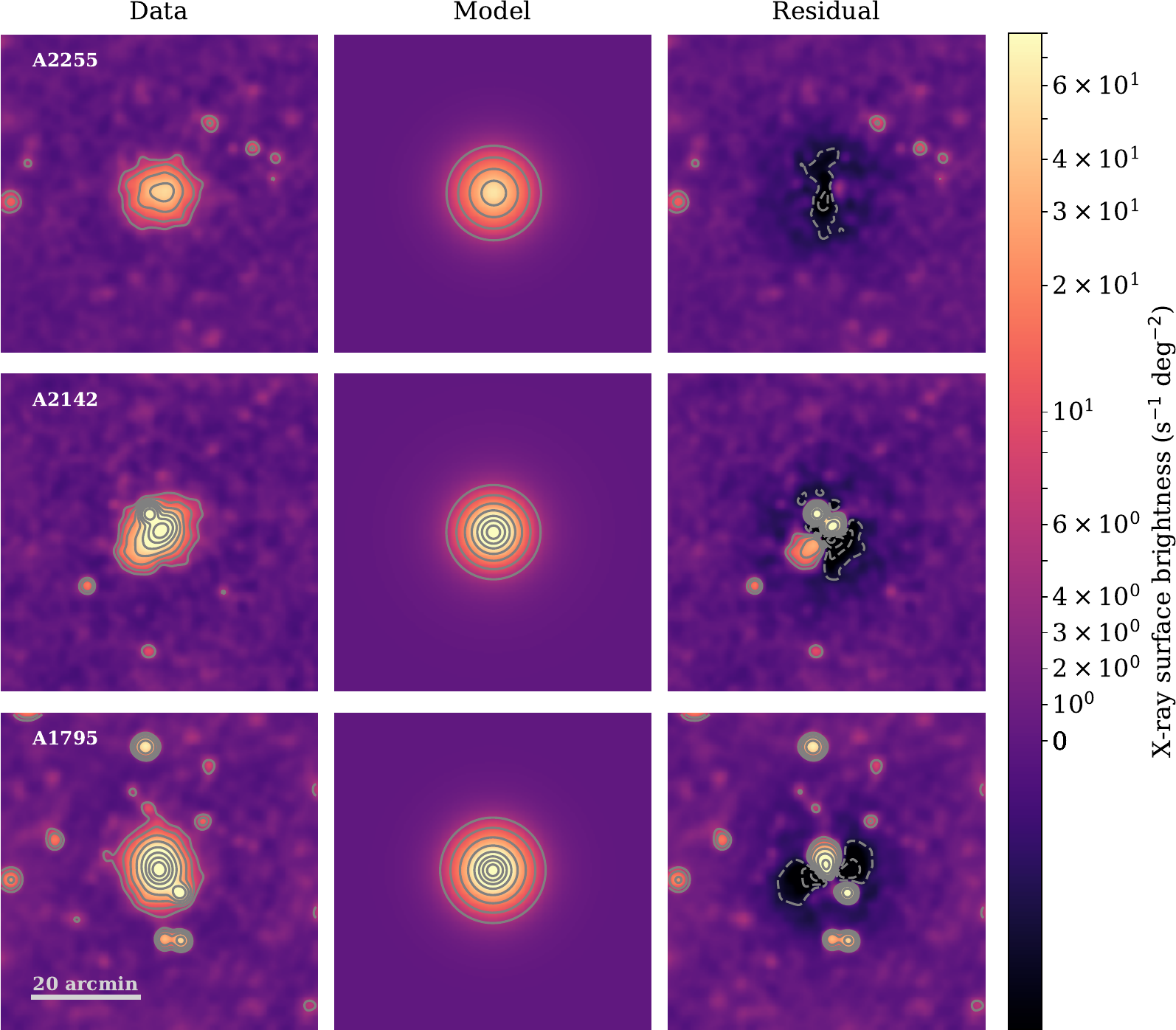}
\hspace{1cm}
\includegraphics[height=7.5cm]{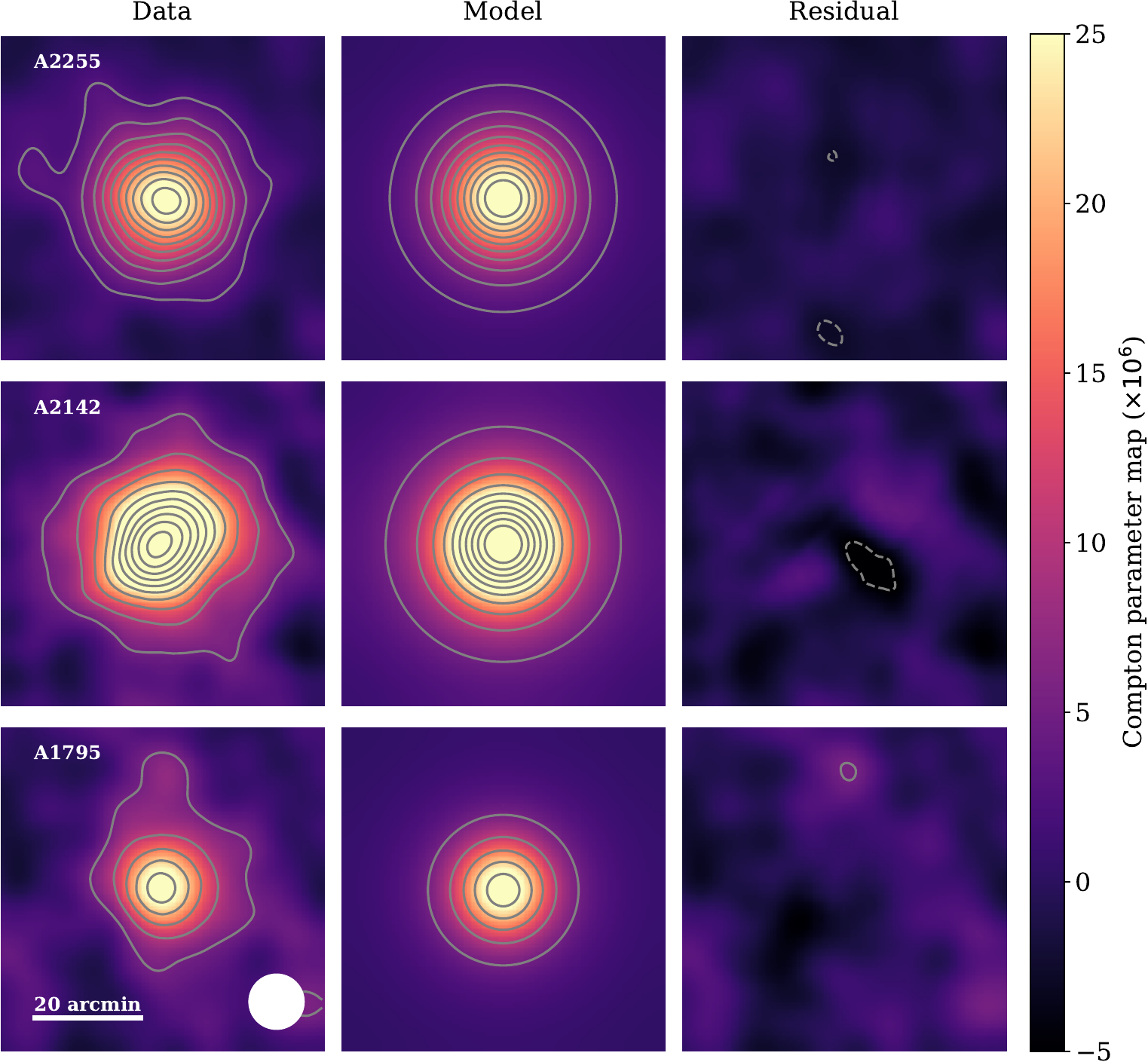}
\caption{Comparison between the thermal gas observables and the {\tt MINOT} model prediction. {\bf Left}: Comparison between \textit{ROSAT} X-ray images and the {\tt MINOT} model prediction. The scale is linear from -10 to 10 s$^{-1}$ deg$^{-2}$ and logarithmic above. Contours are as $\pm 5 \times 2^N$ s$^{-1}$ deg$^{-2}$, with $N$ the contour index starting at 0.
{\bf Right}: Comparison between the \textit{Planck} Compton parameter MILCA image and the {\tt MINOT} model prediction. The contours are multiples of $3 \sigma$, where $\sigma$ is the rms of the residual map. The \textit{Planck} 10 arcmin FWHM beam is shown as a filled white circle. All maps where smoothed (in addition the there intrinsic resolution) with a 2 arcmin FWHM gaussian beam for visual purpose. All the sky patches shown here are 1 degree $\times$ 1 degree.}
\label{fig:data_compare_thermal}
\end{figure*}

First we compared the measured X-ray luminosity given by \cite{Eckert2017} to the luminosity we recover, integrated within $R_{500}$. We used the same values of $R_{500}$ and set the {\tt MINOT} cosmological model to the one used in \cite{Eckert2017} to mitigate differences. The comparison of the obtained rest frame luminosity are given in Table~\ref{tab:table_X_flux}. We obtain comparable luminosities, with differences of a few percent for A1795 and A2142, but differences of up to 19\% for A2255. Because our model is based on the interpolation of the results by the XCOP project, we expect consistency between the two. However, our model has been defined by extrapolating the thermal plasma density profiles down to small radii. While these profile were measured down to about 10 kpc or even less for A1795 and A2142, it was only measured down to about 30 kpc for A2255. Thus, uncertainties coming from the extrapolation are likely to be larger for this cluster. The differences that we observe in Table~\ref{tab:table_X_flux} are thus likely due to extrapolation uncertainties.

We also compared the tSZ flux computed using {\tt MINOT} to the flux given in the \textit{Planck} PSZ2 catalog \citep{Planck2016XXVII}. Again, our model is based on the XCOP outputs, which are themselves obtained using \textit{Planck} data, so that we expect consistency. Table~\ref{tab:table_sz_flux} shows that the total integrated Compton parameter (i.e., computed within $5 R_{500}$) agrees well for A1795 and A2142, and it differs by 2.3 $\sigma$ for A2255. Because A2255 is a strongly merging cluster, the flux arising from the cluster outskirt is likely to be higher than for A1795 and A2142, and the truncation radius involved in {\tt MINOT} (set at 5 Mpc) may not be enough to account for all the tSZ flux. When we increase the truncation radius to $10 R_{500}$ ,  the flux rises to nearly $18 \times 10^{-3}$ arcmin$^2$, which better agrees with the PSZ2 value. As in the case of the X-ray luminosity, the differences that we observe are thus likely due to the interpolation and assumption that we have made in defining the clusters.

\begin{table}
        \caption{Comparison of the rest frame X-ray luminosity in the $[0.5-2]$ keV band, computed with {\tt MINOT} and with the code from \cite{Eckert2017}.}
        \begin{center}
        \begin{tabular}{c|cccc}
        \hline
        \hline
        Cluster & $L_{X,500}$, {\tt MINOT} & $L_{X,500}$, \cite{Eckert2017} \\
                            & \multicolumn{2}{c}{(10$^{44}$ erg s$^{-1}$)} \\
        \hline
        A1795 & 4.10 & $4.43\pm0.01$\\
        A2142 & 8.41 & $8.09\pm0.02$\\
        A2255 & 2.49 & $2.08\pm0.02$ \\
        \hline
        \end{tabular}
        \end{center}
        \label{tab:table_X_flux}
\end{table}

\begin{table}
        \caption{Comparison of the total tSZ flux, $Y_{5 R_{500}}$, obtained with {\tt MINOT} and with the PSZ2 catalog \citep{Planck2016XXVII}.}
        \begin{center}
        \begin{tabular}{c|cc}
        \hline
        \hline
        Cluster & $Y_{5 R_{500}}$, {\tt MINOT} &  $Y_{5 R_{500}}$, PSZ2 \\ 
                    & \multicolumn{2}{c}{(10$^{-3}$ arcmin$^{2}$)} \\
        \hline
        A1795 & 13.62 & $11.66\pm1.38$  \\
        A2142 & 33.31 & $33.62\pm2.70$  \\
        A2255 & 16.68 & $20.25\pm1.53$  \\
        \hline
        \end{tabular}
        \end{center}
        \label{tab:table_sz_flux}
\end{table}

In addition to the fluxes, we also directly compared the X-ray and tSZ images to existing data. We used the publicly available \textit{ROSAT} \citep{Truemper1993} X-ray pointed data\footnote{We used ObsID  rp800105n00, rp800096n00, and rp800512n00 for A1795, A2142, and A2255, respectively, see \url{https://heasarc.gsfc.nasa.gov/docs/rosat/rhp_archive.html}} obtained for our three targets to produce maps in the $[0.1,2.4]$ keV energy band. The maps were subtracted from the background and normalized by the exposure. The \textit{ROSAT} PSPC response matrices are accounted for in {\tt MINOT} through the use of {\tt XSPEC}, as well as the hydrogen column density taken at the location of each cluster, as obtained by the LAB survey \citep{Kalberla2005}. The model is projected on the same header as the original \textit{ROSAT} data and accounts for the \textit{ROSAT} effective area in units of counts per unit of time and solid angle. We accounted for the point spread function (PSF) by smoothing our model with a mean effective Gaussian function with a full width at half maximum (FWHM) of 30 arcsec, which is the typical number expected for \textit{ROSAT} pointed observations. While the detailed analysis of the X-ray data including all instrumental effects is beyond the scope of this work, this comparison already provides a useful qualitative comparison to our modeling. The data, model, and residual images are displayed in the left panel of Fig.~\ref{fig:data_compare_thermal}. The data and the model are shown on a log scale, and the residual is shown on a linear scale. While the overall agreement is good for all three clusters, several features are evident. First, many point sources that are not accounted for here affect the residual. Then, the real clusters are not perfectly azimuthally symmetric, which is shown for the residual with a positive and negative butterfly shape in the core of all targets. Finally, the model of A2255 slightly overpredicts the signal, which agrees with the prediction of the model luminosity, which is too high, as discussed above.

We also used the MILCA \citep{Hurier2013} Compton parameter map obtained from \textit{Planck} \citep{PlanckXXII2016} to compare our tSZ model to real data. We extracted a 1 degree $\times$ 1 degree sky patch centered on the individual clusters, and projected the {\tt MINOT} Compton parameter map on the same grid for comparison. We smoothed the model with a 10 arcmin FWHM Gaussian beam to account for the angular resolution of \textit{Planck} . In the right panel of Fig.~\ref{fig:data_compare_thermal} we show the data, the model, and the residual. The data and the model are shown on a log scale, and the residual is shown on a linear scale. The model and the data agree well for all three clusters. Nevertheless, a low excess in the central part of the A2142 model is evident, which can be explained by the fact that this cluster is slightly elongated, as is also seen in the \textit{ROSAT} image.

In conclusion, we have compared the prediction from {\tt MINOT} to the X-ray luminosity and tSZ fluxes, finding an overall good consistency between the two. The differences are likely explained by uncertainties in the model extrapolation. Similarly, {\tt MINOT} is able with dedicated functions to predict the X-ray and tSZ images associated with a cluster model. The comparison to \textit{ROSAT} and \textit{Planck} maps has shown an overall good consistency for the targets tested here. In the context of modeling the nonthermal component of galaxy clusters, it is therefore possible to use {\tt MINOT} together with X-ray or tSZ data to calibrate the thermal part of the model.

\subsection{Comparison to \textit{Fermi}-LAT constraints}
\begin{table*}
        \caption{Comparison of the $\gamma$-ray (hadronic) flux prediction obtained by \cite{Ackermann2014} and this work, using the same value for the CR-to-thermal energy ratio. We have converted the pressure ratio into an energy ratio as defined in this work. Fluxes are given for energies $E>500$ MeV, in units of s$^{-1}$ cm$^{-2}$. The integration radius was set to the truncation radius (total volume).  $U_{\rm th}$ was rescaled using Eq.~\ref{eq:rescaling}.}
        \begin{center}
        \begin{tabular}{c|c|cccccc}
        \hline
        \hline
        Cluster & $X_{\rm CRp}(R_{200})$ & \cite{Ackermann2014} & Reference model & Applying $U_{\rm th}$ rescaling & Gas density to $\beta$-model & \\
                     & -- & \multicolumn{4}{c}{(10$^{-10}$ cm$^{-2}$ s$^{-1}$)} \\
        \hline
        A1795 & 0.022 & 3.01 & 0.81 & 3.71 & 3.46 \\
        A2142 & 0.028 & 3.45 & 1.60 & 4.11 & 4.14 \\
        A2255 & 0.022 & 0.85 & 0.37 & 0.83 & 0.88 \\
        \hline
        \end{tabular}
        \end{center}
        \label{tab:table_flux_gamma}
\end{table*}

\begin{figure}
\centering
\includegraphics[width=0.42\textwidth]{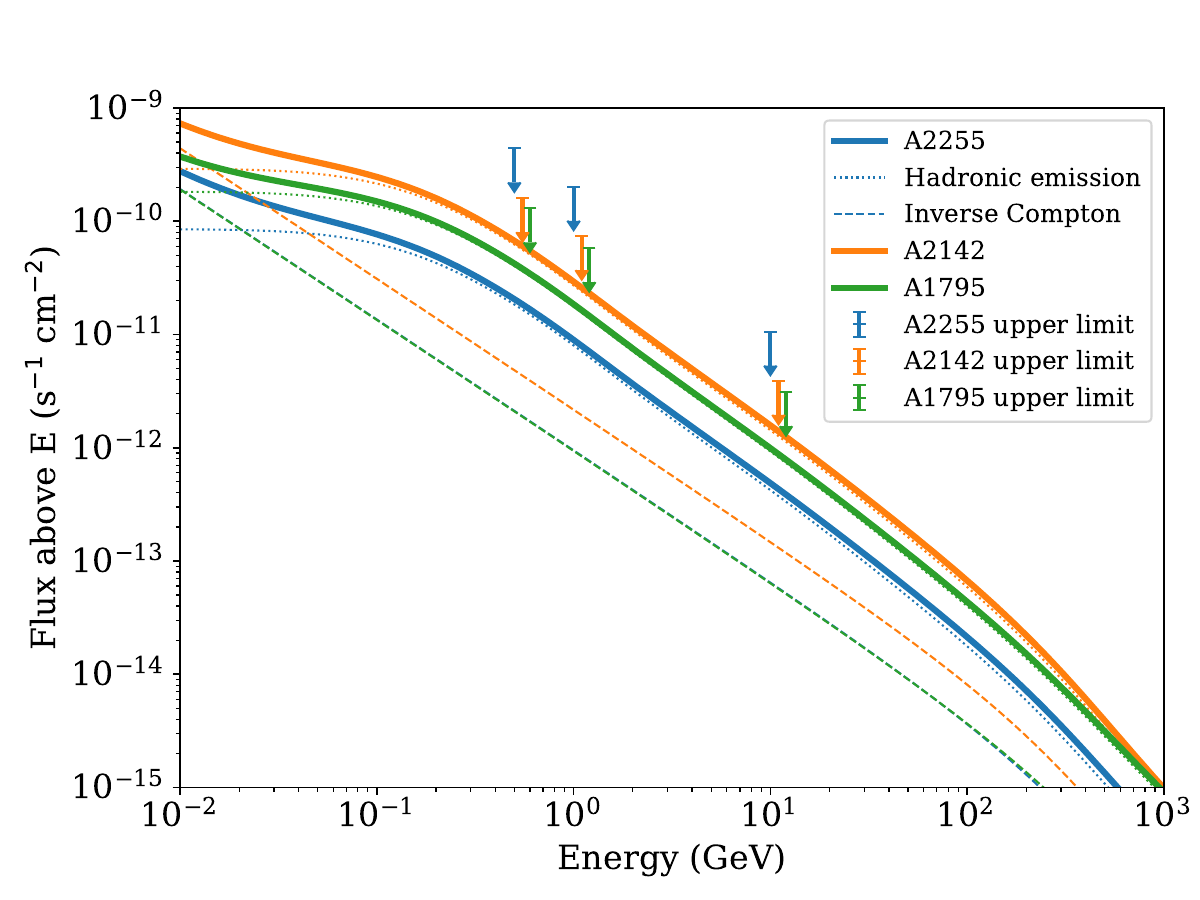}
\caption{Comparison between the cluster model flux predictions and the \textit{Fermi}-LAT upper limit from \cite{Ackermann2014}.}
\label{fig:data_compare_fermi}
\end{figure}

$\gamma$-ray constraints have been obtained by \cite{Ackermann2014} for A1795, A2142 and A2255 as part of a larger sample using \textit{Fermi}-LAT data \citep{Atwood2009}. For all clusters, they predicted the expected fraction of CRp pressure over the thermal pressure based on \cite{Pinzke2010} and \cite{Pinzke2011}, given the mass of the clusters. They used these predictions together with a model for the spatial distribution of the CRs to compute flux prediction for these clusters. In order to compare our predictions to theirs, we first set the value of $X_{\rm CRp}(R_{200})$ to that of \cite{Ackermann2014}, taking into account the fact that the pressure ratio is twice the energy ratio (which we used for our parametrization). Our baseline CRp density model matches their baseline well \citep[based on simulations,][]{Pinzke2010} because the CRs are tied to the gas, as they neglect CR transport.

Table~\ref{tab:table_flux_gamma} shows that our predictions are lower than those of \cite{Ackermann2014} by a factor of about 3. However, we note that the masses of our selected clusters used in \cite{Ackermann2014} that were taken from \cite{Chen2007}, are higher by a factor of 1.5-2.1 than ours. In order to account for this differences, we therefore rescaled our thermal energy according to self-similarity expectations as 
\begin{equation}
U_{\rm th} \rightarrow U_{\rm th} \left(\frac{M_{500, {\rm \ this \ work}}}{M_{500, {\rm \ Ackermann \ 2014}}}\right)^{-5/3}.
\label{eq:rescaling}
\end{equation} 
After we applied this rescaling, our fluxes agreed far better, within 20\%. The differences may arise from the $\gamma$-ray production rate modeling, scatter in the thermal pressure, or differences in the thermal gas distribution that are not necessarily the same. To determine the effect of the latter, we also computed our fluxes by changing our density profiles to the best-fit $\beta$-model of the true density. We find that this change leads to differences in the $\gamma$-ray flux of up to 7\% in the case of these clusters, which is significant.

In Fig.~\ref{fig:data_compare_fermi} we compute the energy-integrated flux as a function of energy in the case of our reference model (Section~\ref{sec:baseline_cluster_models}) and compare it to the upper limit set by \cite{Ackermann2014}. While the predictions are relatively close to the upper limit, they remain below it for all three clusters.

In conclusion, we have shown that our model gives comparable predictions for the $\gamma$-ray flux compared to what is used in the literature. However, significant differences may arise as a result of the inner structure modeling of the clusters, which is generally ignored when large samples are used.

\subsection{Comparison to radio observations}
\begin{figure*}
\centering
\includegraphics[width=0.90\textwidth]{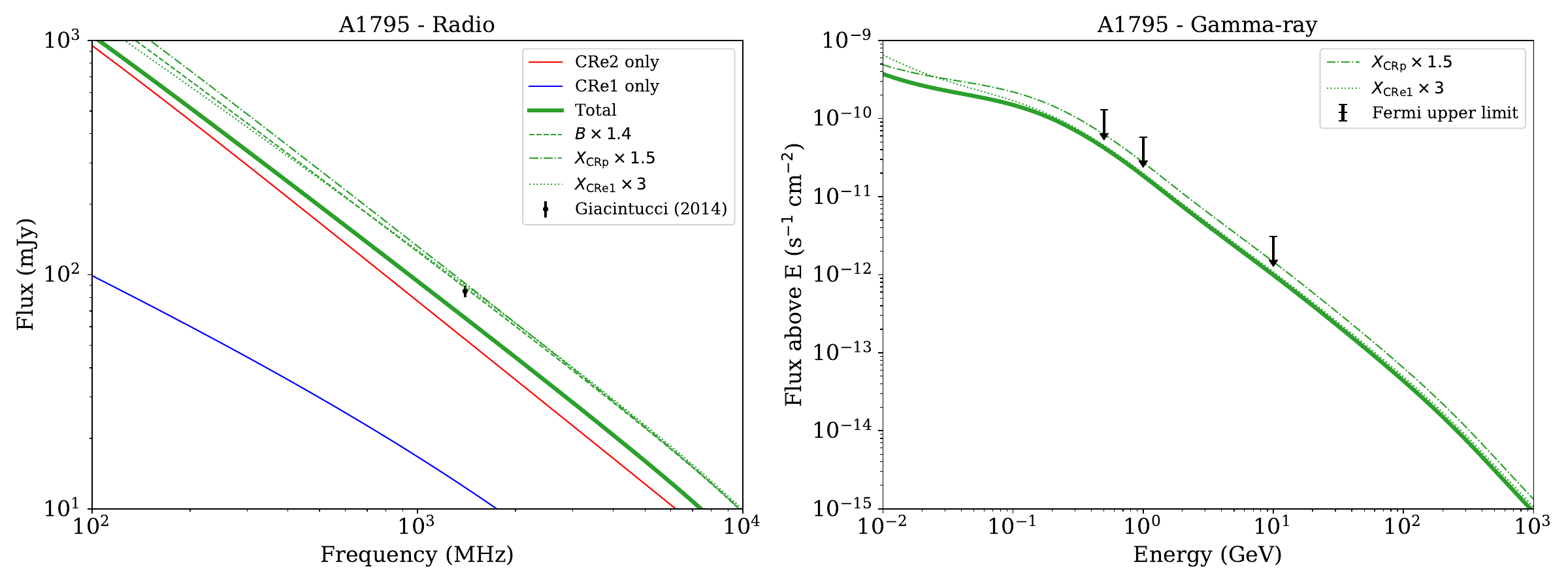}
\includegraphics[width=0.90\textwidth]{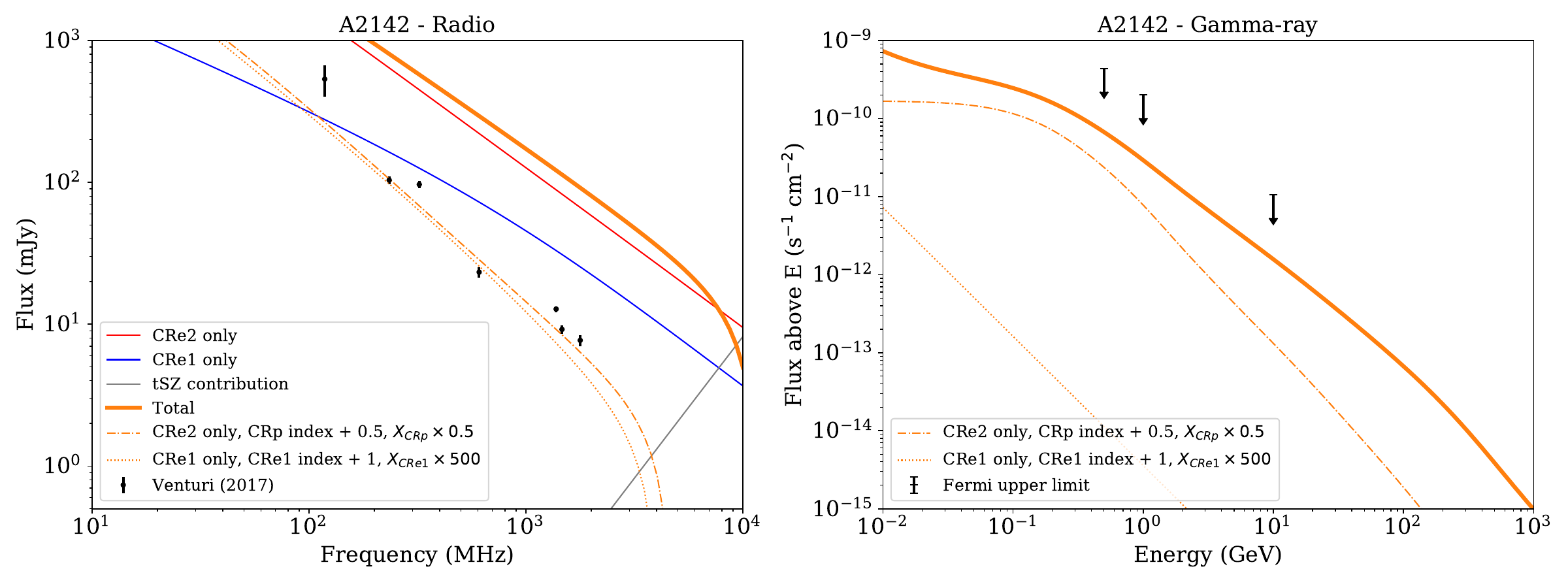}
\includegraphics[width=0.90\textwidth]{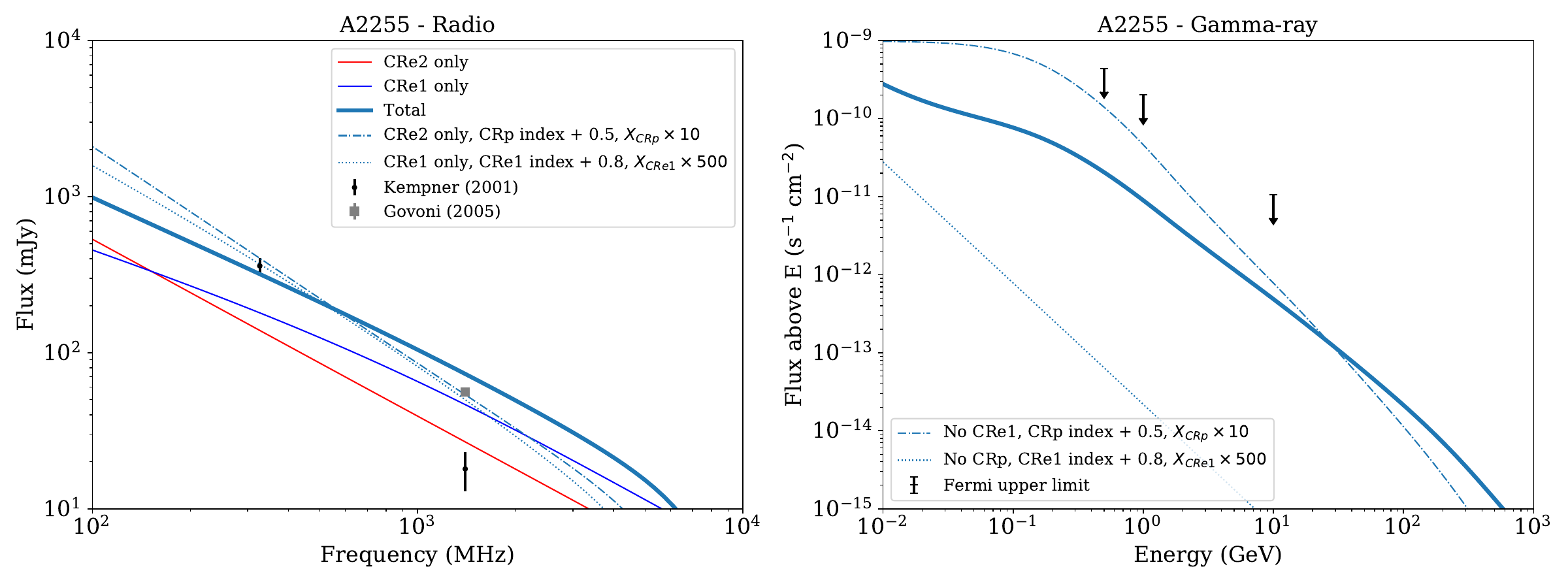}
\caption{Comparison of our model prediction and the radio flux observed in our three clusters (left), and the effect of the radio models on the $\gamma$-ray constraints (right). 
{\bf Top}: A1795.
{\bf Middle}: A2142.
{\bf Bottom}: A2255.
}
\label{fig:data_compare_radio}
\end{figure*}

In this section, we quantitatively compare the radio predictions of our model to measurements available in the literature. To do so, we used the database at {\url{https://galaxyclusters.hs.uni-hamburg.de/}}, in which available radio data for many clusters are listed. All of our target clusters present diffuse radio emission: a radio mini-halo for A1795 \citep{Giacintucci2014}, a giant radio halo for A2142 \citep{Giovannini2000,Venturi2017}, and a giant radio halo plus a relic for A2255 \citep{Kempner2001,Govoni2005}. In the case of A2142, we note that two components were distinguished in \cite{Venturi2017}, and they likely arise from different physical processes. However, in this qualitative comparison, we only considered the global emission and sum the contribution from the two components.

In Fig.~\ref{fig:data_compare_radio} we compare our models (defined in Section~\ref{sec:baseline_cluster_models}) to the flux measured in the literature. In order to compute the model flux emission, we used an aperture radius that matched the signal from the respective articles (100, 500, and 930 kpc, for A1795, A2142, and A2255, respectively) and performed cylindrical integration of the synchrotron emission. For each cluster, we illustrate how it is possible to qualitatively change our model parameters to match the radio data, and also show how these changes translate into the $\gamma$-ray prediction and its comparison to the upper limits of \cite{Ackermann2014}.

In the case of A1795, our default model underpredicts the radio flux by a factor of about 50\%. Increasing the magnetic field by a factor of 1.4 would solve the difference without affecting the $\gamma$-ray prediction. Alternatively, the number of CRp might be inceased by a factor of about 1.5, but a the cost of increasing the $\gamma$-ray prediction by a similar amount and approaching the \textit{Fermi}-LAT limit. Finally, increasing the number of CRe$_1$ by a factor of 3 would also solve the difference, with changes in the $\gamma$-ray prediction only at energies below 0.1 GeV, which is  barely accessible for \textit{Fermi}-LAT.

In the case of A2142, our default model overpredicts the radio emission by almost an order of magnitude, and our spectrum appears too flat compared to the data. First, we considered a purely hadronic scenario (i.e., no CRe$_1$), for which a slope of the CRp spectrum needs to be set to about 2.9 and the amplitude reduced by a factor of two in order to match the data. Another option is to consider only CRe$_1$, in which case the spectrum slope should be set to about 3.3 and the normalization increased by a factor of 500 to match the data. A combination of the two scenarii might also be used, especially because the radio emission presents two distinct components \citep[see][for discussions]{Venturi2017}. In both scenarii, the predicted $\gamma$-ray emission also decreases, and thus remains in agreement with the \textit{Fermi}-LAT limit.

In the case of A2255, we observe a disagreement between the 1400 GHz flux from \cite{Kempner2001} and \cite{Govoni2005}. For our purpose, we chose to use the value of \cite{Govoni2005} as a reference (flux of the halo alone, without the relic). The amplitude of our default model broadly agrees with the observation, but our spectral index is too low. In the purely hadronic scenario, increasing the slope of the CRp spectrum to 2.9 together with increasing its normalization by a factor of 10 would bring agreement between the model and the data. Alternatively, in a model with only CRe$_1$, we would need to increase the CR spectrum slope to 3.1 and multiply the normalization by 500 in order to reach agreement. In the two cases, the $\gamma$-ray flux remains below the \textit{Fermi}-LAT upper limit.

In this section, we showed that we can change the model parameters to match the radio data in a qualitative way. We focused on the slope and normalization of the CR content of the cluster, but opening the parameter space to spatial distributions or a functional form of the spectra might also play an important role. In all the considered cases, it is not possible to rule out any model using $\gamma$-ray data because the limits remain too high. However, purely hadronic model predictions are just a factor of a few below the \textit{Fermi}-LAT limits in the case of these clusters.

\section{Conclusions and summary}\label{sec:conclusions}
We have provided an exhaustive description of {\tt MINOT}, a new software dedicated to the modeling the nonthermal components of galaxy clusters and predicting associated observables. While the software was originally developed to describe the $\gamma$-ray emission from galaxy clusters, {\tt MINOT} also accounts for most of the emission associated with the diffuse ICM component: X-rays from thermal bremsstrahlung, tSZ signal in the millimeter band, $\gamma$-rays and neutrino emission form hadronic processes, $\gamma$-rays from inverse Compton emission, and radio synchrotron emission. Because the $\gamma$-ray emission is connected to the same underlying cluster physical properties as these other observables, {\tt MNOS} provides a useful self-consistent modeling of the signal, and these additional observables can be used, for example, to calibrate a $\gamma$-ray model. However, {\tt MINOT} can also be used to independently model observables in the different bands.

The software is made publicly available at \url{https://github.com/remi-adam/minot}, and this paper aims at providing a reference for any user of the code. To this aim, we have discussed the structure of the code and the interdependencies of the different modules in Section~\ref{sec:overview}, while Sections~\ref{sec:physical_modeling}, \ref{sec:particle_interactions}, and~\ref{sec:observables} provided details about the physical processes considered, how they are accounted for, and the way observables are computed. The different functions were illustrated with the use of a reference cluster model. It allowed us to show the dependence of each wavelength on the physical properties of the cluster. In Section~\ref{sec:comparison} we also compared the predictions from {\tt MINOT} to data available in the literature in order to show how the code can be used to model real data. We used \textit{Planck} and \textit{ROSAT} data for the thermal component, and \textit{Fermi}-LAT plus various radio data for the nonthermal component. Finally, we note that the {\tt MINOT} code is well documented, and many examples are provided in the public repository.

The different assumptions made in the code were discussed. In particular, the modeling relies on primary base quantities that are used to derive secondary properties of the cluster and generate observables under the assumption of spherical symmetry. The primary quantities are the thermal electron pressure and density profiles, the CRe$_1$ and CRp profiles and spectra, and the profile of the magnetic field strength. Regarding the CRe$_2$, they are processed assuming no diffusion in a steady-state scenario. However, other electron populations can be accounted for using the CRe$_1$. In order to set the base physical properties of the cluster, several predefined models are available, but it is also possible to provide any user-defined quantity.

The accuracy of the modeling was addressed. Regarding the thermal component, modeling uncertainties associated with X-rays are below the percent level, and the uncertainties of the tSZ signal are at the percent level when a relativistic correction at high temperature is considered and much smaller otherwise. The  hadronic processes ($\gamma$-rays, neutrinos, and secondary electrons), on the other hand, present uncertainties at a level of typically 25\% above the considered energy range when the latest models available in the literature are considered. In addition, we stress that the effect of helium is about 50\% of the signal and should be accounted for (as done in {\tt MINOT}). The computation of inverse Compton and synchrotron emission is based on analytical approximations whose precision is expected to remain within 1\% and 0.2\%, respectively. Nevertheless, we stress that the main limitations of the modeling is not the accuracy of the computation, but the underlying assumptions discussed above. In particular the use of spherical symmetry and the assumption of stationarity to compute the distribution of secondary electrons are likely to be the dominant sources of mismodeling.

The {\tt MINOT} software can be used for a wide variety of operations, and we list just a few examples here:
\begin{itemize}
\item The joint modeling of the nonthermal emission in galaxy clusters for which detailed multi-wavelength data are available. The parameters of the model can be fit jointly to such data to constrain different scenarios while accounting for uncertainties in the different components.
\item The prediction of the expected signal, based on ancillary data, for observation proposals. For instance, it is possible to predict the tSZ emission associated with that of an X-ray observed cluster, assuming that the pressure follows a universal profile.
\item The prediction of the background CR induced $\gamma$-rays in the context of dark matter searches.  {\tt MINOT} provides an easy way to model the CR background, which needs to be marginalized over to obtain constraints on the nature of dark matter.
\item The simulation of sky maps associated with the observables considered here, given a halo catalog.
\item Pedagogical purposes. Because it includes most of the ICM associated processes, {\tt MINOT} can be used to understand the effect of some given parameters on the observable.
\end{itemize}

Historically, the understanding of the physical properties of galaxy clusters has strongly benefited from multiwavelength observations and analysis. With the current and upcoming facilities aiming at exploring the nonthermal component of galaxy clusters, in particular in the radio and $\gamma$-ray bands, it has become very useful to have an easy-to-use self-consistent modeling software that allows us to predict the expected signal based on some assumptions. The {\tt MINOT} software provides such a tool.

\begin{acknowledgements}
We are thankful to the anonymous referee for useful comments that helped improve the quality of the paper.
We would like to thank R. Alves Batista for useful comments.
The work of JPR and MASC was supported by the Spanish Agencia Estatal de Investigaci\'on through the grants PGC2018-095161-B-I00 and IFT Centro de Excelencia Severo Ochoa SEV-2016-0597, and the {\it Atracci\'on de Talento} contract no. 2016-T1/TIC-1542 granted by the Comunidad de Madrid in Spain.
We acknowledge the use of HEALPix \citep{Gorski2005}.
This research made use of Astropy, a community-developed core Python package for Astronomy \citep{Astropy2013}, in addition to NumPy \citep{VanDerWalt2011}, SciPy \citep{Jones2001} and Ipython \citep{Perez2007}.
Figures were generated using Matplotlib \citep{Hunter2007}.
Several modules of {\tt MINOT} are based on the {\tt Naima} software \citep{Zabalza2015}.
\end{acknowledgements}

\bibliographystyle{aa}
\bibliography{minot_biblio}

\end{document}